\documentclass[twocolumn]{aastex61}

\usepackage{amsmath}
\usepackage{color}
\usepackage{graphicx}
\usepackage{verbatim}

\newcommand{\eprvec}{EPRV3 Evidence Challenge}
\newcommand{\sw}[1]{\textsc{#1}}
\newcommand{\MULTINEST}{\sw{MultiNest}}

\newcommand{\logZ}{\ensuremath{\log\mathcal{\widehat{Z}}}}
\newcommand{\medianlogZ}{\ensuremath{\langle \logZ \rangle}}
\newcommand{\sigZ}{\ensuremath{\sigma_\mathcal{\widehat{Z}}}}
\newcommand{\siglogZ}{\ensuremath{\sigma_{\log\mathcal{\widehat{Z}}}}}
\newcommand{\DZ}{\ensuremath{D_{\mathcal{\widehat{Z}}}}}
\newcommand{\DlogZ}{\ensuremath{D_{\log\mathcal{\widehat{Z}}}}}

\shorttitle{Evidence for a Planet}
\shortauthors{Nelson et al.}

\begin{document}

\title{Quantifying the Bayesian Evidence for a Planet in Radial Velocity Data}


\author{Benjamin E. Nelson}
\affil{Center for Interdisciplinary Exploration and Research in Astrophysics (CIERA), Department of Physics and Astronomy, Northwestern University, 2145 Sheridan Road, Evanston, IL 60208, USA}
\affil{Northwestern Institute for Complex Systems, 600 Foster Street, Evanston, IL 60208, USA}

\author{Eric B. Ford}
\affil{Center for Exoplanets and Habitable Worlds, The Pennsylvania State University, 525 Davey Laboratory, University Park, PA, 16802, USA}
\affil{Department of Astronomy \& Astrophysics, The Pennsylvania State  University, 525 Davey Laboratory, University Park, PA 16802, USA}
\affil{Institute for CyberScience, The Pennsylvania State  University, 525 Davey Laboratory, University Park, PA 16802, USA}
\affil{Center for Astrostatistics, The Pennsylvania State  University, 525 Davey Laboratory, University Park, PA 16802, USA}

\author{Johannes Buchner}
\affil{Millenium Institute of Astrophysics, Vicu\~{n}a MacKenna 4860, 7820436 Macul, Santiago, Chile}
\affil{Pontificia Universidad Católica de Chile, Instituto de Astrofísica, Casilla 306, Santiago 22, Chile}
\affil{Excellence Cluster Universe, Boltzmannstr. 2, D-85748, Garching, Germany}
\affil{Max Planck Institute for Extraterrestrial Physics, Gie{\ss}enbachstra{\ss}se 1, D-85748 Garching bei M{\"u}nchen, Germany}

\author{Ryan Cloutier}
\affil{Department of Astronomy \& Astrophysics, University of Toronto. 50 St. George Street, Toronto, Ontario, M5S 3H4, Canada}
\affil{Centre for Planetary Sciences, Department of Physical \& Environmental Sciences, University of Toronto Scarborough. 1265 Military Trail, Toronto, Ontario, M1C 1A4, Canada}
\affil{Institut de recherche sur les exoplan\`etes, D\'epartement de physique, Universit\'e de Montr\'eal. 2900 boul. \'Edouard-Montpetit, Montr\'eal, Quebec, H3T 1J4, Canada}

\author{Rodrigo F. D\'iaz}
\affil{Universidad de Buenos Aires, Facultad de Ciencias Exactas y Naturales. Buenos Aires C1428, Argentina.}
\affil{CONICET - Universidad de Buenos Aires. Instituto de Astronom\'ia y F\'isica del Espacio (IAFE). Buenos Aires C1428, Argentina.}

\author{Jo\~ao P. Faria}
\affil{Instituto\,de\,Astrof\'isica\,e\,Ci\^encias\,do\,Espa\c{c}o,\,%
       Universidade do Porto, CAUP, Rua das Estrelas, 4150-762 Porto, Portugal}
\affil{Departamento\,de\,F\'isica\,e\,Astronomia,\,Faculdade\,de\,Ci\^encias,\,%
       Universidade\,do\,Porto,\,Rua\,do\,Campo\,Alegre,\,4169-007\,Porto,\,Portugal}

\author{Nathan C. Hara}
\affil{Observatoire de Gen\`eve, Universit\'e de Gen\`eve, 51 ch. des Maillettes, 1290 Versoix, Switzerland}
\affil{NCCR PlanetS CHEOPS Fellow, Switzerland}   

\author{Vinesh M.~Rajpaul}
\affil{University of Cambridge, Astrophysics Group, Cavendish Laboratory, J.\ J.\ Thomson Avenue, Cambridge CB3 0HE, UK}

\author{Surangkhana Rukdee}
\affil{Pontificia Universidad Cat\'olica de Chile, Center of Astro-engineering UC-AIUC,  Avda. Vicu\~{n}a Mackenna 4860, Macul, Santiago, Chile}



\newcommand{\BENELSON}[1]{{\textbf{\textcolor{black}{BENELSON: #1}}}}
\newcommand{\REWRITE}[1]{#1}
\newcommand{\REWRITEg}[1]{#1}

\begin{abstract}
We present results from a data challenge posed to the radial velocity (RV) community: namely, to quantify the \REWRITE{Bayesian} ``evidence'' for $n=\{0, 1, 2, 3\}$ planets in a set of synthetically generated RV datasets containing a range of planet signals.
Participating teams were provided the same likelihood function and set of priors to use in their analysis.
They applied a variety of methods to estimate $\widehat{\mathcal{Z}}$, the marginal likelihood for each $n$-planet model, including cross-validation, the Laplace approximation, importance sampling, and nested sampling.
We found the dispersion in $\widehat{\mathcal{Z}}$ across different methods grew with increasing $n$-planet models: $\sim\,$3 for 0-planets, $\sim\,$10 for 1-planet, $\sim\,$$10^2$-$10^3$ for 2-planets, and $>\,$$10^4$ for 3-planets.
Most internal estimates of uncertainty in $\widehat{\mathcal{Z}}$ for individual methods significantly underestimated the observed dispersion across all methods.
Methods that adopted a Monte Carlo approach by comparing estimates from multiple runs yielded plausible uncertainties.
Finally, two classes of numerical algorithms (those based on importance and nested samplers) arrived at similar conclusions regarding the ratio of $\widehat{\mathcal{Z}}$s for $n$ and ($n+1$)-planet models.
One analytic method (the Laplace approximation) demonstrated comparable performance.
We express both optimism and caution: we demonstrate that it is practical to perform rigorous Bayesian model comparison for $\leq$3-planet models, yet robust planet discoveries require researchers to better understand the uncertainty in $\widehat{\mathcal{Z}}$ and its connections to model selection. 
%
\end{abstract}




\section{Introduction} \label{sec:intro}
Early Doppler surveys of nearby solar-like stars provided the first census of exoplanet systems.
Relatively massive and short orbital period planets with strong radial velocity (RV) signals made up most of this sample, but instrumental upgrades and extended monitoring facilitated the detection of lower mass and longer period planets.
State-of-the-art RV instruments can reach precisions better than 1 m/s, and continued improvements in spectrograph technologies and stellar modeling \REWRITEg{\citep[see review by][]{Fisher16review}} hope to achieve a precision sufficient to detect an exo-Earth, an Earth-mass planet orbiting at the habitable zone distances from their host stars.
This is roughly 10 cm/s for a Solar mass star.

The journey to this milestone has been fraught with methodological and astrophysical hurdles.
One of the most notable are new stellar processes that emerged at the $\sim$1 m/s level, including but not limited to starspots rotating in and out of view, plages, granulation, stellar oscillations, and long-term stellar activity cycles \citep{bastien2014, cegla2014, haywood2014}.
Some of these nuisance signals have been previously mistaken as low mass and/or long period planets, until follow-up photometric or spectroscopic activity measurements could explain the observed periodicities otherwise \citep[e.g.,][]{robertson2014, kane2016}.
In some cases, false positive detections can arise from aliases in the RV time series itself \citep[e.g.,][]{dawsonfabrycky2010, rajpaul2016}.

In light of these challenges, the RV community needs to improve their analysis of RV data.
\citet{dumusque2016} and \citet{dumusque2017} posed a data challenge to the RV community, in which teams had to disentangle planetary signals from other nuisance signals using a set of synthetically generated RV data and activity indicators (bisector span, full width at half maximum of the cross-correlation function, the calcium activity index $\log{R}^\prime\{hk\}$) and whatever methods they deemed appropriate.
Methods that performed best took into account activity indicators, incorporated correlated noise models, and imposed some kind of Bayesian framework.
In the longer term, many groups have strayed from a traditional frequentist framework, which attempts to reject the null hypothesis of a no-planet model being compatible with the RV data, and experimented with various algorithms to compute a quantitative evidence for $n$ versus $n+1$ planets.
\REWRITE{The Bayesian ``evidence'' refers to the fully marginalized likelihood, i.e.,}
\begin{equation}
\mathcal{Z} \equiv p(\vec{d}|\mathcal{M}) = \int p(\vec{d}|\vec{\theta},\mathcal{M})p(\vec{\theta}|\mathcal{M})d\vec{\theta}
\label{eq-fml}
\end{equation}
where $\vec{d}$ is a set of real velocity data, $\mathcal{M}$ is the underlying physical and noise model, and $\vec{\theta}$ is the set of model parameters that describe $\mathcal{M}$.
For two models $\mathcal{M}_1$ and $\mathcal{M}_2$, one can update $p(\mathcal{M}_1)/p(\mathcal{M}_2)$ (the ratio of model prior beliefs) with $p(\vec{d}|\mathcal{M}_1) / p(\vec{d}|\mathcal{M}_2)$ (the Bayes factor) to calculate $p(\mathcal{M}_1|\vec{d}) / p(\mathcal{M}_2|\vec{d})$ (the posterior odds ratio\REWRITEg{, POR}).

\REWRITE{The art of exoplanet detection ultimately comes down to a decision on whether or not the data support the existence of a planet. The Bayes factor can be interpreted against empirical scales \citep{jeffreys1961theory,kassraftery1995}, see for example \cite{gregory2007}. However, to make decisions with known false positive and false negative rates, thresholds on $B$  (or correspondingly POR) need to be calibrated with extensive simulations.}

\REWRITE{In this work, we focus on the preliminary step towards such Bayesian exoplanet inference: numerically reliable computation of $\mathcal{Z}$.
In particular, we would like to know if different methods converge to similar conclusions about the evidence for $n$-planets, given the exact same datasets and assuming the exact same noise model and prior beliefs?
Some examples in RV of methods for computing the Bayesian evidence include thermodynamic integration \citep{gregory2007}, nested sampling \citep{ferozhobson2014}, geometric path Monte Carlo \citep{hou2014}, transdimensional MCMC \citep{brewerdonovan2015}, and importance sampling \citep{nelson2016, jenkins2017}.
The above studies were applied to real RV data for systems with suspect planets. The methods were not developed in the same context; each study considered a different RV dataset, noise model, and set of $n$-planet hypotheses, so the relative strengths of these model comparison algorithms are largely unknown.
\citet{fordgregory2007} compared several methods for 0 and 1-planet models and \citet{guo2012} applied some promising methods to multi-planet systems. }

Inspired by these previous studies, we designed a data challenge for the RV community to compare different algorithms and implementations for performing model comparison.
Participants were given six synthetic RV datasets and a set of $n$-planet models, where $n=\{0, 1, 2, 3\}$.
They were asked to compute quantitative estimates for $\mathcal{Z}$ ($\widehat{\mathcal{Z}}$, henceforth) for each model and their respective uncertainties using whatever computational methods and simplifying assumptions that they choose.
This challenge took place in association with a breakout session at The Third Workshop on Extremely Precise Radial Velocities at The Pennsylvania State University in 2017, August 14 to 17 (EPRV3, henceforth).
Some teams participated remotely, while others exchanged ideas during the breakout sessions.

There are four questions we hope to answer for the \eprvec:
\begin{itemize}
\item What is the dispersion in reported $\widehat{\mathcal{Z}}$s (i.e., \DZ) and how does it change with increasing model complexity (i.e., number of planets)?
\item Does each method's reported uncertainty in $\widehat{\mathcal{Z}}$ (i.e., \sigZ) accurately reflect the observed dispersion?
\item How does \DZ\, and \sigZ\, affect our ability to favor $n$ versus ($n+1$)-planet models for different datasets?
\item Within the context of this study, which methods should be recommended, avoided, and/or further developed?
\end{itemize}

This paper summarizes the results of the data challenge.
In \S \ref{sec:models}, we present the assumed observational and statistical models.
In \S \ref{sec:methods}, we present brief summaries of the different methods that teams employed.
In \S \ref{sec:results}, we compare everyone's results across many parameters of interest.
Finally in \S \ref{sec:discussion}, we discuss the relative strengths of these methods in the context of the challenge.
We reserve a set of variable names to be used throughout the paper, described in Table \ref{tbl-notation}.

\section{Observational and Statistical Models} \label{sec:models}
Each participating team used a standardized set of assumptions for the physical and statistical models.
Here, we describe the process used to generate the datasets in detail.

We provided six simulated datasets.
The datasets were generated with a set of consistent properties: 1. each dataset was an RV time series, including the times of observations ($\vec{t}$), the ``measured'' RVs ($\vec{v}$), and the measurement uncertainties ($\vec{\sigma}$); 2. the number of observations was fixed at 200; 3. the data were drawn over an observing baseline of 600 days; and 4. each dataset included a single velocity offset and correlated Gaussian noise to model stellar activity.
We also injected two planets into each dataset with a wide range of orbital and mass properties to be described in \S \ref{models-phys}.

\begin{deluxetable}{c|c}
\tablecaption{Common variable names used throughout the manuscript. \label{tbl-notation}}
\tablehead{
\colhead{Variable} & \colhead{Description}
}
\startdata \hline
$\mathcal{M}_n$ & The Radial Velocity Model with $n$ planets \\ \hline \hline
$\vec{d}$ & The Radial Velocity Data \\ \hline
$\vec{t}$ & times \\
$\vec{v}$ & radial velocities \\
$\vec{\sigma}$ & radial velocity uncertainties \\ \hline \hline
$\vec{\theta}$ & The Model Parameters \\ \hline
$P_i$ & orbital period for $i$th planet \\
$K_i$ & RV semi-amplitude for $i$th planet \\
$e_i$ & eccentricity for $i$th planet \\
$\omega_i$ & argument of pericenter for $i$th planet \\
$M_i$ & mean anomaly for $i$th planet at a fixed epoch \\
$C$ & RV zero-point offset \\
$\sigma_{J}$ & RV jitter \\
$\alpha$ & amplitude of $\kappa$ \\
$\lambda_e$ & scale length of exponential component of $\kappa$ \\
$\lambda_p$ & scale length of periodic component of $\kappa$ \\
$\tau$ & period of periodic component of $\kappa$ \\ 
$n$ & number of planets \\ \hline \hline
 & Statistical Parameters \\ \hline
$\mathcal{Z}$ & the fully marginalized likelihood \\
$\mathcal{L(\vec{\theta})}$, $p(\vec{d}|\vec{\theta})$ & the likelihood function \\
$p(\vec{\theta})$ & the joint prior probability distribution \\
$\Sigma$ & covariance matrix in likelihood function \\
$\kappa$ & quasi-periodic kernel defined by $\alpha$, $\lambda_e$, $\lambda_p$, $\tau$ \\ \hline \hline
 & Meta-Analysis Parameters \\ \hline
$\widehat{\mathcal{Z}}$ & estimate of the fully marginalized likelihood \\ 
\sigZ, \siglogZ & uncertainty in each $\mathcal{\widehat{Z}}$ and $\logZ$ respectively \\
\DZ, \DlogZ & dispersion in $\mathcal{\widehat{Z}}$ and $\logZ$ respectively \\
\enddata
\end{deluxetable}

\subsection{Physical Model}
\label{models-phys}
In each dataset, the RV of the star was computed via $n$-body integrations using Newtonian gravity, one star and two planets.
While the full model formally included mutual planetary interactions, we fully expect that it would be well-described by the linear super-position of two Keplerian orbits plus a constant velocity offset and a noise term.
We estimate the difference between these two assumptions to be less than a couple cm/s across all datasets.

The simulation returned a set of line-of-sight velocities of the star $\vec{v}_{\mathrm{pred}}(\vec{t}|\vec{\theta})$ for a set of input times $\vec{t}$ and mass/orbital parameters $\vec{\theta}$.
For the sake of computational efficiency, we restricted the range of injected planet orbital periods to between 10 and 2,400 days.
Table \ref{tbl-planets} describes the orbital and mass properties of each pair of planets, along with each dataset's input zero-point offset and jitter.
Note that Datasets 3 and 6 have the exact same injected planets, but the zero-point offset, time series, and noise realizations are different.

\begin{deluxetable*}{c|c|c|c|c|c|c|c|c}
\label{tbl-planets}
\tablecaption{Simulated planet properties. Each dataset contains two planets with a variety of orbits and masses, which we also summarize with their supposed level of detectability (to be referenced again in Figures \ref{fig:results-logZ} and \ref{fig:results-logPOR}). Note that Datasets 3 and 6 have the same injected planets.
}
\tablehead{
\colhead{Dataset Number} & \colhead{Detectability} & \colhead{$P$ (days)} & \colhead{$K$ (m/s)} & \colhead{$e$ (unitless)} & \colhead{$\omega$ (rad)} & \colhead{$M$ (rad)} & \colhead{$C$ (m/s)} & \colhead{$\sigma_J$ (m/s)}
}
\startdata \hline
1 & easy & 12.1 & 1.86 & 0.08 & 0.0 & 0.87 & 1.46 & 0.6 \\
& easy & 42.4 & 2.44 & 0.04 & 2.0 & 2.99 & & \\ \hline
2 & easy & 15.96 & 2.12 & 0.05 & 0.1 & 0.18 & 6.33 & 0.6 \\
& difficult & 120.5 & 1.36 & 0.31 & 1.3 & 0.82 & & \\ \hline
3 & difficult & 40.4 & 1.25 & 0.1 & 3.0 & 4.16 & -8.28 & 0.6 \\
& difficult & 91.9 & 1.19 & 0.1 & 0.3 & 0.33 & & \\ \hline
4 & easy & 169.1 & 1.58 & 0.22 & 2.1 & 0.06 & -6.23 & 0.6 \\
& impractical & 23.45 & 0.74 & 0.04 & 6.5 & 4.37 & & \\ \hline
5 & difficult & 31.1 & 0.75 & 0.04 & 0.2 & 3.31 & -4.55 & 0.6 \\
& impractical & 10.9 & 0.67 & 0.02 & 6.2 & 4.14 & & \\ \hline
6 & difficult & 40.4 & 1.25 & 0.1 & 3.0 & 4.16 & -10.7 & 0.6 \\
& difficult & 91.9 & 1.19 & 0.1 & 0.3 & 0.33 & & \\ \hline
\enddata
\end{deluxetable*}

We designed these six datasets with a range of planet detectability in mind.
Some planetary signals were relatively easy to identify ($K/\sigma > 1$), which may facilitate efficient computation of $\widehat{\mathcal{Z}}$.
Some were relatively difficult ($K/\sigma \sim 1$) or nearly impractical ($K/\sigma < 1$) to find, which could lead to challenging $\widehat{\mathcal{Z}}$ calculations.
To reiterate, the main purpose of this challenge is to determine how accurately different algorithms can compute the evidence of $n$-planets in RV data, not their ability to disentangle real planets from astrophysical noise.
However, we are interested in how the variation in teams' calculations of $\widehat{\mathcal{Z}}$ depends on the strength of a supposed planetary signal.

\begin{figure*}
\centering
\begin{tabular}{cc}
\includegraphics[scale=0.33]{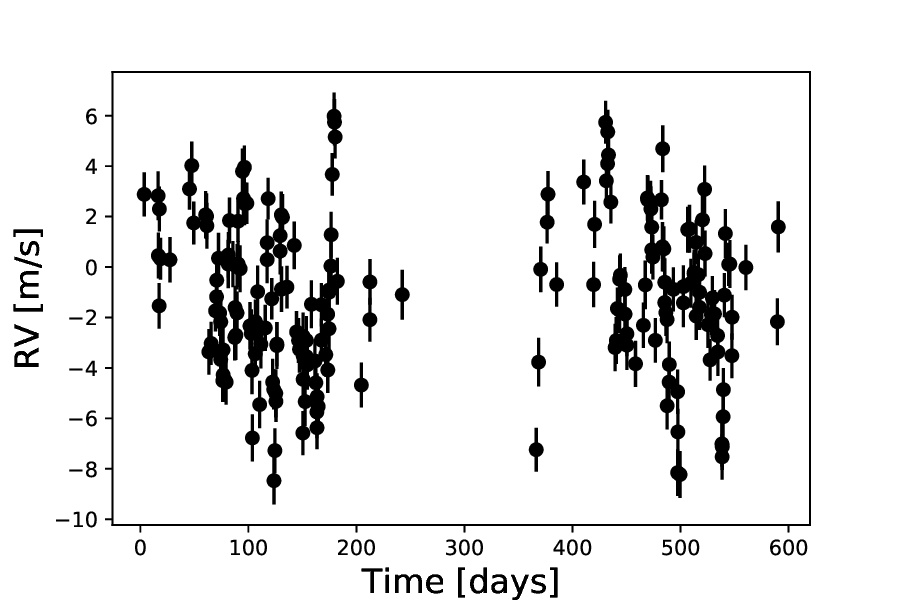} & \includegraphics[scale=0.33]{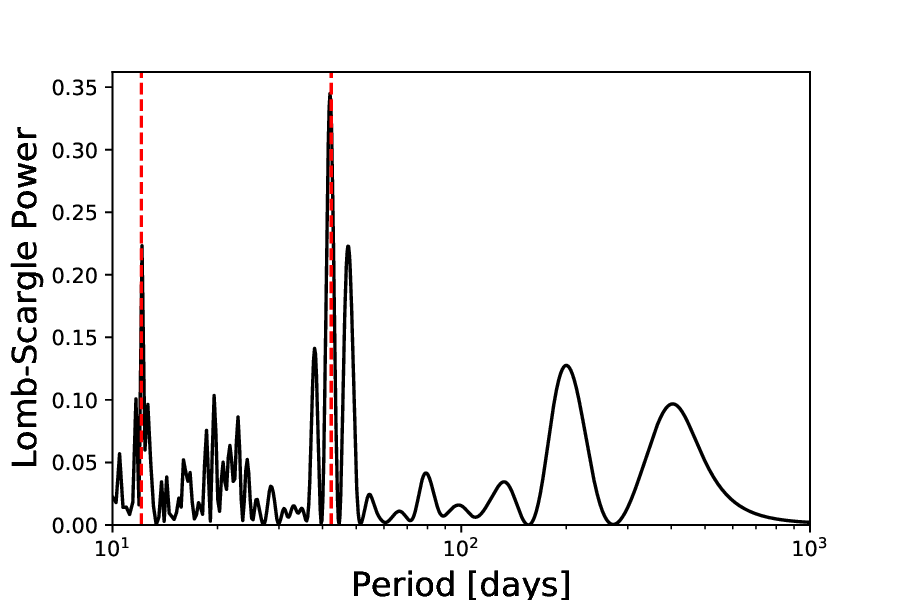} \\
\includegraphics[scale=0.33]{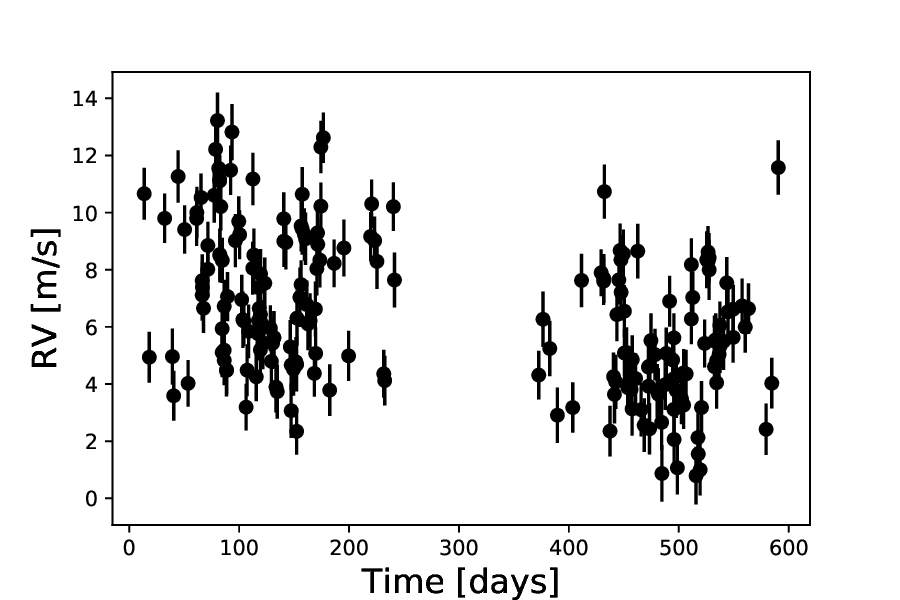} & \includegraphics[scale=0.33]{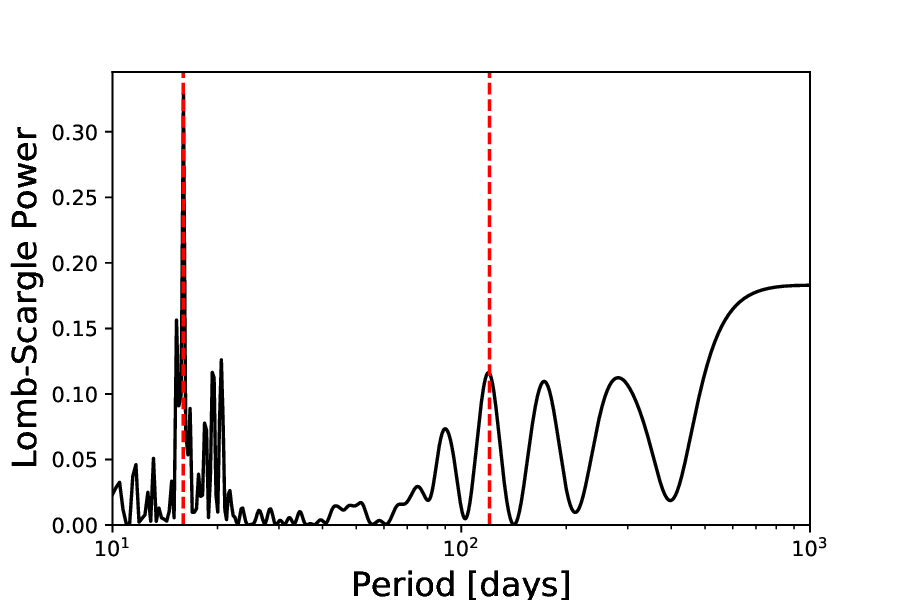} \\
\includegraphics[scale=0.33]{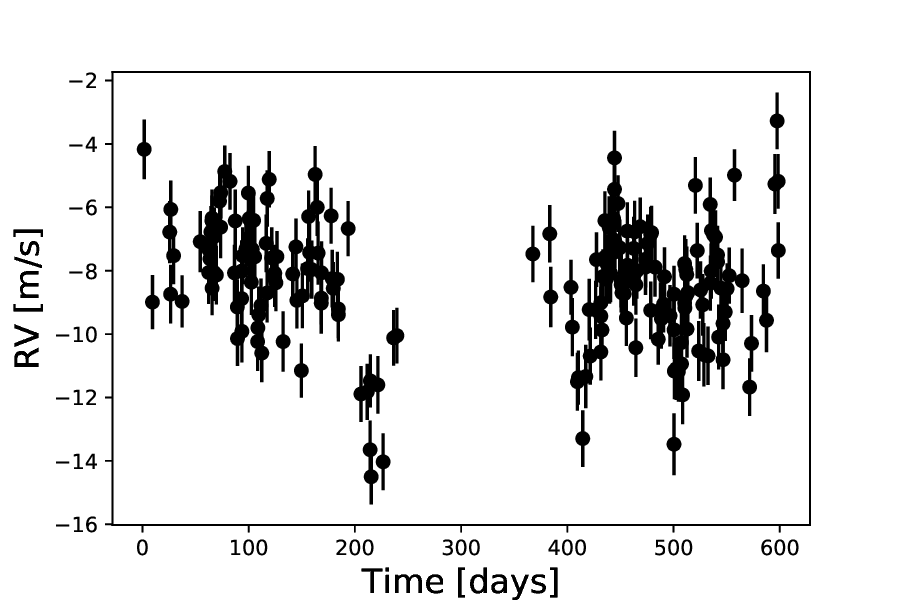} & \includegraphics[scale=0.33]{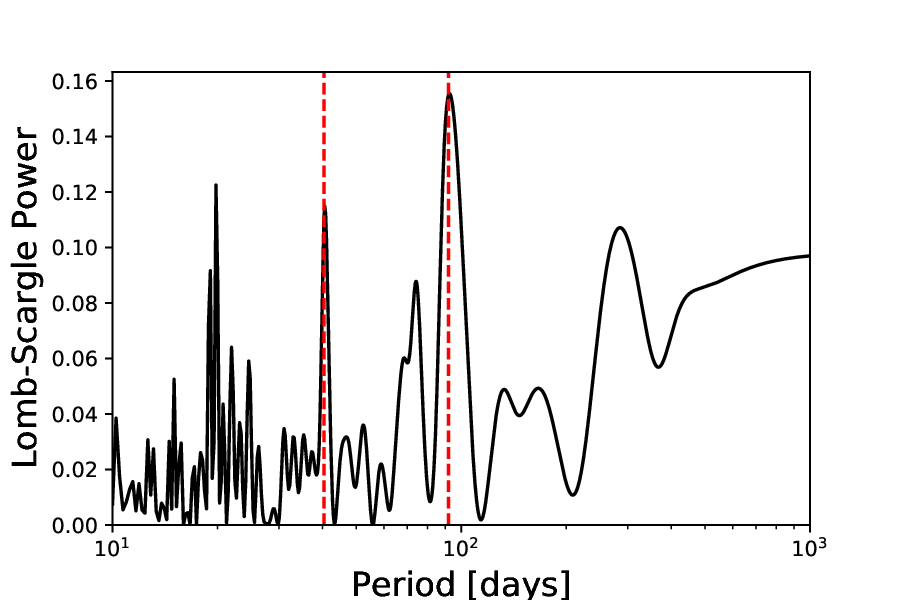} \\
\includegraphics[scale=0.33]{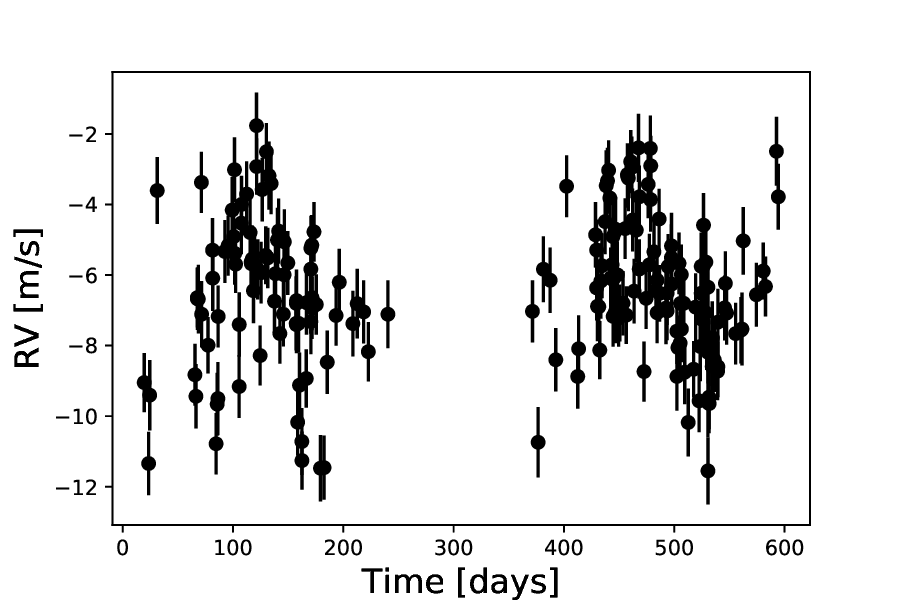} & \includegraphics[scale=0.33]{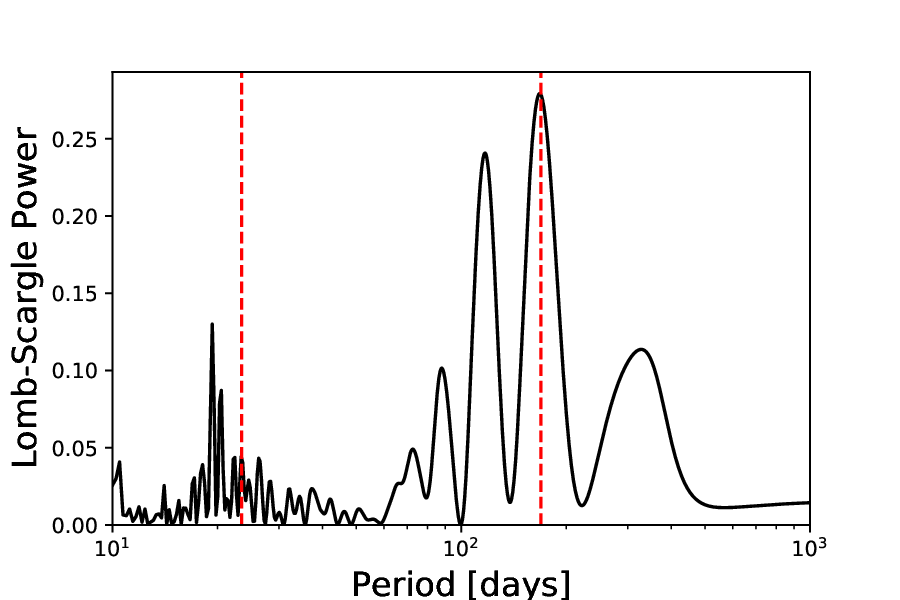} \\
\includegraphics[scale=0.33]{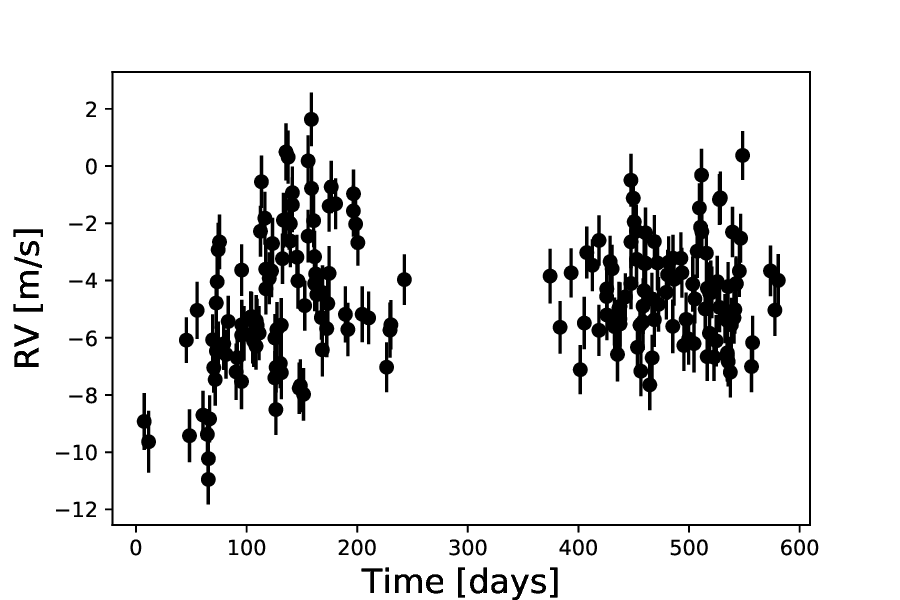} & \includegraphics[scale=0.33]{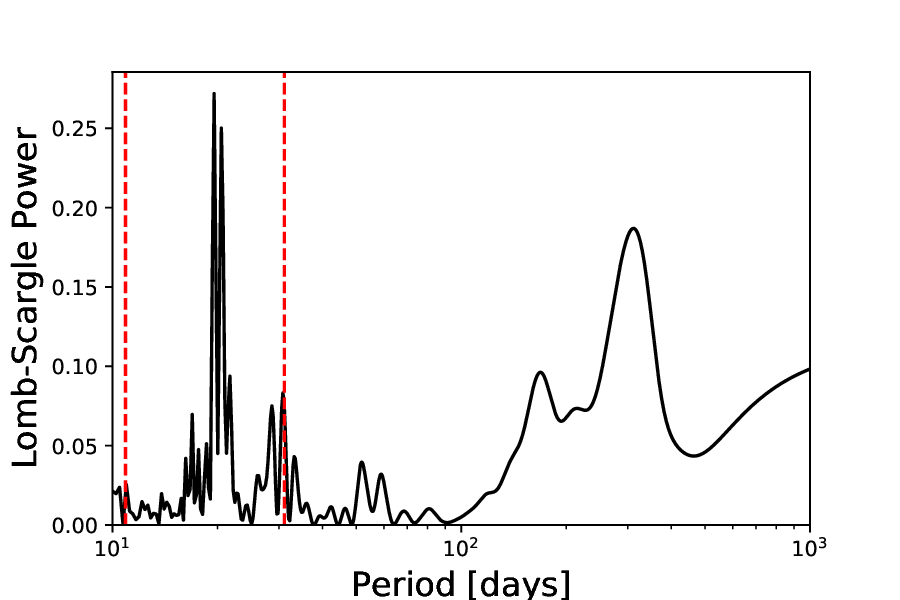} \\
\includegraphics[scale=0.33]{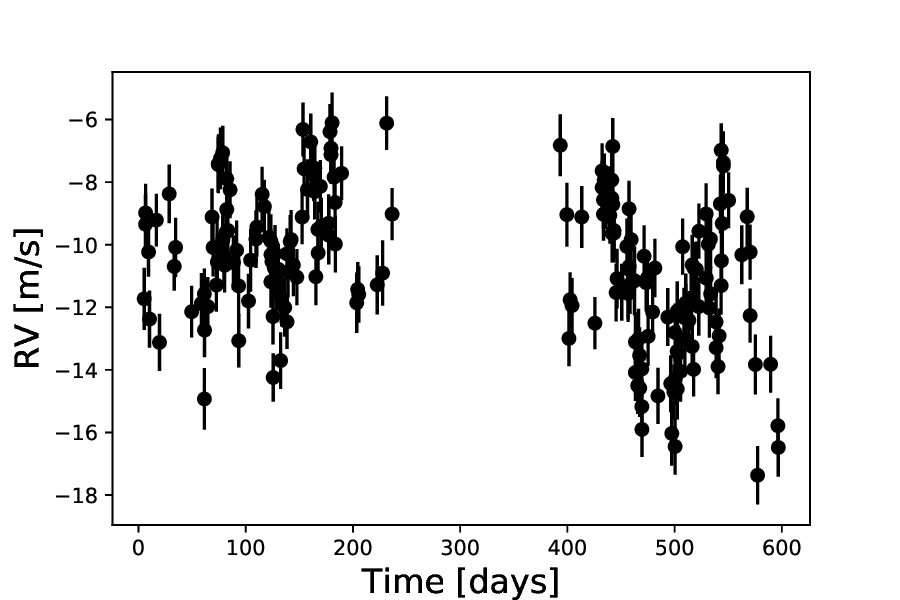} & \includegraphics[scale=0.33]{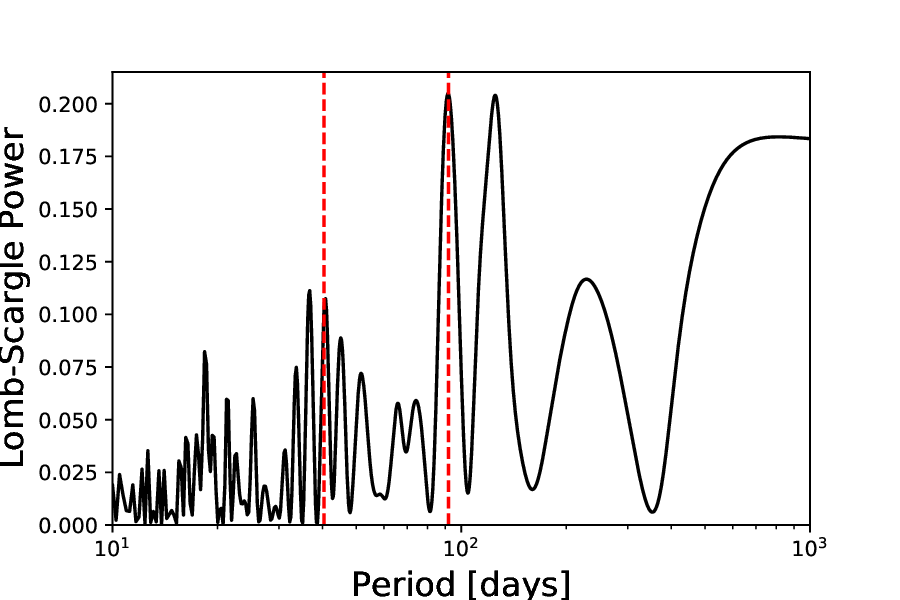} \\
\end{tabular}
\caption{\footnotesize For the Evidence Challenge, we generate six radial velocity datasets (left). Lomb-Scargle periodograms (right) show the relative strengths of periodic signals in the datasets, with the orbital periods of injected planets indicated (vertical red dashed lines).}
\label{fig:datasets}
\end{figure*}

\subsection{Statistical Model}
A likelihood function ($\mathcal{L}(\vec{\theta})$ = $p(\vec{d}|\vec{\theta},\mathcal{M})$) and prior probability distribution on the model parameters ($p(\vec{\theta})$) are needed to compute the integral in Equation \ref{eq-fml}.
Below, we specify both of these distributions.


\subsubsection{Likelihood}
Each simulated data point was generated according to
\begin{equation}
v_i = v_{\mathrm{pred}}(t_i|\vec{\theta}) + \epsilon_i,
\end{equation}
where $v_i$ is a component of $\vec{v}$, $t_i$ is a component of $\vec{t}$, and $\epsilon_i$ is the perturbation to the measurement due to noise.  
The noise vector was drawn from a multivariate normal distribution with covariance matrix $\Sigma$, i.e., $\vec{\epsilon} \sim \mathcal{N}(0,\Sigma)$.
Therefore, the appropriate likelihood is a multi-variable normal distribution, centered on the predictions of the model (parameterized by $\vec{\theta}$),
\begin{equation}
\begin{split}
\log \mathcal{L}(\vec{\theta}) = -\frac{1}{2} (\vec{v}-\vec{v}_{\mathrm{pred}}(\vec{\theta}))^{T} \Sigma^{-1} (\vec{v}-\vec{v}_{\mathrm{pred}}(\vec{\theta})) \\
-\frac{1}{2} \log \left| \mathrm{det} \Sigma \right| -\frac{n_{\mathrm{obs}}}{2} \log (2\pi).
\end{split}
\label{eq:lnL}
\end{equation}

The Gaussian noise is correlated from one observation to the next.  
$\Sigma$ is given by 
\begin{equation}
\label{equation-cov}
\Sigma_{i,j} = \kappa_{i,j} + \delta_{i,j} \left(\sigma_i^2 + \sigma_{J}^2 \right),
\end{equation}
where $\kappa_{i,j}$ is a quasi-periodic kernel, $\delta_{i,j}$ is the Kronecker delta, and
$\sigma_{J}^2$ is the amplitude of an additional unknown noise term (often casually referred to as RV ``jitter'').

As argued by \cite{haywood2014} and \citet{rajpaul2015}, we expect some degree of periodicity in stellar activity, modulated by the rotation of the star, which motivates our choice of a quasi-periodic kernel.
It is defined by
\begin{equation}
\label{equation-kernel}
\kappa_{i,j} = \alpha^2 \exp\left[-\frac{1}{2}\left\{ \frac{\sin^2[\pi(t_i-t_j)/\tau]}{\lambda_p^2} + \frac{(t_i-t_j)^2}{\lambda_e^2}\right\}\right], 
\end{equation}
where the hyperparameters are fixed at the following values: $\alpha = \sqrt{3}$ m/s, $\lambda_e = 50.0$ days, $\lambda_p = 0.5$ (unitless), and $\tau = 20.0$ (days).
These values were given for the Evidence Challenge, so teams did not need to marginalize over these hyperparameters.

\subsubsection{Priors}
\label{sec-priors}

\REWRITE{In the Bayesian framework, the prior probability density function specifies the state of information prior to taking the observation. It could thus vary from system to system, or as additional information becomes available (for example from transits). 
}
\REWRITE{To enable direct comparisons of results across teams, we asked that they adopt a common set of priors described below. We use a prior that is plausible and convenient to implement, albeit not necessarily informed by orbital mechanics or the latest exoplanet statistics.}

We assumed a prior that factorizes in terms of each planet's orbital period ($P_i$), RV semi-amplitude ($K_i$), eccentricity ($e_i$), argument of pericenter ($\omega_i$) and mean anomaly at epoch ($M_i$), as well as the RV offset ($C$) and the white-noise term ($\sigma_J$).
Note that for the purpose of computing evidences, teams adopted an orbital period prior ranging from 1.25 to $10^4$ days.

\begin{itemize}
\item For each planet's orbital period, we assumed a truncated Jeffreys prior, $p(P) \, dP = \frac{dP}{P} \times  \frac{1}{\log(P_{\max}/P_{\min})}$ for $P_{\min} \le P \le P_{\max}$.  For the primary analysis, we assumed $P_{\min}=1.25$ day and $P_{\max}=10^4$ days for each of the planets. For an alternative analysis, we provided specific values of $P_{\min,i}$ and $P_{\max,i}$ for each planet and dataset to be described in \S \ref{sec-altpriors}.
\item For each planet's RV semi-amplitude, we assumed a truncated modified Jeffreys prior, $p(K) \, dK = \frac{dK}{K_0(1+K/K_0)} \times  \frac{1}{\log(1+K_{\max}/K_0)}$ for $0<K\le K_{\max}$, where $K_0=1$ m/s and $K_{max}=999$ m/s.
\item For each planet's eccentricity, we assumed a truncated Rayleigh distribution, $p(e) \, de = \,\,\,$ $\frac{e\, de}{\sigma_e^2} \exp\left(-\frac{e^2}{2\sigma_e^2}\right)/ \left[ 1-\exp\left(-\frac{e_{\max}^2}{2\sigma_e^2}\right) \right]$ from $0 \leq e < e_{\max} = 1$ and zero for $e\ge e_{\max} = 1$, where $\sigma_e = 0.2$.
\item For each planet's argument of pericenter, we assumed a uniform distribution, $p(\omega) \, d\omega = \frac{d\omega}{2\pi}$ from $0 \leq \omega < 2\pi$ radians.
\item For each planet's mean anomaly, we assumed a uniform distribution, $p(M) \, dM = \frac{dM}{2\pi}$ from $0 \leq M < 2\pi$ radians.
\item For the additional white-noise term, we assumed a truncated modified Jeffreys prior, $p(\sigma_J) \, d\sigma_J = \frac{d\sigma_J}{\sigma_{J,0}(1+\sigma_J/\sigma_{J,0})} \times  \frac{1}{\log(1+\sigma_{J,\max}/\sigma_{J,0})}$ for $0<\sigma_{J,0} \le \sigma_{J,\max}$, where $\sigma_{J,0}=1$ m/s and $\sigma_{J,\max}=99$ m/s.
\item For the RV velocity offset, we assumed a uniform distribution, $p(C) \, dC = \frac{dC}{2C_{\max}}$ from $-C_{\max} \le C \le C_{\max}$, where $C_{\max} = 1,000$ m/s.
\end{itemize}
Here, the $\log$ refers to the natural logarithm.
The combined prior for a given $n$-planet model is
\begin{equation}
\begin{split}
p\left(\left\{P_i,K_i,e_i,\omega_i,M_i\right\}_{i=1..n}, \sigma_J, C \right) = \\
p(\sigma_J) p(C) \prod_{i=1}^{n} p(P_i) p(K_i) p(e_i) p(\omega_i) p(M_i).
\end{split}
\end{equation}
\subsubsection{Two Sets of Priors for Orbital Periods}
\label{sec-altpriors}
We previously described a prior where $P_{\min}=1.25$ day and $P_{max}=10^4$ days for each of the planets (the \emph{broad prior}, henceforth).
Note that even for a very well-behaved dataset (i.e., one dominant posterior mode if we assume $P_1<P_2<P_3$), the posterior would have $n!$ modes corresponding to the number of permutations for ordering $n$ planets.
If a team only explores one mode, they would have to renormalize their orbital period prior by a factor of $n!$.
However for the challenge, we imposed an order restriction so teams will neglect this degeneracy when computing $\widehat{\mathcal{Z}}$.

Based on preliminary results reported at the EPRV3 breakout sessions, we noticed that different groups sometimes focused their exploration of parameter space on different regions, particularly in terms of the orbital periods.
This made it difficult to directly compare methods. 
We decided to impose a second choice of priors for orbital period that force all groups to explore the same regions of parameter space in orbital period (the \emph{narrow prior}, henceforth).
That is we specified different values of $P_{\min,i}$ and $P_{\max,i}$ for each planet and each dataset.
The values (in days) are as follows for each dataset:
\begin{itemize}
\item Dataset 1: $P_{\min,1}=39.8107$, $P_{\max,1}=44.6684$, $P_{\min,2}=11.4815$, $P_{\max,2}=12.8825$, $P_{\min,3}=10.0$, $P_{\max,3}=10.7152$
\item Dataset 2: $P_{\min,1}=15.4882$, $P_{\max,1}=16.2181$, $P_{\min,2}=14.7911$, $P_{\max,2}=17.0608$, $P_{\min,3}=158.489$, $P_{\max,3}=251.189$
\item Dataset 3: $P_{\min,1}=81.2831$, $P_{\max,1}=107.152$, $P_{\min,2}=38.0189$, $P_{\max,2}=42.658$, $P_{\min,3}=16.5959$, $P_{\max,3}=17.5792$
\item Dataset 4: $P_{\min,1}=138.038$, $P_{\max,1}=204.174$, $P_{\min,2}=15.1356$, $P_{\max,2}=16.5959$, $P_{\min,3}=398.107$, $P_{\max,3}=1000.0$
\item Dataset 5: $P_{\min,1}=29.5121$, $P_{\max,1}=32.3594$, $P_{\min,2}=10.7152$, $P_{\max,2}=11.4815$, $P_{\min,3}=18.197$, $P_{\max,3}=19.9526$
\item Dataset 6: $P_{\min,1}=79.4328$, $P_{\max,1}=141.254$, $P_{\min,2}=31.6228$, $P_{\max,2}=50.1187$, $P_{\min,3}=316.228$, $P_{\max,3}=398.107$
\end{itemize}
These $P_{\min,i}$ and $P_{\max,i}$ values do not necessarily bound true orbital parameters used to generate the datasets.
These merely represent a set of reasonable period ranges for each dataset to facilitate more direct comparison of different methods. 
They were chosen without knowledge of the true planet parameters.

\subsubsection{Prior over models}

Participants submitted their $\mathcal{Z}$ estimates for the evidence for each $\mathcal{M}_n$, assuming that is the correct $n$-planet model.
In case some participants performed a non-Bayesian analysis, it would be useful to have something that can be compared between Bayesian and non-Bayesian estimates.
For those analyses that could not report the marginalized likelihood, we compared the posterior odds ratio to whatever they provide that they think is analogous to a posterior odds ratio.
To estimate posterior odds ratios, we must define a prior over $\mathcal{M}_n$,
\begin{equation}
	p(\mathcal{M}_n)=\left\{
    \begin{array}{cl}
		\beta^{n} \, & \mathrm{for} \, n={1,2,3} \\
		1-\sum_{i=1}^3 \beta^i & \, \mathrm{for} \, n=0
	\end{array}
    \right.
\end{equation}
and set $\beta=\frac{1}{3}$.  
Any participants submitting non-Bayesian estimates were instructed to take this into consideration, so that they could calibrate their estimates appropriately.

\begin{deluxetable*}{c|c|c}
\tablecaption{Evidence Challenge Teams and Methods.\label{tbl-methods}}
\tablehead{
\colhead{Method Class} & \colhead{Team Name} & \colhead{Method Name}
}
\startdata \hline
computationally cheap & Feng & Bayesian Information Criterion  \\
 & Feng & Chib's approximation \\ 
& Ford & Laplace approximation \\
& Hara & $\ell_1$ periodogram + Laplace approximation \\ \hline
importance samplers & D\'iaz & Perrakis Estimator \\
& Nelson & Ratio Estimator (MCMC+Importance Sampling) \\
& Team PUC & Variational Bayes with Importance Sampling  \\ \hline
nested samplers & Rajpaul & MCMC Nested Sampling \\ 
& Team PUC & \MULTINEST\,(Nested Sampling) \\
& Team PUC & \MULTINEST\,(Importance Nested Sampling) \\
& Team PUC & Multirun-\MULTINEST\,(Nested Sampling) \\
& Team PUC & Multirun-\MULTINEST\,(Importance Nested Sampling) \\
& Faria & Diffusive Nested Sampling \\ 
\hline
prediction-based & Cloutier & Leave-One-Out Cross Validation \\
& Cloutier & Time Series Cross Validation \\ \hline
\enddata
\end{deluxetable*}

\section{Methods for calculating the Marginal Likelihoods}
\label{sec:methods}

In this section, we will briefly list and describe each method used in the \eprvec.
They are described in greater detail in Appendix \ref{app:methods}.
Table \ref{tbl-methods} provides a list of the teams and methods they employed.
\REWRITE{Most of the submissions used a unique sampling technique, but some were simply different tunings for the same sampling algorithm. For example, Team PUC submitted \MULTINEST\, results using nested sampling and importance nested sampling approaches.
For each of those algorithms, they also submitted a variety of different \MULTINEST\, tunings (i.e., adjusting the number of live points or the efficiency parameter). When describing each method, we specifically refer to the particular choice of \emph{algorithm} as opposed to every algorithm and tuning combination.}

\subsection{Bayesian computationally cheap methods}  
\begin{itemize}
\item Bayesian Information Criterion (§\ref{sec:bic}): The BIC is defined as -2$\log{\mathcal{L}_{\rm max}}+k\log{N}$, where $\mathcal{L}_{\rm max}$ is the value of the maximum likelihood, $k$ is the number of free parameters, and $N$ is the number of data points. 
\REWRITE{Smaller BIC values suggest higher model probability. Two competing models $\mathcal{M}_1$ and $\mathcal{M}_2$ can be compared with 
$\exp[-(\textup{BIC}_{\mathcal{M}_2}-\textup{BIC}_{\mathcal{M}_1})/2]$, similar to a Bayes factor.
The BIC is derived under very strong simplifying assumptions. Under infinite data, $N \rightarrow \infty$, the evidence integral is assumed to become a single, infinitely narrow peak, independent of any prior. In realistic data sets, the posterior has finite width, so the BIC is at best a poor approximation of a Bayesian evidence into question.}

\item Chib's Approximation: Chib's approximation is based on the fact that the evidence is the normalization constant of the posterior density at a given point in the parameter space.
To estimate the evidence, we choose a point with high posterior probability, and calculate the evidence using the one-block sampling of parameter space (Eqn. 9 and 10 in \citet{chib01}).
We divide the Markov chain Monte Carlo (MCMC) chain into 100 sub-samples, and calculate the distribution of the evidence.  
\item Laplace Approximation (§\ref{sec:laplaceapprox}): The Laplace approximation computes the required integral analytically by approximating the target distribution as a Gaussian.  For this challenge, we numerically integrate over the orbital period (grid search) and jitter parameter (Gauss-Legendre quadrature) and apply the Laplace approximation to approximate the remaining model parameters.  For this challenge, we used either a circular or epicyclic approximation for the planetary motion to facilitate rapid computation.  
\item $\ell_1$ periodogram (§\ref{sec:hara}): This method relies on the basis pursuit de-noising algorithm~\citep{chen1998}, and is detailed in~\cite{hara2017}. It is an alternative to the Lomb-Scargle periodogram or its generalizations, and can be read similarly, but mitigates the problem of aliasing. We here use two ways to assess the significance of its peaks: the false alarm probabilities (FAPs) as provided by~\cite{baluev2008} and a Laplace approximation of the evidence of the model given by its $n$ tallest peaks.  
\end{itemize}

\subsection{Bayesian importance samplers}
Importance sampling is a integration technique that draws from a simple, normalised distribution that approximates the target distribution, the posterior. If the two distributions are close matches, the integral estimator is accurate and efficient.

\begin{itemize}
\item Perrakis estimator (§\ref{sec:perrakisis}): 
In the Perrakis estimator \citep{perrakis2014}, the importance sampling function is constructed from the product of marginal posterior densities. Samples are drawn by shuffling the vector elements of joint posterior samples (e.g., from a previous MCMC run) across samples. Additionally, the estimator requires an estimation of the marginal posterior densities of each parameter, which are approximated from a normalised histogram of the marginal samples.
\item Ratio estimator (MCMC + importance sampling) (§\ref{sec:mcmcis}): 
This importance sampling technique adopts for the sampling distribution a truncated Gaussian with mean and covariance estimated from a previous MCMC run.
For each model and dataset, we perform 20 separate MCMC runs, apply this algorithm for each case, and calculate $\widehat{\mathcal{Z}}$ using the median and standard deviation based on the 20 different estimates.
\item Variational Bayes with importance sampling (§\ref{sec:VarBayes}): A mixture of Gaussians is used for the importance sampling proposal distribution.
For the initial guess of the mixture, multiple global maxima searches are performed.
Variational Bayes is a iterative procedure that optimally updates the Gaussians to match the target distribution better.
It samples from the mixture proposal distribution, evaluates the target distribution and adjust the parameters of the Gaussians.
As with the above techniques, importance sampling estimates the integral.
\end{itemize}

\subsection{Bayesian nested samplers}
Nested sampling (NS) is an efficient technique for estimating Bayesian evidence integrals (and numerical quadrature more generally).
It computes the geometric size at various likelihood $\mathcal{L}$ thresholds.
That threshold is continuously increased, such that the volume decreases exponentially.
The gradual increase overcomes the difficulty to handle multimodal posterior distributions (compared to, e.g., MCMC).
Nested sampling allows both parameter estimation and model comparison.
$\mathcal{Z}$ is the integral over likelihood and volume at each likelihood threshold. 

Internally, however, nested sampling requires a method for drawing a new random point from the prior with the condition that its likelihood is higher than the current likelihood threshold.

\begin{itemize}
\item MCMC nested sampling (§\ref{sec:mcmcns}): Rajpaul's implementation used a semi-adaptive MCMC scheme for this purpose; this was chosen as a foil to \MULTINEST\ (below), which instead makes use of a more sophisticated ellipsoidal rejection scheme and clustering algorithm for drawing new points.
\item MultiNest (§\ref{sec:method-multinest}): A robust nested sampling technique, which draws a new uniformly random point with higher likelihood through an ellipsoidal rejection sampling scheme \citep{Shaw2007,Feroz2009}. Existing live points are clustered into multiple ellipsoids, from which points are drawn. Studying the algorithm parameters, we vary the number of live points ($\texttt{nlive}$=400-2000) and the target efficiency (inverse of the ellipsoid expansion factor) from 0.3 to 0.01.
\item MultiNest using importance nested sampling (INS): An alternative summation of MULTINEST draws that interprets the ellipsoid draws as a importance sampling process \citep{Cameron2013,Feroz2013}.
While the standard NS technique may reject many drawn points failing the likelihood constraint $(\mathcal{L}>\mathcal{L}_{i})$, INS uses all the points drawn to improve the estimation. The uncertainty on \logZ\, can become very small, with up to an order of magnitude higher accuracy than typical NS \citep{Feroz2013}.
However, applying INS in this exoplanet problem, we found that INS estimator leads to overly small uncertainties.
This is  shown in the Appendix, Figures~\ref{fig:evi01} and \ref{fig:evi04}.
\item Multirun-MultiNest (with NS and INS): Examining MULTINEST \logZ\, estimates, we find scatter far exceeding the reported uncertainties (in both NS and INS, to be discussed in detail in Sections \ref{sec:results-uncertainty} and \ref{methods-teampuc-multinestscatter}).
To obtain robust estimates with realistic uncertainties, we define quantities over multiple runs. We define the multirun evidence estimate as the median \logZ\, across runs.
For an estimate of the uncertainty on \logZ, we add in quadrature the median absolute deviations (scatter) and the median reported uncertainty. The multirun results are also shown in Figures~\ref{fig:evi01} and \ref{fig:evi04}.

\item Diffusive nested sampling (§\ref{sec:dnest}):
The Diffusive Nested Sampling algorithm (DNS; \citealt{Brewer2011}) 
is a Monte Carlo method based on NS.
Unlike classic NS, which samples from the prior subject to a hard likelihood constraint,
DNS explores a mixture of successively nested distributions,
each occupying about $e^{-1}$ times the enclosed prior mass of the previous one.
Using a mixture of distributions allows DNS to ``go back'' 
to a lower likelihood threshold. 
After an inital phase where these distributions are created,
DNS starts sampling from the complete mixture with uniform weights, 
which means that the prior is also included in the target distribution,
improving the sampling efficiency in multimodal posteriors.

\end{itemize}

\REWRITE{\subsection{Prediction-based methods}}
\begin{itemize}
\item Leave-One-Out Cross Validation (§\ref{sect:loocv}): In general, cross-validation techniques are commonly used in the field of machine-learning to evaluate model performance and inform model selection as an alternative to calculating the fully marginalized likelihood. Cross-validation techniques are used to evaluate the predictive power of a model by splitting the input dataset into $N$ training and testing sets. Competing models are then fit to each training set with an objective function (i.e. Eq.~\ref{eq:lnL}) being evaluated on the testing set with the optimized model; the score. This formalism helps to avoid over-fitting of data as models that appear to provide excellent fits to training data will exhibit poor scores on previously unseen testing data if they are actually over-fitting. The relative scores between competing models are used for model selection. Leave-one-out cross-validation refers to a particular strategy for train/test splitting wherein $N$ unique splits of the RV time-series $\bar{v}$ are made. Each training set contains $N-1$ of the RV measurements with the remaining measurement being used for testing.
\item Time Series Cross Validation (§\ref{sec:timeseriescv}): The principle behind time series cross-validation is equivalent to that of leave-one-out cross-validation but differs in the method of train/test splitting. As is the case with RV time series featuring temporally correlated signals---from planets or possibly from stellar activity---removing a single random measurement fails to remove all of signal associated with that measurement. Time series cross-validation works to alleviate this bias by constructing training sets from subsets of the sequential measurements containing at least $N_{\text{min}}=20$ measurements. Each unique training set will then contain $N_{\text{min}}+i$ measurements for $i=0,\dots,N-N_{\text{min}}-1$. In the single-step forecasting method used here, the corresponding testing sets are the next sequential measurement; i.e. $N_{\text{min}}+i+1$. 
\end{itemize}

\section{Results}
\label{sec:results}

The four main goals associated with the Evidence Challenge are: (1) to better understand the dispersion of estimates of the marginal likelihood (\DZ) and how much this varies with the number of planets in the model, (2) to see if the reported uncertainty of $\logZ$ (\sigZ) accurately reflects the empirical \DZ, (3) to understand how \DZ\ and \sigZ\ affect our ability to compare the evidence for $n$ versus ($n+1$)-planet models, and (4) to identify promising methods for use and refinement in future studies.
In this section, we will address the first three questions and leave the fourth for \S \ref{sec:discussion}.

The methods used to estimate $\mathcal{Z}$ are labeled in the figures based on their directory names in the Evidence Challenge's Github repository\footnote{\url{https://github.com/EPRV3EvidenceChallenge}}.

First, we compare $\logZ$ (always in base 10) \REWRITE{from Bayesian methods that compute it, i.e., without the prediction-based methods in Table~\ref{tbl-methods}}. We are most interested in the differences and dispersion in $\logZ$s, not necessarily their absolute values, so we plot each method's $\logZ - \medianlogZ$, where \medianlogZ\, is the median $\logZ$ among the methods being considered.

Note that Team PUC submitted roughly half of the total analyses considered.
Most of these were different variations on \MULTINEST\, in which they varied algorithm settings (number of live points  [\texttt{nlive}] and efficiency [\texttt{eff}]) and sampling techniques (nested sampling vs. importance nested sampling, a single run vs. multiple runs).
This study focuses on comparing methods for estimating $\logZ$, rather than the choice of algorithm settings for any one method.
Therefore in this section, we include results provided by one set of \MULTINEST\ runs (those with \texttt{nlive}=2000 and \texttt{eff}=0.3) which appears to perform well.
By including \MULTINEST\ results based on a single set of settings when calculating the median $\logZ$, we prevent the results from appearing heavily biased towards the \MULTINEST\ results in the figures that follow.
An analysis of all \MULTINEST\ results is presented in Section \ref{methods-teampuc-multinestscatter}.
All results submitted to the Evidence Challenge are available for further analysis at the Github repository.  

\subsection{Dispersion in $\logZ$ ($\DlogZ$)}

\begin{figure}
\centering
\includegraphics[angle=270, width=0.83\columnwidth]{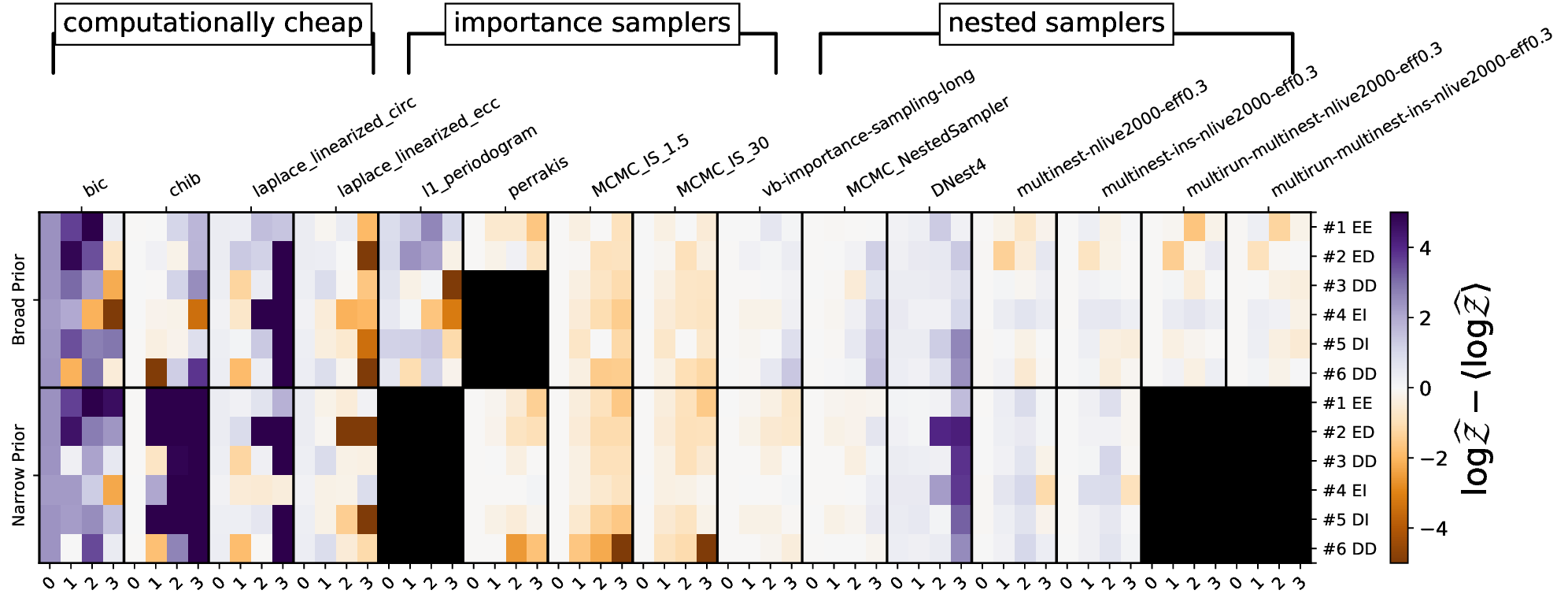}
\caption{\footnotesize{Summary of $\logZ$ results across all datasets and models. A row of pixels corresponds to an $n$-planet model, where $n=\{0, 1, 2, 3\}$. Columns correspond to one of the six datasets, each simulated with  two planets of varying levels of detectability (``easy''=``E'', ``difficult''=``D'', impractical=``I''). Rows of pixels are grouped with black outlines by method.
The left (right) grouped columns correspond to the model with narrow (broad) period priors. The color of each pixel shows $\logZ$ with respect to the median $\logZ$ ($\medianlogZ$) for that particular dataset and model, in order to emphasize the level of scatter seen in all computed $\logZ$s. Any $|\logZ - \medianlogZ|$ greater than 5 is set to a color at the end of the colorscale. Black pixels are unreported values.}}.
\label{fig:results-logZ}
\end{figure}

Figure \ref{fig:results-logZ} summarizes the Bayesian results submitted to the Evidence Challenge.
Each pixel corresponds to one estimate of \logZ\, based on a particular method, orbital period prior, dataset, and number of planets included in the model.
The color is $\logZ - \medianlogZ$ and the colorscale spans 10 orders of magnitude in $\widehat{\mathcal{Z}}$.
Black pixels are unreported values.
We grouped methods into three different classes based on the sample of methods submitted: ``computationally cheap'', ``importance samplers'', and ``nested samplers.''
In essence, paler colors correspond to $\logZ$ values closer to \medianlogZ\, and more saturated colors stray further from the median.
Purple colors are biased toward larger $\logZ$ with respect to \medianlogZ\, and orange colors are biased toward smaller values.
We do not consider reported uncertainties ($\sigZ$) here but present that information in Figures \ref{fig:results-logZ-errorbar-broad} and \ref{fig:results-logZ-errorbar-narrow}.

In most cases, we do not know the true value of $\logZ$.
Thus, it is difficult to quickly evaluate the accuracy of each estimate.
For the 0-planet model ($\mathcal{M}_0$, 2 parameters), multiple teams performed brute force calculations via a very fine grid or large number of Monte Carlo samples to provide a comparison point.
However, brute force was not practical for $\geq 1$-planet models (7+ parameters).
Therefore, we focus our attention on $\logZ$ estimates relative to $\medianlogZ$ and $\DlogZ$, emphasizing that $\medianlogZ$ should not be regarded as the ``true'' \logZ.
The dispersion in results across methods can be seen by comparing the color of pixels across rows in Figure \ref{fig:results-logZ}.
All Bayesian methods provided very similar estimates for $\logZ$s for $\mathcal{M}_0$, with less than a factor of $\DlogZ \sim 0.5$ in variation or $\DZ \sim 3$.
However, \DlogZ\, grows to $\sim 1$ for $\mathcal{M}_1$, $\sim 2-3$ for $\mathcal{M}_2$, and $>$3 for $\mathcal{M}_3$.

We also observe differences among the classes of algorithms.
Computationally cheap methods have the greatest variability and appear \REWRITE{to estimate systematically higher $\logZ$ values than the results provided by the importance and nested samplers}.
In practice, this would imply that the computationally cheap methods are typically more confident in the evidence for additional planets.
Overall, the importance samplers seem slightly biased to smaller $\logZ$ relative to the nested samplers which tend to report larger values of $\logZ$.
In consideration of this, we reanalyzed the $\logZ$ results excluding the computationally cheap methods, recalculated the {\medianlogZ}s, and found that the patterns in $\logZ-\medianlogZ$ did not significantly change.

Different teams could be computing the evidence for planets at different orbital periods, which may contribute to a substantial fraction of the dispersion seen here.
However, we see similar dispersion when teams were instructed to use the narrow period prior.
Interestingly, some methods seem to have greater dispersion for the narrow priors, denoted by the more saturated pixels in the left column of Figure \ref{fig:results-logZ}.
We found that some teams renormalized their orbital period prior when they imposed this narrower range while others did not.  
We corrected for this as noted in the $\tt{\_renormalized.txt}$ files in the Evidence Challenge repository, but significant dispersion remained.
In particular, Chib's approximation and the Laplace approximation for circular orbits calculate a $\widehat{\mathcal{Z}}$ for $\geq$1-planet models that can be over 5 orders of magnitude different than the other methods.

\subsection{Uncertainty in $\logZ$ (\siglogZ) }
\label{sec:results-uncertainty}
\begin{figure*}
\centering
\begin{tabular}{cc}
\includegraphics[width=\columnwidth]{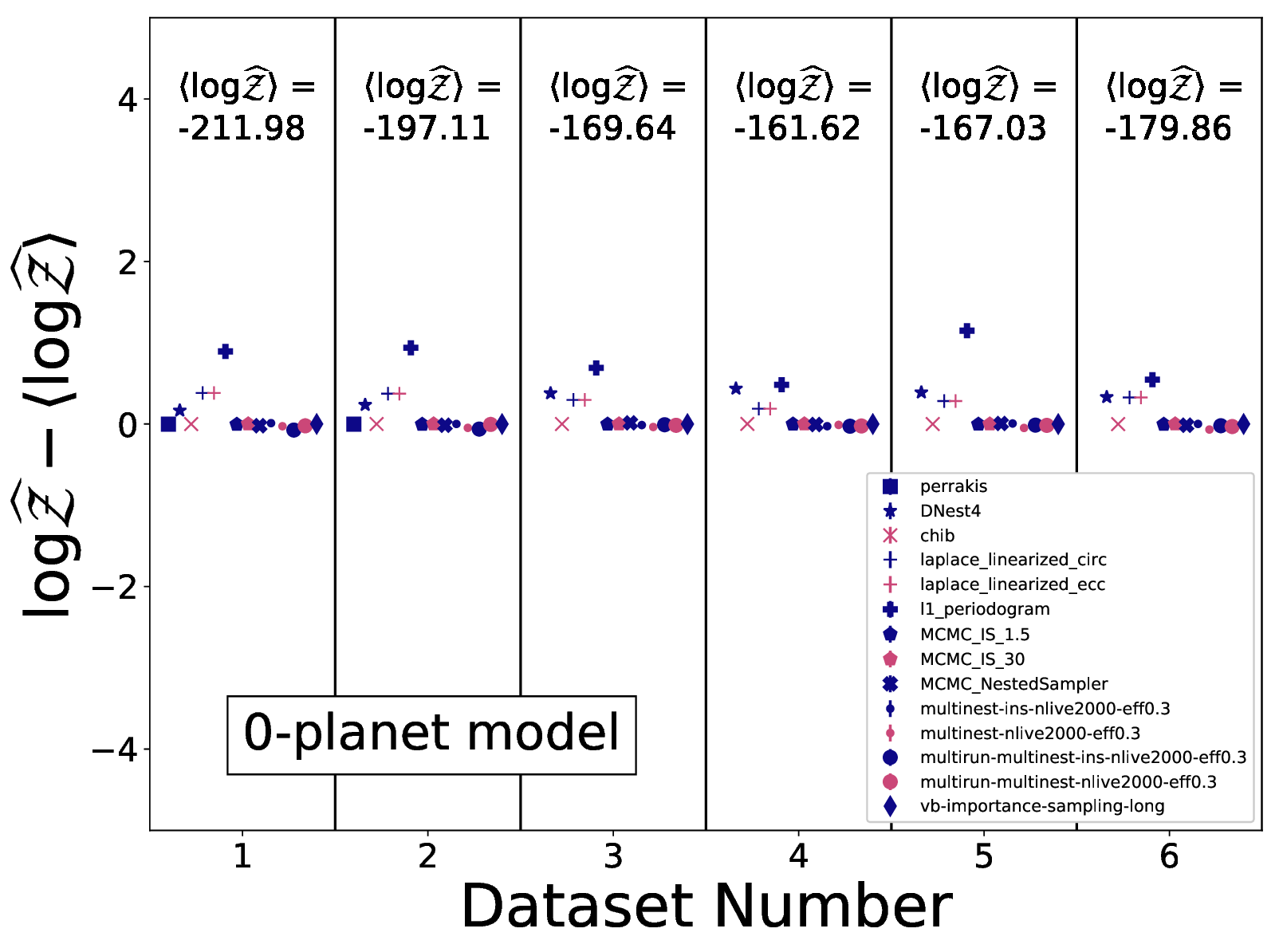} & \includegraphics[width=\columnwidth]{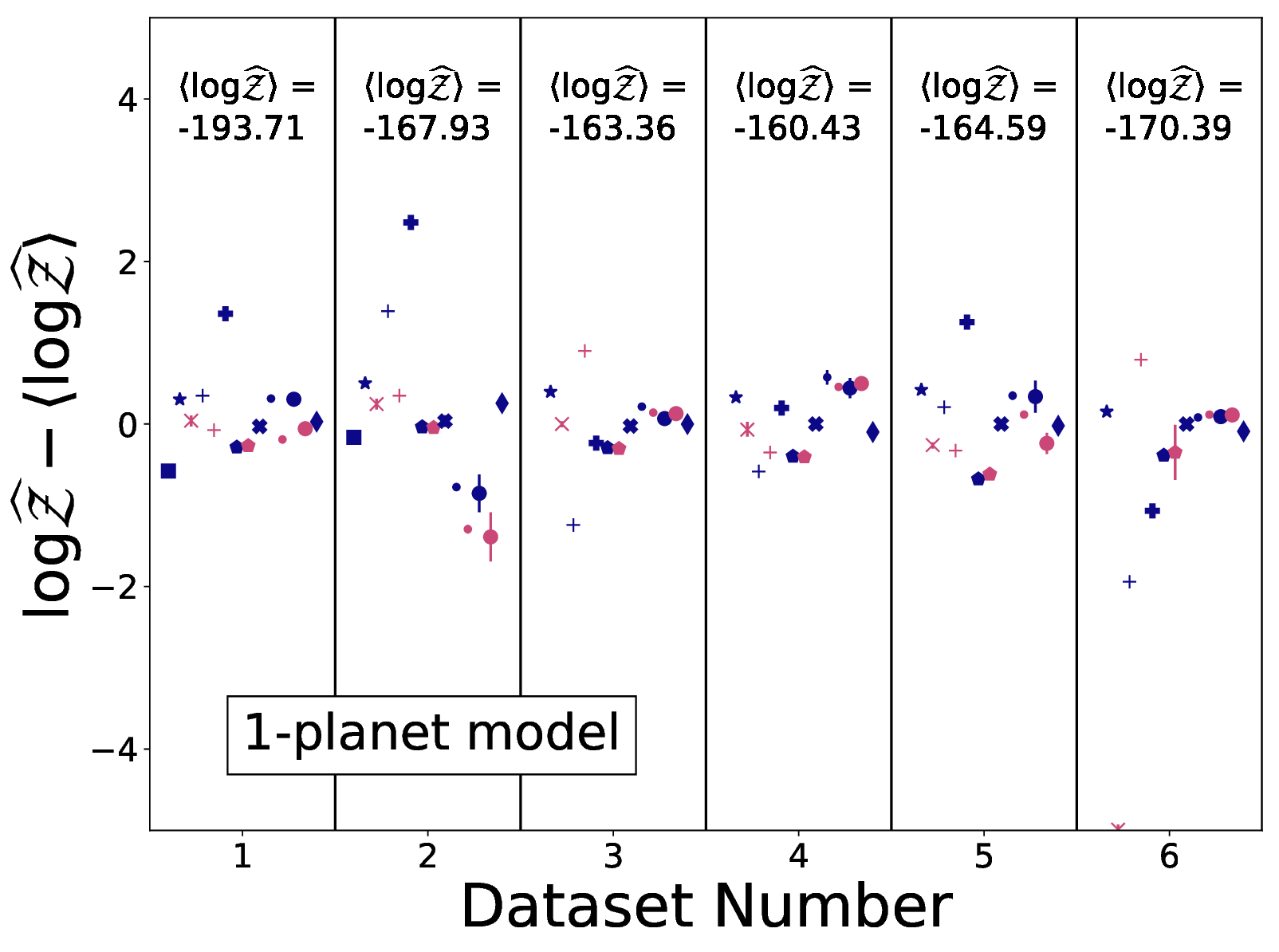} \\
\includegraphics[width=\columnwidth]{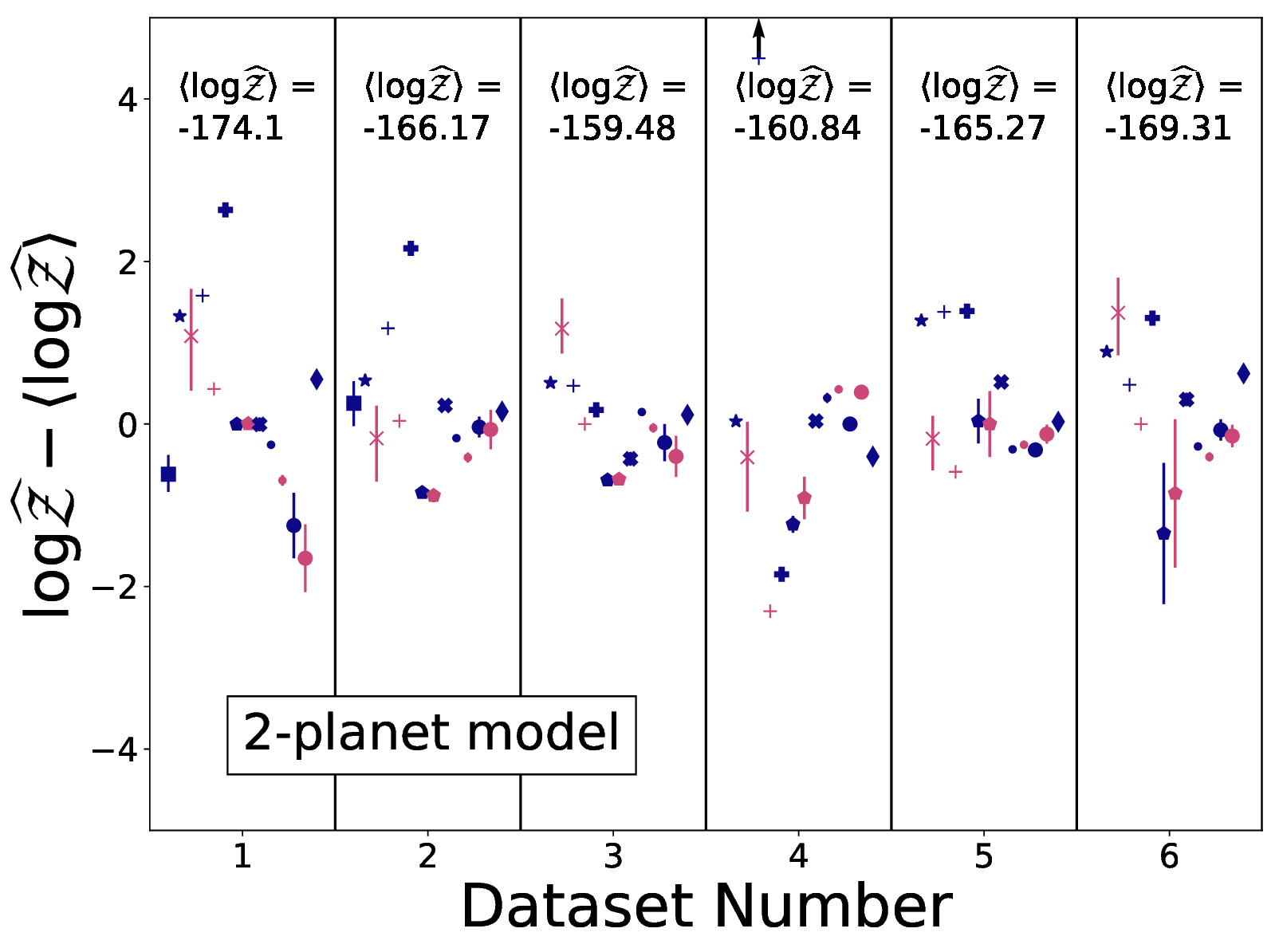} & \includegraphics[width=\columnwidth]{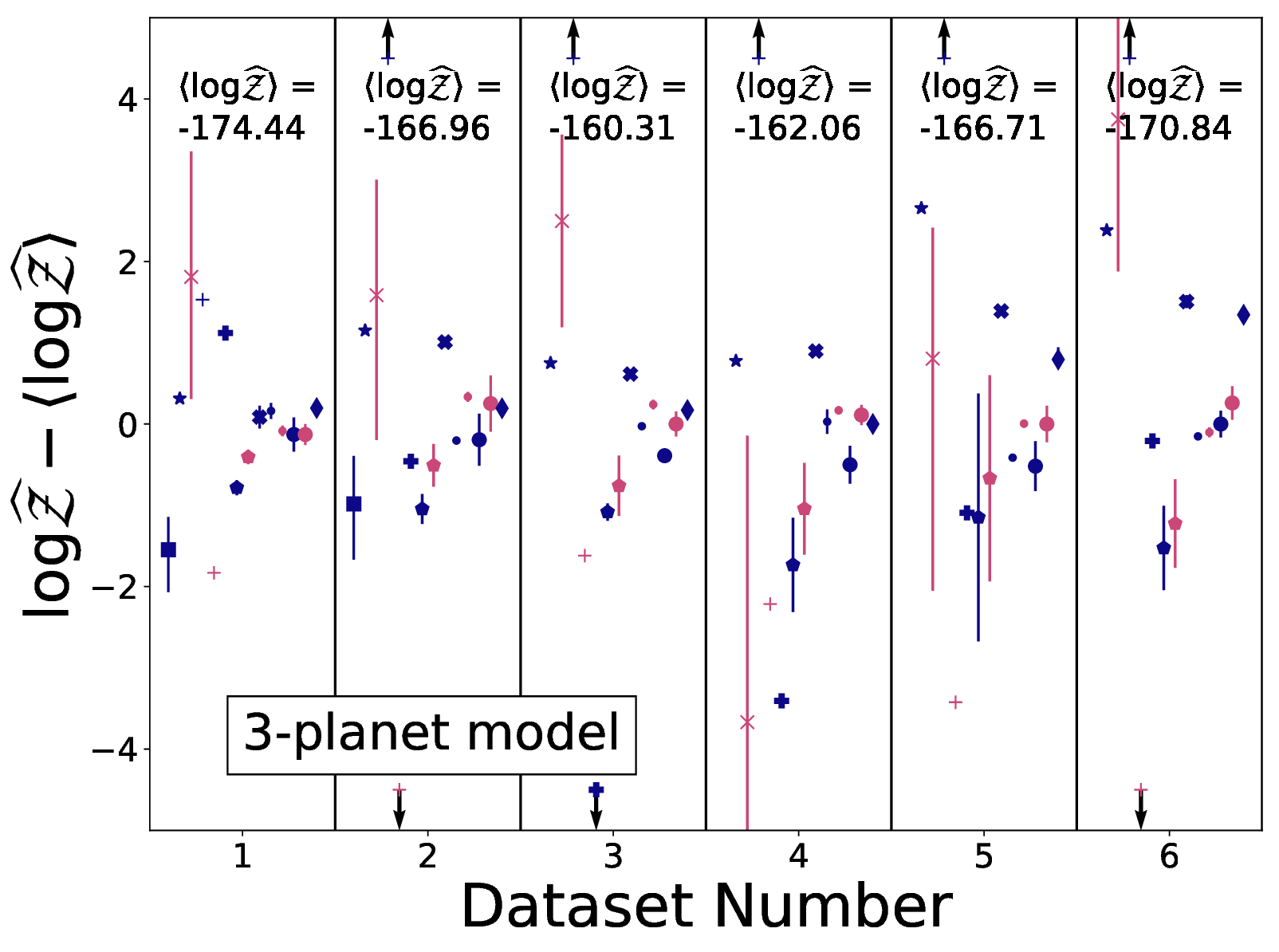} \\
\end{tabular}
\caption{\footnotesize $\logZ$ estimates for $\mathcal{M}_0$ (upper left), $\mathcal{M}_1$ (upper right), $\mathcal{M}_2$ (lower left), and $\mathcal{M}_3$ (lower right) models assuming broad orbital period priors.
All figures show $\logZ$ with respect to the median value for each dataset and model, \medianlogZ\, displayed at the top of each figure.
The symbols correspond to different methods and colors correspond to different implementations (e.g., input parameters or assumptions) of the same method.
Error bars show $1$-$\sigma$ equivalent uncertainties in $\logZ$, some of which are too small to resolve.
Methods reporting $|\logZ-\medianlogZ| > 5$ are denoted with arrows pointing outside of the figure bounds.}
\label{fig:results-logZ-errorbar-broad}
\end{figure*}

\begin{figure*}
\centering
\begin{tabular}{cc}
\includegraphics[width=\columnwidth]{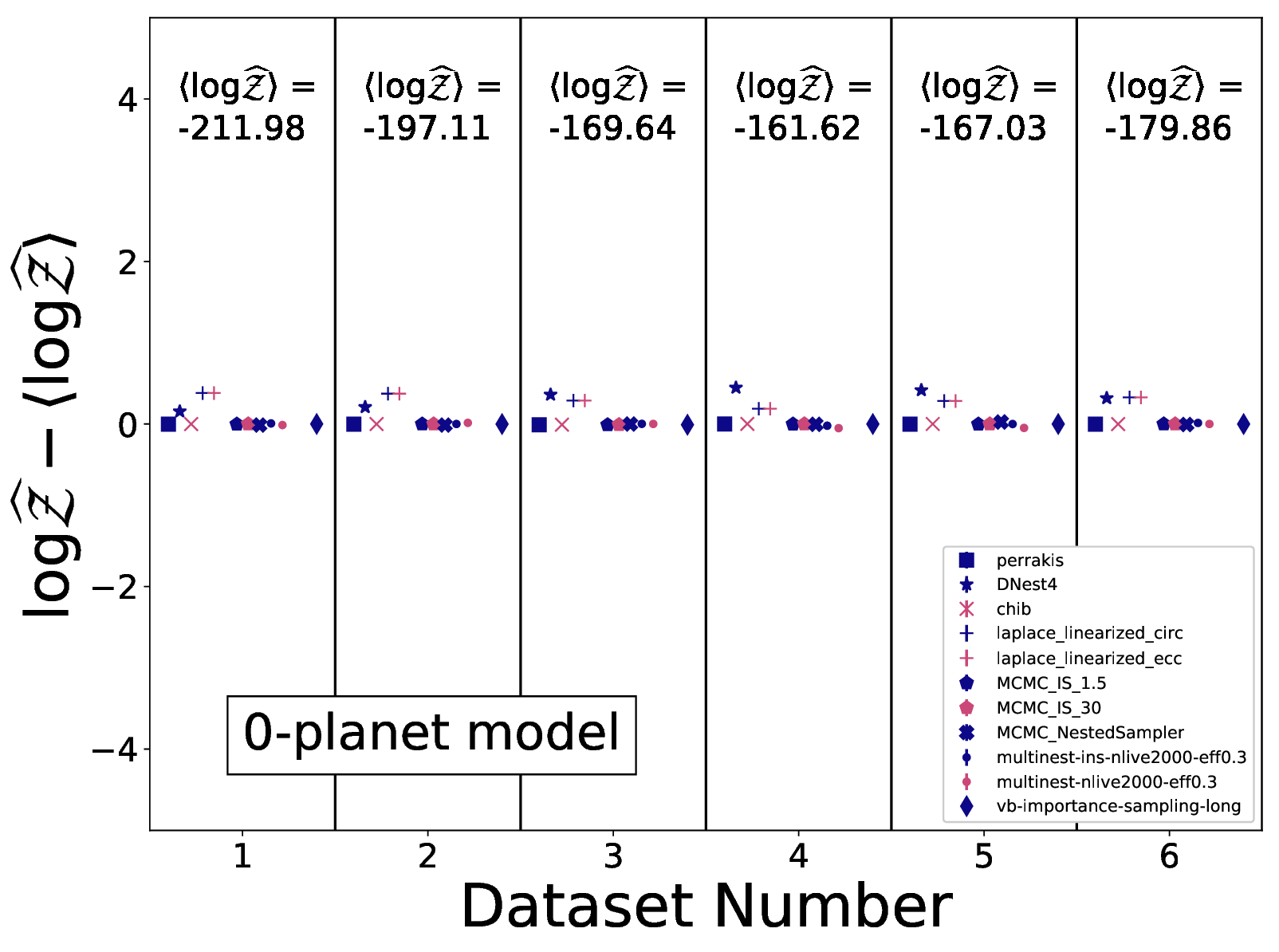} & \includegraphics[width=\columnwidth]{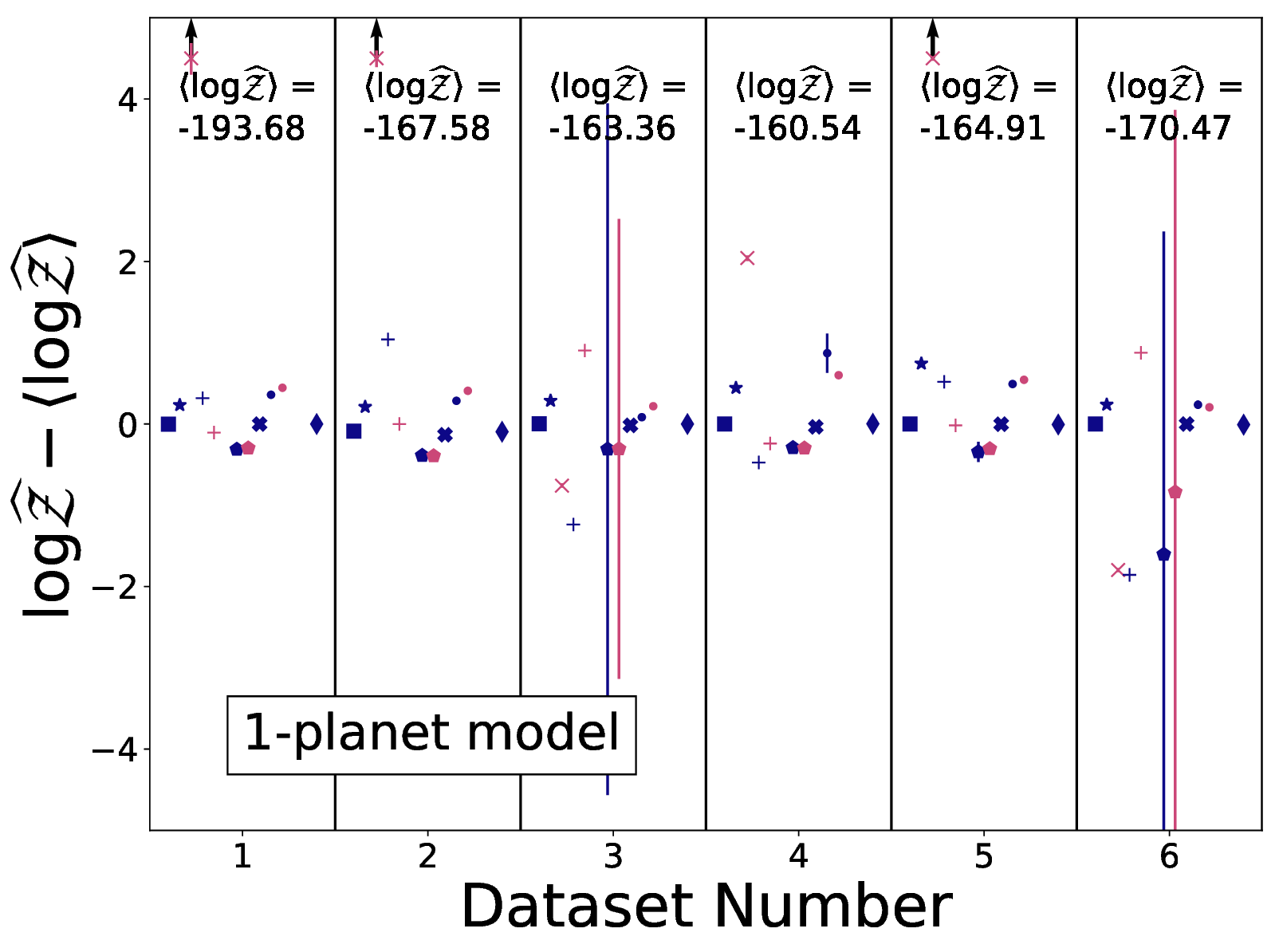} \\
\includegraphics[width=\columnwidth]{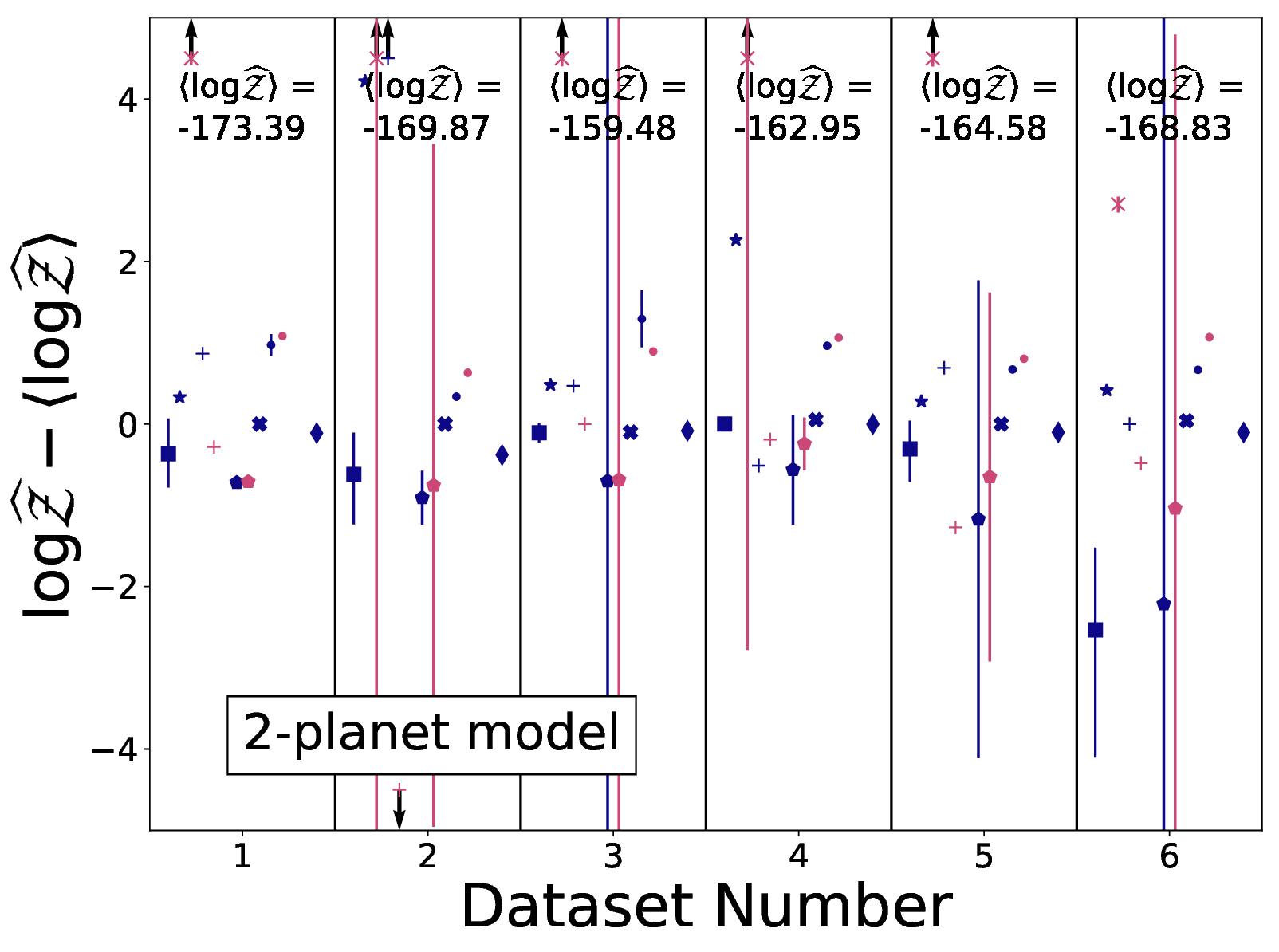} & \includegraphics[width=\columnwidth]{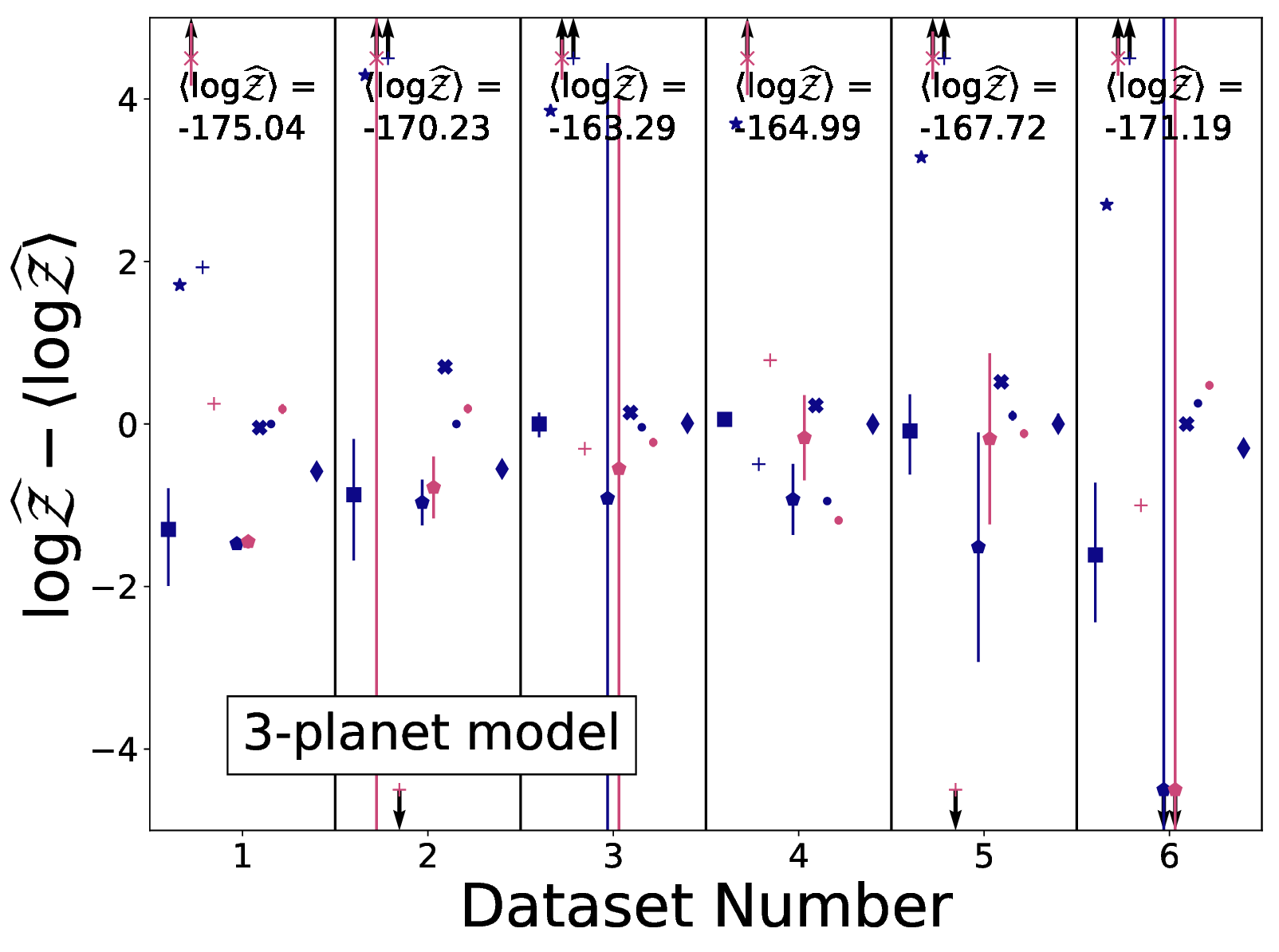} \\
\end{tabular}
\caption{\footnotesize Following the same format as Figure \ref{fig:results-logZ-errorbar-broad}, $\logZ$ estimates of each $n$-planet model but assuming narrow orbital period priors.
}
\label{fig:results-logZ-errorbar-narrow}
\end{figure*}

Figure \ref{fig:results-logZ-errorbar-broad} displays the $\logZ$ results assuming the broad priors and includes the uncertainties in $\logZ$ ($\siglogZ$).
Every panel corresponds to a different $n$-planet model, and each panel is divided into six subpanels for the six different datasets.
Each subpanel plots every method's $\logZ - \medianlogZ$, and we display \medianlogZ\, for that dataset and model near the top.
Figure \ref{fig:results-logZ-errorbar-narrow} is in the same format as Figure \ref{fig:results-logZ-errorbar-broad} but displays the results for the narrow period prior.
These figures are designed to emphasize \DlogZ\, across all datasets and how it compares to each reported \siglogZ.

For both priors, we find most methods claim a high degree of precision in \logZ\, that does not reflect the observed scatter in estimates of \logZ\,(\DlogZ).
In other words, the estimates are mutually exclusive to an extreme degree.
Analytic methods like the Laplace approximation did not report estimates for the uncertainty \siglogZ.
However, a handful of methods appear to report reasonable \siglogZ: the MCMC + importance sampling ratio estimator and variations of multirun-\MULTINEST.  
One common feature among these methods is that \siglogZ\, was based on comparing the estimates of \logZ\, from multiple runs of the same method, rather than an internal estimate of uncertainty based upon a single run.
Despite being more computationally expensive, this Monte Carlo approach seems to provide more plausible uncertainty estimates.
The MCMC + importance sampling ratio estimator shows particularly large errorbars for some datasets in Figure \ref{fig:results-logZ-errorbar-narrow}.
This is likely due to many MCMC runs not converging for those models, thus providing a poor importance sampling density for the estimator. 
Team PUC directly compared \siglogZ\, across multiple \MULTINEST\ runs in \S \ref{methods-teampuc-multinestscatter}.

\subsection{How \DlogZ\, Affects Odds Ratios}

We see significant dispersion in \logZ\, across methods even when assuming the same statistical model.
How does this affect our interpretation of $n$ versus ($n+1$)-planet models?
In practice, the evidence is rarely used by itself.  
Instead, we compare \logZ\, for different models by taking ratios of their respective $\widehat{\mathcal{Z}}${s} to compute a Bayes factor or posterior odds ratio (POR) for assessing the evidence of the $n^{\rm th}$ planet.
Methods that initially appear to generate biased estimates of \logZ\, could provide an accurate odds ratio if the apparent bias cancels out.  

\begin{figure}
\centering
\includegraphics[angle=270, width=0.99\columnwidth]{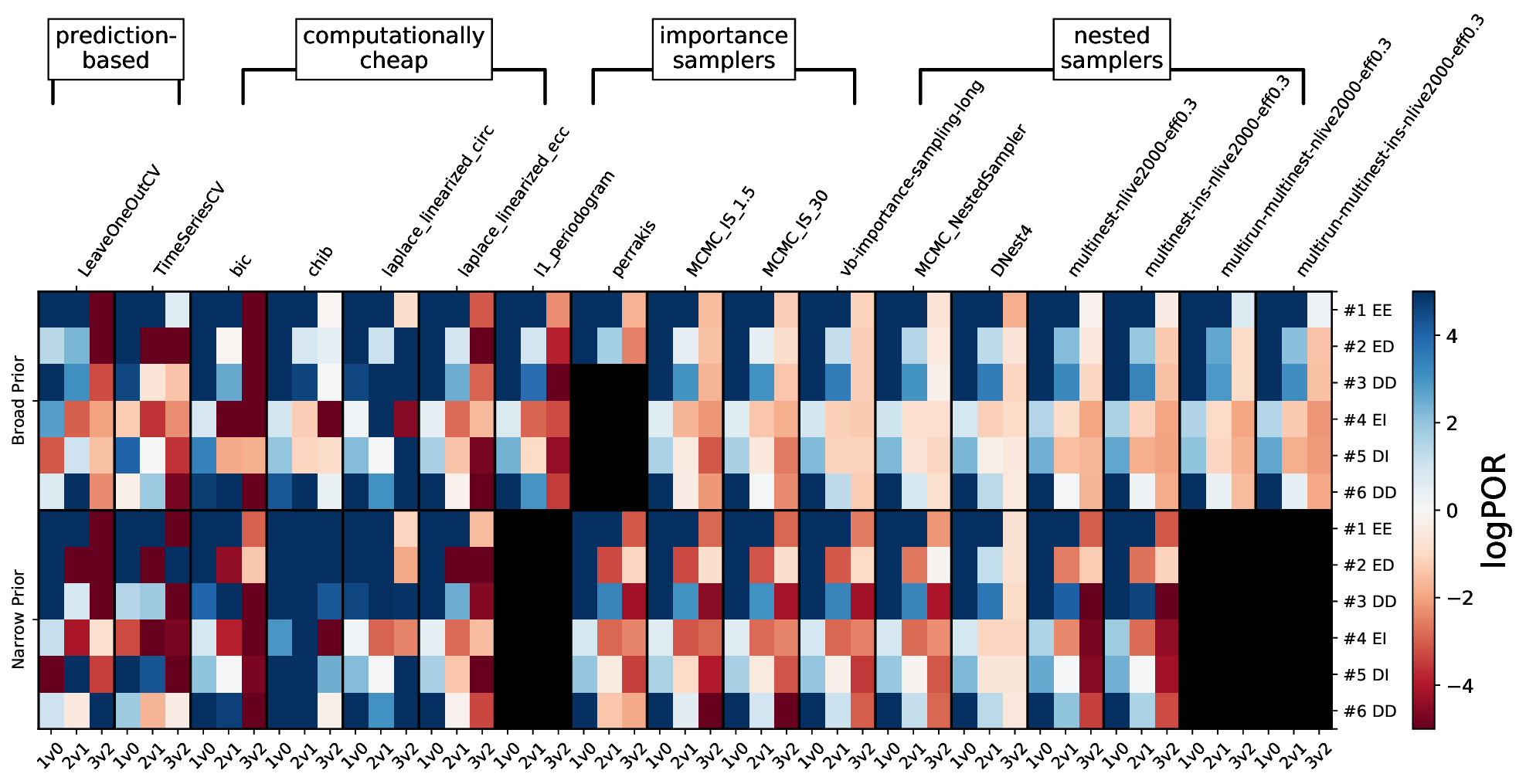}
\caption{\footnotesize{Summary of $\log$POR results across all datasets and models. A row of pixels corresponds to an odds ratio of an $n$ versus ($n+1$)-planet model comparison (i.e., 1v0, 2v1, 3v2). Pixel columns correspond to one of the six datasets, and we also denote the detectability of the two injected planets (easy=``E'', difficult=``D'', impractical=``I''). Rows of pixels are grouped by method with black outlines. The left (right) grouped column corresponds to the model with narrow (broad) period priors. Pixel colors indicate the $\log$POR value for that particular dataset and model pair to be compared: blue pixels favor the ($n+1$)-planet model, red pixels favor the $n$-planet model. Any $|\log$POR$|$ value greater than 5 is set to a color at the end of the colorscale. Black pixels are unreported values.}}.

\label{fig:results-logPOR}
\end{figure}

Figure \ref{fig:results-logPOR} shows the POR results for each method and dataset in a format very similar to that of Figure \ref{fig:results-logZ}.
However instead of results for each individual $n$-planet model, each pixel corresponds to the POR for a particular pair of models to be compared (for a given method, prior, and dataset).
For instance, a pixel corresponding to the 1-planet vs 0-planet model comparison is denoted as simply ``1v0.''
The color of each pixel is $\log$10 of the POR and the colorscale spans 10 orders of magnitude in POR.
In essence, the bluer pixels favor the ($n+1$)-planet model, redder pixels favor the $n$-planet model, and pale pixels find roughly similar evidence for the $n$ and ($n+1$)-planet models.
Black pixels are unreported values.

In addition to the Bayesian methods shown in the previous figures, Figure \ref{fig:results-logPOR} \REWRITE{also includes two results based on prediction-based methods:  Leave-One-Out Cross Validation, and Time-Series Cross Validation. }
In each case, the team was asked to report a quantity that would be as analogous to a POR as practical given their method.

We discuss several trends in the computed odds ratios across datasets, priors, and method class.  
After results were submitted, we revealed that each dataset contained two planets with different levels of detectability (see \S\ref{models-phys}).
Note that there was an error in the evidence calculations of the $\ell_1$ periodogram, and these were revised after the true answers were revealed.

\subsubsection{Initial Observations for Posterior Odds Ratio Estimates}

\REWRITE{Nearly all methods found odds ratios favoring $\mathcal{M}_1$ over $\mathcal{M}_0$.
There was more variability across methods when comparing the evidence for $\mathcal{M}_2$ and $\mathcal{M}_1$.
Aside from a few exceptions, methods generally did not find odds ratios favoring $\mathcal{M}_3$, across all datasets.}

\subsubsection{Results for Posterior Odds Ratio by Method Class}
We previously identified four classes of algorithms based on everyone's submissions: \REWRITE{ Bayesian computationally cheap methods, Bayesian importance samplers, and Bayesian nested samplers, and prediction-based methods.}
The latter two classes of methods require large numbers of model evaluations ($>{10}^3$) to compute $\mathcal{Z}$.
The former two are comprised of (semi-)analytic methods or methods that require relatively fewer model evaluations.

We find the numerical Bayesian methods qualitatively agree on the strength of the evidence for $n$ versus $(n+1)$ planets for nearly all datasets and model comparison permutations considered.  
Even when they do favor detecting an additional planet, these numerical methods tend to report less extreme PORs than the computationally cheap methods, as denoted by the paler pixels for 2v1 and 3v2 comparisons.
However, \REWRITE{the computationally cheap methods and prediction-based often} do not agree on the sign or strength of the evidence for finding an additional planet.
Furthermore, they tend to have a much stronger interpretation for either $n$ or ($n+1$)-planet models, as denoted by the more saturated pixels.

Of the computationally cheaper methods, the Laplace approximation using a linear approximation for eccentric orbits also displayed qualitative agreement with the more computationally expensive methods.
We address this importance in Section \ref{sec:discussion-promisingmethods}.

\subsubsection{Results for Posterior Odds Ratio by Dataset and Priors}

Here, we assess the reported odds ratios in light of the expected difficulty to detect the planets in each dataset. 
Dataset 1 contained two easily detectable planets.
Dataset 2 contained an easily detectable planet and two planets that we expected would be difficult to detect.
Datasets 3 and 6 contained two planets that we expected would be difficult to detect.
These two datasets used the same planet masses and orbits, but different zero-point offsets, observation times, and realization of measurement noise.
Datasets 4 and 5 had ``easy-impractical'' and ``difficult-impractical'' planets respectively.

For the broad prior, most methods found decisive evidence for at least one planet in Datasets 1, 2, 3, and 6.
The notable expectations were the \REWRITE{prediction-based methods}, which disagreed on the evidence for one-planet in Datasets 2, 5, and 6.
In particular, Leave-One-Out CV found marginal evidence for a planet in Datasets 2 and favored no planets in Dataset 5.
All of the remaining methods reported qualitatively similar results for the 1v0 case.
For the narrow prior, we see the Cross-Validation methods had similar disagreements for the 1v0 case in the same datasets.
Moreover, Chib's approximation had a much stronger 1v0 interpretation for Datasets 4 and 5 than other methods.

For both priors, there is more interesting variability in the POR for the 2v1 and 3v2 cases.
There are only two planets in each dataset, so the ``correct'' result is unlikely to have a POR strongly favoring $\mathcal{M}_3$, but could have a POR either near unity or strongly favors $\mathcal{M}_1$ or $\mathcal{M}_2$.

For Dataset 1, all methods found strong evidence for at least 2-planets. \REWRITE{Overall, this matches well with the planets' expected level of detectability. The only exception was Chib's approximation, which found strong evidence for 3-planets when the narrow prior was imposed.}
For Dataset 2 and the broad prior, all methods found strong evidence for 1-planet and most found weak to marginal evidence for 2-planets.
For the narrow prior, most methods did not find evidence for a second planet, but the narrow prior interval did not \REWRITE{bracket} the true orbital period for the second planet (120.5 days).
For Dataset 3, all of the Bayesian methods found evidence for both planets using either set of priors.
The narrow prior \REWRITE{bracketed} the true orbital period values (40.4 and 91.9 days).
For Dataset 4, methods typically found weak evidence for 1-planet and no evidence for more planets.
The supposedly easy-to-detect planet had $P=169.1$ days, $K=1.58$ m/s, and $e$=0.22.
Perhaps having a $P$ near half the Earth's orbital period and this particular noise realization made it more difficult to detect than expected.
For Dataset 5, methods typically found weak evidence for 1-planet and comparable to no evidence for a second planet, similar to Dataset 4.
In this case, the narrow prior did \REWRITE{bracket} the true orbital period values (31.1 and 10.9 days).
For Dataset 6, methods found strong evidence for at least 1-planet and mostly weak evidence for 2-planets.
These conclusions are moderately different than those for Dataset 3, which contained the exact same planets.

Comparing results for the narrow and broad priors, most methods reported less decisive evidence against 3-planets when they were allowed to choose a planet at any orbital period (i.e., paler red pixels in the right grouped column than the left).
When the narrow prior was imposed, methods typically found evidence for fewer planets.

Note that these odds ratios calculations are based on a physical model that assumes Keplerian orbits.
In some cases, one of the three planets was very closely spaced to another planet (e.g., as imposed by the narrow priors for Dataset 1 and 2). 
We suspect that these scenarios would likely break the Keplerian assumption, and if teams had been instructed to apply an n-body model, then evidence calculations might be affected.

\section{Discussion}
\label{sec:discussion}

The Evidence Challenge was envisioned as an opportunity to empirically characterize the accuracy, precision, and robustness of various methods for computing the marginal likelihood of realistic RV datasets.  

\subsection{Scatter in estimates}
Upon characterizing the dispersion in $\logZ$, we find reasons for both caution and optimism.

On one hand, estimates for \logZ\ often differed by one to two orders of magnitude for the test cases considered.
This dispersion is seen across different classes of methods and even within some individual methods.
Furthermore, the internal estimates of uncertainty in \logZ\ often significantly underestimated the observed dispersion of estimates.  
For the methods that estimated the uncertainty in \logZ\ based on multiple runs, the Monte Carlo uncertainties sometimes spanned $>$1 orders of magnitude, particularly for multi-planet models.
Therefore, we recommend caution when claiming strong evidence for multiple planets based on an estimated posterior odds ratio within a few orders of magnitude of unity.

On the other hand, it is reassuring to find that the computationally intensive Bayesian methods provided posterior odds ratios that would lead to similar qualitative conclusions (i.e., favoring $n$-planet or ($n+1$)-planet model by at least $>10^4$, or too close to call).
For datasets with many high-precision observations (such as considered here), the posterior odds ratio is likely to deviate from unity by many orders of magnitude, allowing for robust conclusions despite the limitations of existing methods for estimating marginal likelihoods. 
However, we caution that the posterior odds ratio is more likely to be within a few orders of magnitude of unity for smaller datasets and/or datasets with reduced measurement precision. 
Additionally, the observed dispersion in marginalized likelihoods increases with the number of planets in the model.
Therefore, we caution that even greater estimated posterior odds ratios are likely necessary to support strong claims for the evidence of more than three planets in a given system, if they are derived with different methods.

Conventional wisdom would suggest that computationally cheap methods are not as robust at estimating $\logZ$ or $\log$POR as the more computationally intensive methods.
Indeed, most of the computationally cheap methods often disagreed with the computationally intensive methods, especially for cases where the latter found an odds ratio within a two orders of magnitude of unity.
Furthermore, the likelihood shows complex, multimodal shapes in some datasets, which are missed when only characterising the best fit location.

\subsection{Non-Bayesian methods}

\REWRITE{Here we discuss some alternatives to the Bayesian evidence for deciding the detection of a planet.}

\REWRITE{Among the submitted results, the prediction-based methods often resulted in a different qualitative conclusion about the evidence for a second or third planet. This is not surprising, since these methods are not estimating the posterior odds ratios.}  
\REWRITE{Future improvements to these methods might reduce the number of false positives and false negatives, including via calibration of the algorithm.}

\REWRITE{Information criteria, rooted in information theory, quantify if the additional complexity of models is worth storing. One example is the Akaike Information Criterion (\cite{AIC}, see also \cite{watanabe2013widely}). For our sample size ($N=200$), the AIC punishes complex models more severely than the BIC (the $2\times k$ term is replaced by $7.39 \times k$). Considering the results of the BIC, this would introduce several false negatives (the white pixels in the BIC results of Figure~\ref{fig:results-logPOR}).}

\REWRITE{A frequentist approach to distinguish models would be to identify the maximum likelihood $\mathcal{L}_\mathrm{max}$, and investigate whether this statistic is substantially higher in the more complex model than in a simpler one ($\mathrm{LR}=\mathcal{L}_{i+1,\mathrm{max}} / \mathcal{L}_{i,\mathrm{max}}$). Because the simpler model is embedded in the parameter edge of the more complex model ($K_(i+1)=0$), analytic formulas do not hold to judge the LR. Instead, the significance (p-value) of the LR improvement has to be found by generating random data sets assuming the best-fit parameters of the simpler model, fitting both the simpler and more complex model (parametric bootstrap). This is however computationally expensive, and even more so when $\mathcal{Z}$ would be considered as the statistic.}

\subsection{Caveats and Limitations}

This Evidence Challenge considered only six datasets, which is not enough to represent the full diversity seen in real RV datasets (e.g., number of observations, observing baselines, planet SNRs, time series, etc.).
Therefore, it is unclear how robust our conclusions are to a wider range of RV data quality.
These datasets were designed considering the expected future of the RV field: prioritizing low mass planets (low RV SNR) with hundreds of observations over multiple observing seasons.
On one hand, these specific concerns about the accuracy and precision of marginal likelihood estimates demonstrated here are not necessarily problematic for the vast majority of previously RV discovered planets, since most of these planets are relatively more massive (i.e., higher RV SNR) and often had complementary follow-up observations.  
Furthermore, this analysis was based on RV observations alone with no other forms of supporting ancillary, activity-sensitive data (e.g., transits, activity indicators).

The Evidence Challenge provided an idealized scenario where each team was provided a standardized model, set of priors, and the precise noise model that was used to generate the RV data.
When analyzing real data, different teams might reasonably choose to impose different sets of priors.
\REWRITE{In such cases, if teams explicitly state their statistical model and provide posterior samples, other researchers could reweigh the results using another set of priors} (assuming there is sufficient overlap between the posteriors under the two priors).
Unfortunately, the exact noise model that generates real data will not be available.
Therefore, conclusions about the strength of the evidence for an $n^{\rm th}$ planet must be tempered by uncertainty in the noise model.
In the spirit of starting simple, each team was provided the exact values of the other hyperparameters in Equation \ref{equation-kernel} (e.g., stellar rotation period, correlation lengths) and instructed to hold these parameters fixed.  
These would need to be estimated or marginalized over for real data \citep[e.g.,][]{faria2016,millholland2018}, ideally at the same time as the planetary parameters.
Marginalizing over additional hyperparameters would have made it more challenging to estimate evidence accurately, due to increased dimensionality and the potential for multi-modal posteriors \citep{dumusque2017}.
In addition to these numerical difficulties, there is an additional challenge of model misspecification, since realistic astrophysical noise is likely more complex than a simple mathematical model.

With recent improvements in the precision, accuracy, and stability of spectrographs, the limitations of current and next-generation RV surveys will often come from stellar astrophysics, rather than random measurement noise.
Astronomers are actively seeking new methods of characterizing intrinsic spectroscopic variability of the target stars due to a wide variety of effects (e.g., star spots, granulation, convection, pulsations).
In principle, one could estimate the evidence for a model which includes a likelihood on $\vec{d}$ including both apparent RV measurements and various stellar activity indicators (e.g., $\log{R^\prime\{hk\}}$).
Multivariate Gaussian process noise models seem a particularly promising approach \citep[e.g.,][]{rajpaul2016,jones2018}.  
However, performing the computations necessary for rigorous statistical inference with such models will be even more challenging than for the simple noise model considered in this Evidence Challenge.
As astronomers develop more powerful statistical models for analyzing spectroscopic time-series, it will likely be useful to perform additional data challenges with such models.

In principle, it is possible that the observed \DlogZ\, overestimates the dispersion if each method were ideally implemented and tuned.
Teams analyzed these datasets independently using a wide variety of codes and tools on platforms with different compilers, libraries, operating systems, and hardware.
We can not eliminate the possibility that some teams may have reported results based on a buggy implementation of an algorithm or chose \REWRITE{algorithm settings} that resulted in less than ideal performance of the algorithm. 
In any case, the observed \DlogZ\, reflects a combination of random and systematic errors intrinsic to each method, finite-precision numerical calculations, and perhaps human errors, similar to that which would arise if these teams had been analyzing real astronomical datasets.

Finally, the evidence estimates submitted do not fully represent the array of statistical methods available to perform quantitative model comparison \citep[e.g.,][]{fordgregory2007}.
In particular, no results were submitted based on methods using thermodynamic integration.
It would also be useful to investigate other computationally cheap methods such as AIC, DIC or WAIC \citep{gelman2014}.  
Other researchers are encouraged to develop and apply alternative methods to the same datasets available in the Evidence Challenge Github repository, as they evaluate methods and implementations.

\subsection{Computational Costs}

\begin{deluxetable*}{l|r|c}
\tablecaption{Number of likelihood evaluations ($n_{\mathcal{L}}$) reported to calculate \logZ\, for Dataset 2 and $\mathcal{M}_2$, assuming broad period priors. Similar methods with different tuning parameters or simplifying assumptions are grouped together. The median \logZ\, for this set of methods is -166.005. \label{tbl-nlike}
}
\tablehead{
\colhead{Method (directory name)} & \colhead{$n_{\mathcal{L}}$} & \colhead{$\logZ\, - \medianlogZ$}
}
\startdata \hline
$\texttt{chib}$ & 1000000 & -0.342 \\
$\texttt{laplace\_linearized\_circ}$ & 264 & 1.012 \\
$\texttt{laplace\_linearized\_ecc}$ & 319 & -0.128 \\ \hline
$\texttt{vb-importance-sampling}$ & 261979 & -0.449 \\
$\texttt{vb-importance-sampling-long}$ & 2883983 & -0.012 \\
$\texttt{MCMC\_NestedSampler}$ & 8814939 & 0.062 \\ \hline
$\texttt{multinest-nlive400-eff0.3}$ & 173460 & -0.516 \\
$\texttt{multinest-nlive400-eff0.01}$ & 768668 & 0.551 \\
$\texttt{multinest-nlive2000-eff0.3}$ & 1017587 & -0.578 \\
$\texttt{multinest-ins-nlive400-eff0.3}$ & 173460 & 0.018 \\
$\texttt{multinest-ins-nlive400-eff0.01}$ & 768668 & 0.984 \\
$\texttt{multinest-ins-nlive2000-eff0.3}$ & 1017587 & -0.34 \\ \hline
$\texttt{multirun-multinest-nlive400-eff0.3}$ & 1164856 & 0.012 \\
$\texttt{multirun-multinest-nlive400-eff0.01}$ & 5093831 & 0.588 \\
$\texttt{multirun-multinest-nlive2000-eff0.3}$ & 5132502 & -0.234 \\
$\texttt{multirun-multinest-ins-nlive400-eff0.3}$ & 1164856 & 0.107 \\
$\texttt{multirun-multinest-ins-nlive400-eff0.01}$ & 5093831 & 1.106 \\
$\texttt{multirun-multinest-ins-nlive2000-eff0.3}$ & 5132502 & -0.204 \\ \hline
\enddata
\end{deluxetable*}

On top of the reported evidence values, roughly half of the teams also provided benchmarking results for their methods, detailing the number of likelihood evaluations, wall-clock time, and/or number of cores required for the evidence calculation.
This gives a useful, yet incomplete picture on the efficiency of these methods. We will take a qualitative look at these results, focusing on the total number of likelihood evaluations ($n_{\mathcal{L}}$, henceforth) of one particular problem: Dataset 2 and $\mathcal{M}_2$, assuming broad priors.
Table~\ref{tbl-nlike} shows $n_{\mathcal{L}}$ and the evidence estimate \logZ\, relative to the median.

Focusing first on computationally cheap methods (first three rows in Table~\ref{tbl-nlike}), the Laplace approximation required the fewest $n_{\mathcal{L}}$.
These were mainly used in the grid search for the ($n+1$)$^{\text{th}}$ planet, since the integral calculation itself was analytic.
For the other datasets, Ford reported a wide range of $n_{\mathcal{L}}$, from $n_{\mathcal{L}} = 1$ for $\mathcal{M}_0$ up to $n_{\mathcal{L}} \sim\,10^5$ for $\mathcal{M}_3$.
In general, $\widehat{\mathcal{Z}}$ computed via the circular approximation deviates from other methods by one to several orders of magnitude.
For Chib's approximation, Feng used a constant $n_{\mathcal{L}} = 10^6$ across all datasets and models.

The remaining methods listed in Table~\ref{tbl-nlike} are computationally expensive, and include variational Bayes with importance sampling, MCMC-based nested sampling, and variations of \MULTINEST.
For the MCMC nested sampler, Rajpaul used the largest $n_{\mathcal{L}}$ for this particular case.  A future study could investigate whether it is possible for this algorithm to achieve similarly accurate result with fewer $n_{\mathcal{L}}$.
For other datasets and models, the number of model evaluations spanned a large range ($n_{\mathcal{L}}\sim\,10^6 - 10^7$) with no clear pattern across different models or datasets.
For \MULTINEST, $n_{\mathcal{L}}$ increases for larger \texttt{nlive} and smaller \texttt{eff}.
However, interpreting the number of likelihood evaluations requires also to understand the robustness of results. The \logZ\, differences of \MULTINEST{} variations are analysed in detail in Section~\ref{methods-teampuc-multinestscatter}. 
Briefly, low efficiency runs (i.e., the \texttt{-eff0.01} suffix) show consistent estimates, while \texttt{-eff0.3} is unstable.
This could suggest that the true \logZ\, is above the median (+0.5 or +1.0).
In all variants, multiple runs increased the \logZ\, estimate, indicating that substantial parts of the integral are often missed.
This is also seen in the importance sampling technique increasing the estimate when run longer.
With this in mind, $n_{\mathcal{L}} > 10^6$ with low efficiency and/or multiple runs seem to be required.

The same trends also hold for the importance nested sampling estimator, which use the same run.
However additionally, enabling importance nested sampling requires substantially more memory. Unexplained systematic differences between the importance nested sampling and classic importance nested sampling remain (also seen in Table~\ref{tbl-nlike}). These indicate that the \MULTINEST\, integrations is encountering some difficulties.

Some methods like Chib's approximation and the MCMC + importance sampling ratio estimator rely on a set of posterior samples to estimate $\mathcal{Z}$.
If reliable posterior samples were already available (via a database or published along with an RV data analysis), then this would substantially reduce the number of additional likelihood evaluations needed.

\subsection{Promising Methods for Future Studies}
\label{sec:discussion-promisingmethods}

With the aforementioned results and caveats in mind, we now address the fourth question of the Evidence Challenge: which methods should be recommended, avoided, and/or further developed?
In practice, it is difficult to reliably estimate the true value for the odds ratio of high-dimensional (12+ parameter) models.
However, we consider the numerical Bayesian methods (i.e., MCMC+importance sampling, variational Bayes+Importance sampling, the Perrakis estimator, MCMC+Nested Sampler, \sw{DNest4} and multirun-\MULTINEST) to be more reliable since they provided a consistent set of conclusions.
Among this set of evidence estimators, \sw{DNest4} demonstrated the widest deviations from the consensus of the other methods.  
To reiterate, we found that it is important to estimate uncertainties in the evidence based on multiple independent runs of Monte Carlo algorithms, rather than trusting internal uncertainty estimates based on a single run or posterior sample.

We also identify one computationally cheap method that was consistent with the numerical methods: the Laplace approximation with a linearized eccentric model.
This is important because this suggests a (semi)-analytic method has comparable performance to methods that often require orders of magnitude more model evaluations.
Other than the grid search to find plausible planets, the most computationally expensive part of the Laplace approximation is a single log determinant calculation of the Hessian matrix described in \S \ref{methods-ford}.
For this study, the Laplace approximation demonstrates a nice balance between efficiency and robustness, which would be particularly appealing for analyzing a large number of datasets or datasets with expensive model evaluations.
Since this model adopted a linear expansion of the Keplerian motion it would not be appropriate for application to systems with ``high'' eccentricity planets.
For planets near the threshold of detection, the linear approximation can be much more precise than measurement precision even sizable eccentricities (e.g., 0.3), since the error term is of order $\sim K e^2$.
We also note that the BIC results generally shared the same sign, but sometimes claimed much more extreme odds ratios in cases where other methods found more marginal ratios.

\subsection{Areas for future research}
Recently, \citet{butler2017} released RVs for 1642 stars and identified/classified significant signals for each case.  
Having demonstrated the viability of multiple methods for computing evidence for 1, 2 and 3 planet models, one could apply these methods to perform a systematic analysis of these systems.  
Due to the varied number and precision of RV observations, one should estimate the uncertainty for evidence of each combination of model and dataset.  
When interpreting the results of such an analysis, one should also consider the robustness of conclusions to the choice of likelihood function and potential for model misspecification.

Previous studies that have compared methods for computing marginal likelihoods for RV data were limited to relatively few datasets.  
Our study was also limited to six RV datasets and four $n$-planet models, partially because some methods would not scale well to thousands of synthetic datasets.
Regardless, this first step at identifying efficient methods will help drive next-generation RV analyses.

Our results illustrate a few of the challenges in the responsible analysis of RV datasets.
In order to support current and upcoming RV planet surveys, we recommend much broader evidence challenges that would involve analyzing large number of simulated datasets, so as to understand the rate at which different methods favor non-existent planets.
Such studies could:
(1) test the robustness \REWRITE{and false discovery rates} of the algorithms that performed well over a wider range of RV baselines, cadences, and planet SNRs by analyzing thousands of simulated RV datasets;
(2) compare estimates of the evidence for more sophisticated noise models or more sophisticated physical models (i.e., some that impose stability criterion for multi-planet systems); and
(3) compare estimates of the evidence for heterogeneous datasets (i.e., RVs + activity indicators).
Interpreting results from the current and next generation of RV surveys will be increasingly complex (e.g., combining large number of observations, correlated noise models, stellar activity indicators).
Therefore, studies such as those recommended above will be critical to establishing the robustness of RV detections and mass measurements.

\acknowledgments
B.E.N.~ acknowledges support from CIERA and the Data Science Initiative at Northwestern University.
B.E.N.~ also acknowledges support from the computational resources and staff contributions provided for the Quest high performance computing facility at Northwestern University which is jointly supported by the Office of the Provost, the Office for Research, and Northwestern University Information Technology.
E.B.F.~ acknowledges support from the Institute for CyberScience and the Center for Exoplanets and Habitable Worlds, which is supported by The Pennsylvania State University, the Eberly College of Science, and the Pennsylvania Space Grant Consortium. E.B.F.~ also acknowledges support from NSF award 1616086, NASA Exoplanets Research Program grant \#NNX15AE21G, and supporting collaborations within NASA's Nexus for Exoplanet System Science (NExSS).  
J.B.~ acknowledges support from the CONICYT-Chile grants Basal-CATA PFB-06/2007, FONDECYT Postdoctorados 3160439 and the Ministry of Economy, Development, and Tourism's Millennium Science Initiative through grant IC120009, awarded to The Millennium Institute of Astrophysics, MAS. This research was supported by the DFG cluster of excellence `Origin and Structure of the Universe.' 
R.C. thanks the Canadian Institute for Theoretical Astrophysics
for use of the Sunnyvale computing cluster throughout
this work.
R.C. is partially supported in this work by the
National Science and Engineering Research Council of
Canada.
J.P.F. acknowledges support from the fellowship with reference SFRH/BD/93848/2013 funded by 
Funda\c{c}\~ao para a Ci\^encia e Tecnologia (FCT, Portugal) and POPH/FSE (EC),
and also from FCT through national funds and FEDER through COMPETE2020, in the form of grants UID/FIS/04434/2013 \& POCI-01-0145-FEDER-007672 and PTDC/FIS-AST/1526/2014 \& POCI-01-0145-FEDER-016886.
R.F.D. and J.P.F. acknowledge financial support from the organizers to attend the EPRV3 meeting. 
N.C.H. acknowledges the financial support of the National Centre for Competence in Research PlanetS of the Swiss National Science Foundation (SNSF)

\software{emcee \citep{emcee}, george \citep{george}, matplotlib \citep{matplotlib}, MultiNest \citep{Feroz2009,Feroz2013}, PyMultiNest \citep{pymultinest}, pypmc \citep{pypmc}}

\appendix

\section{Appendix information}
\label{app:methods}
In this section, we will describe the methods presented in Section \ref{sec:methods} in greater detail.
As mentioned previously, some variables in the following subsections may share common notation as other variables seen in the main manuscript.
For such conflicts, we recommend the reader treat these variables as ``locally defined'' within that method's subsection.

\subsection{Feng, BIC}
The BIC measures the plausibility of a model through the Laplace approximation of a Gaussian likelihood distribution. \REWRITE{It further assumes that under $N\rightarrow \infty$, the posterior becomes dominated by a infinitely narrow peak, which is insensitive to the prior and linear data terms \citep{Konishi2008}.} Despite these strong simplifications, the BIC is frequently used because the posterior density for many inference problems is dominated by a single Gaussian-like distribution and is not sensitive to prior distribution.
To compare with other evidence estimators, we follow \citep{kass95} to approximate the evidence by using $E=e^{-{\rm BIC}/2}$, where $BIC=-2\ln{\mathcal{L}_{\rm max}}+k\ln{N}$, $\mathcal{L}_{\rm max}$ is the maximum likelihood, $k$ is the number of free parameters, and $N$ is the number of data points.
Considering such approximation, we use the evidence ratio to assess the performance of the BIC.

The maximum likelihood is calculated through MCMC posterior sampling using DRAM, an adaptive Metropolis algorithm \citep{haario06}.
The Gelman-Rubin criteria is used to judge whether a chain approximately converges to a stationary distribution \citep{gelman92}.
We draw one million posterior samples using DRAM, drop the first half of the chain as burn-in part, divide the rest sample into one hundred sub-samples, and calculate the distribution of $\mathcal{L}_{\rm max}$ and BIC from these sub-samples.
\label{sec:bic}

\subsection{Cloutier, Cross-Validation}
\label{methods-cloutier}
\label{sec:method-cv}
In general, cross-validation (CV) is a technique used to evaluate the predictive power of a particular model on an input dataset.
CV is a commonly used to assess model over-fitting as overly complex models can often be fine-tuned to produce a high data likelihood while not necessarily generalizing to unseen data (e.g. future observations) and thus demonstrating poor predictive power.

\REWRITE{In CV, the first step is splitting the input dataset in a training and a test data set. The model parameters are optimized with the training data set. The predictive power of this model with the best-fit parameters is then evaluated on the (previously unseen) test data. The `score' is a user-defined objective function that measures the quality of the prediction.
This procedure is often repeated for multiple possible splits of the data. To summarize the model's predictive power, the scores are averaged over the split to give a single score.
For selecting a model, competing models can be compared by their scores. Generally, overly complex models over-fit the training data and poorly predict the test data, giving low scores. Overly simple models produce low scores in general, because they poorly fit both training and test data. Good models generalize well from the training data to the test data and have the highest scores.}

We note that numerous flavors of CV exist and the exact nature of the train/test splitting can vary depending on the flavor of CV used. 
A general summary of the various CV techniques can be found in \cite{arlot10a}.

\subsubsection{Leave-One-Out CV}
\label{sect:loocv}
Leave-one-out CV (LOOCV) represents a common form of train/test splitting in CV. When considering the set of $N$ radial velocity observations $\vec{v}$, LOOCV first splits the data into $N$ training/testing sets.
\REWRITE{In each split one observation is left out as the test data and the training set contains the other $N-1$ observations.}
\REWRITE{For each split, the best-fit parameters $\bar{\theta}$ for each model $\mathcal{M}_n$ under consideration are optimized using a user-defined technique such as least-squares minimization or gradient descent methods. As such, the resulting best-fit parameters may be the maximum a-posteriori point estimate, the maximum likelihood parameters, or similar depending on the employed objective function. We have adopted to identify the maximum a-posteriori model parameters via MCMC ensemble sampling \citep{emcee} to search for global maxima in the posterior parameter space. For each model we employed 100 chains that are run until the chain lengths are $\gtrsim 10$ times the average autocorrelation time among the chains. Each $P_i$ is initialized to the period value of a significant peak in the Lomb-Scargle periodogram of the RVs (in descending order of power) while all other orbital parameters are assigned random initial values drawn from their respective priors. In subsequent MCMC simulations on the same dataset but under a unique planet model, all parameters for planets featured in both models were initialized to their MAP values from the previous MCMC.}

The predictive power of the model is then calculated as the value of the objective function of the testing set under the optimized model. Here this is the likelihood $\cal{L}$ evaluated only on the left out data point. The final score for each model's predictive power is obtained from the median score among the $N$ splits. We quantify the score dispersion with the median absolute deviation. 



\REWRITE{We now discuss how to quantify the preference for one model over another within the framework of this challenge. The median score describing a model's predictive power clearly is calculated from the median  $\ln{\mathcal{L}(\bar{\theta})}$ of a single data point. This differs from the Bayesian evidence, which integrates the value $\mathcal{L}(\bar{\theta})$ computed on the full input data set. Therefore, these values cannot be compared directly. However, a useful analogy may be made between the score ratio obtained from LOOCV and the evidence odds ratio obtained from Bayesian techniques. Recall that the training set in each LOOCV split is a single measurement. Therefore, in order to compare scores to Bayesian evidences one must account for the difference in scale between individual observations and the full $N$ data set. An applicable correction is applied by multiplying the score per observation ---obtained in each split from LOOCV--- by $N$. The ratio of median scaled scores can then be used to compute the odds ratio from LOOCV. It is worth re-emphasizing that odds ratios derived in this way are not the same as Bayesian odds ratios. }


\subsubsection{Time-Series CV}
\label{sec:timeseriescv}
In general, LOOCV (see Sect.~\ref{sect:loocv}) is only applicable when the measurements within in the input dataset are independent.
\textbf{In the case when the input dataset features correlated observations, standard CV techniques such as LOOCV need to be modified as removing a single random observation fails to remove all associated information due to temporal correlations with the remaining observations.}
Radial velocity time-series are often highly correlated in time due to the presence of periodic planetary signals and correlated noise arising from stellar activity \citep[e.g.][]{astudillodefru17, bonfils18, cloutier17}.
The latter signal is present in all of the simulated time-series used throughout this study and consequently warrants an alternative form of CV.

One such form of CV used when treating temporally correlated datasets is known as time-series CV (TSCV).
TSCV is a variant of LOOCV that measures the predictive power of competing models on a set of observations that are known to be correlated in time.
The procedure follows almost identically to LOOCV but differs in the method of train/test splitting. In TSCV, training sets are \REWRITE{constructed from a chronologically ordered input data set $\bar{v} = v_1,\ldots,v_N$. The training set $t$ ($t \in [N_{\text{min}},N-1]$) contains the data $v_1,\ldots,v_t$, and the corresponding testing set is $v_{t+1}$, the chronologically next observation.}
For each train/test split the value of the index $t$ is increased from a minimum training set size $N_{\text{min}}$, which we fix to 20, to the full size of the input data set minus one (i.e. $N-1$).
Therefore, just like in LOOCV, the testing set in each split is always a single observation\REWRITE{, and} the scale of each split's calculated score is consistent with the values obtained from LOOCV.
TSCV features \REWRITE{only $N-N_{\text{min}}-1$ splits, compared to the $N$ splits computed in LOOCV.}
\textbf{Quantifying each model's predictive power proceeds identically to LOOCV via the median score and its median absolute deviation over the $N-N_{\text{min}}-1$ splits.
The odds ratio comparing competing models is again computed after scaling each model's score per observation by $N$
, before computing the score ratios for each pair of competing models.}

\subsection{Ford, Laplace Approximation}
\label{methods-ford}
\label{sec:laplaceapprox}

The Laplace approximation can provide a fast and accurate method for approximating the integral of a function with a single dominant mode that is well separated from the boundary of integration domain.
In particular, consider the integral $\int dx\, \exp f(x)$ and insert the second-order Taylor series for $f(x)$, expanding about $x_o$ the location of the global mode.
Then 
\begin{equation}
f(x) \simeq f(x_o) + \frac{1}{2} \sum_{a,b} \frac{\partial^2 f}{\partial x_a \partial x_b} (x-x_o)^2, 
\end{equation}
and the first term can be brought outside the integral.
The remaining integral can be approximated analytically if one extends the limits of integration to infinity.
Then
\begin{equation}
\int dx\, \exp f(x) \simeq \left[ \frac{\left(2\pi\right)^2}{\left|\det H(x_o)\right|} \right]^{1/2} \exp f(x_o), 
\end{equation}
where $H(x_o)$ is the Hessian matrix, $\frac{\partial^2 f}{\partial x_a \partial x_b}$, evaluated at $x_o$.
The Laplace approximation can be understood as proportional to the maximum value of $\exp f(x)$ times the width of the global mode.
The maximum a posteriori value, the Akaike information criterion (AIC), and ``Bayesian'' Information Criterion which are sometimes used as heuristics for model comparison include the maximum posterior value, but do not properly account for the width of the posterior mode.
\REWRITE{In comparison to the BIC, the Laplace approximation here exploits information about both the priors and the posterior width.  Thus, it is expected to be more reliable when the number of observations is finite, and particularly for RV datasets where the number of observations is not very large.}

The accuracy of the Laplace approximation depends on the posterior density.
For the application to RV survey data, formally the posterior for models with $n\ge~1$ planets his highly multi-modal, particularly in terms of the orbital period.
Fortunately, the posterior for RV datasets is often dominated by a single posterior mode.
Indeed, one could adopt a criterion for ``detecting'' a planet based on the posterior probability distribution being dominated by a single mode.
Therefore, we anticipate that the Laplace approximation is likely to be accurate for a dataset with $n$ planets if all $n$ planets have been strongly detected, but is likely to be inaccurate for calculating the marginal likelihood to a model with $n+1$ planets.  

In practice, the most difficult part of approximating the marginal likelihood via the Laplace approximation is identifying the dominant posterior mode.
This is non-trivial for a full Keplerian model.
Further, it is possible for the formal posterior mode to occur at a very high eccentricity and to correspond to a such a narrow spike that the marginal likelihood is actually dominated by the integral around another mode.
While it is possible for the marginalized likelihood to strongly favor an $n$ planet model even if the posterior has multiple significant modes, this implies that there is significant uncertainty in the orbit of the object.
This has occurred in the literature for actual exoplanet datasets when aliasing issues cause there to significant uncertainty in the orbital period of planet \citep[e.g., 55 Cnc e,][]{dawsonfabrycky2010}.  
In principle, one could apply the Laplace approximation around multiple posterior modes to estimate the marginal likelihood.
For this study, we instead apply the Laplace approximation to a simplified model, so as to avoid this difficulty.
In particular, we construct one of two linearized models for the RV perturbation due to each planet.
In the first model, we assume that each planet follows a circular orbit and induces a stellar RV of $v_{\mathrm{pred}}(t|A,B,P) = A\cos(2\pi t/P) + B\sin(2\pi t/P)$.
In the second model, we adopt an epicycle approximation to each planet's orbit, in which case the RV perturbation can written as $v_{\mathrm{pred}}(t|A,B,P) = A_1 \cos(2\pi t/P) + B_1*\sin(2\pi t/P) + A_2 \cos(4\pi t/P) + B_2 \sin(4\pi t/P)$.
If the orbital period and the covariance matrix are fixed, then there is a single global mode and one can find the values of $A$ and $B$ which maximize the likelihood via linear algebra.
Once the posterior mode (conditioned on orbital period and parameters to the covariance matrix) is identified, one can rapidly evaluate the model and the Hessian at the posterior mode.  

To find the orbital periods corresponding to the posterior mode, we adopt an iterative approach adding one planet at a time.
When evaluating the marginal likelihood for the $n$ planet model, we perform a brute force grid search over the period of the $n^{\rm th}$ planet, while holding the orbital period of planets 1 through $n-1$ fixed at the values which maximized the posterior probability under the $n-1$ planet model.
The grid is uniformly spaced in orbital frequency with a density proportional to the frequency range being searched, the timespan of observations and the root mean square of the velocity residuals under the best-fit $n-1$ planet model.
To avoid local maxima due to aliases with previous planets, we exclude orbital periods periods within 20\% of the orbital period of one of the first $n-1$ planets identified.
We apply the Laplace approximation with either the circular or epicycle model to compute the posterior probability marginalized over all model parameters other than the orbital periods and the parameters in the covariance matrix.

For each set of orbital periods, we compute the posterior probability given the orbital period and marginalized over the covariance matrix (i.e., $\sigma_J$) using 40-point Gauss-Legendre quadrature, as provided by the Julia FastGaussQuadrature.jl package\footnote{\url{https://github.com/ajt60gaibb/FastGaussQuadrature.jl}}.  Initially, we attempted to perform integration over $\sigma_J$ via the Laplace approximation, but found that this often introduced a non-trivial error due to the cubic term in the expansion about the modal $\sigma_J$.  This approach is conceptually similar to the  Integrated Nested Laplace Approximations (INLA) technique for latent Gaussian models \citep{INLA}.

Finally, we integrate the posterior probability over the orbital period of the $n^{\rm th}$ planet via the Laplace approximation to arrive at the marginalized posterior probability given an $n^{\rm th}$ planet model, where orbits are approximated as circular or epicycles.
The orbital period of the $n^{\rm th}$ planet that maximizes the marginalized posterior probability is adopted for future calculations involving $n+1$ planets.  

The Laplace approximation combined with the circular model can be interpreted as a Bayesian periodogram, i.e., a brute force search/integration over orbital period combined with a fast approximate model conditional on the orbital periods.
This method has the advantage of performing a global search of parameter space for each planet.
We anticipate that the Laplace approximation will underestimate the marginal likelihood for models that include more planets that are justified by the data.
In these cases, multiple small posterior modes would contribute significantly to the marginalize probability, but our particularly implementation only includes one mode.
In principle, this could be addressed by summing over multiple posterior modes, but such generalizations are beyond the scope of this study.
In practice, this is not a serious limitation, since there is relatively little scientific value in precisely calculating the marginal probability for a model which is not dominated by a single mode (i.e., there are qualitative uncertainties in the orbit of at least one planet).  

We anticipate that our  Laplace approximation method will be accurate for planetary systems with weak to modest detections, as the posterior would be dominated by a single model and the RV amplitude is small enough that the deviations from circular orbit are small compared to the measurement precision.
In order to address this limitation, we performed a similar calculation using the epicycle approximation, so the physical model error is reduced from $O(K e)$ to $O(K e^2)$.
We anticipate that this will improve the Laplace approximation for planets with strong detections and modest eccentricities.
Unfortunately, this also comes with the risk of the model finding spurious posterior modes at high or even unphysical eccentricities.
We address the issue of unphysical eccentricities (i.e., $e\ge1$) when using the epicycle model by drawing 100 samples for the inferred $A$ and $B$ coefficients given the modal values of orbital periods and $\sigma_J$ and computing what fraction of those samples correspond to an eccentricity less than unity.
We multiplied the marginal posterior probability for that set of orbital periods by the fraction of accepted samples.
While this eliminated totally unphysical models, it does not make the physical model accurate in the high eccentricity regime.
For systems with high-eccentricity planets, our linearized models will introduce a non-random error term.
Curiously, it is also possible that the high computational efficiency of this method may result in it finding a narrow posterior mode that other methods may have overlooked due to the difficulty of performing a global search with a non-linear model.
Therefore, when there are significant differences between the marginal likelihood computed via the Laplace approximation and other methods using a Keplerian model, it may not be obvious whether the differences are primarily due to the limitations of the Laplace approximation, the use of an approximate physical models, or the more comprehensive search of parameter space possible with the Laplace approximation.  

\subsection{Hara, $\ell_1$ periodogram}
\label{method-hara}
\label{sec:hara}

\subsubsection{Overview}

In the present work, most of the presented techniques aim at approximating closely the evidence of a model with a given number of planets, in order to perform model comparison.
The method presented in this section is similar in that it aims at finding how many planets are orbiting a given star, but differs in that its initial goal is not to compute evidences.
Its aim is to provide a quick and reliable search for periodicities in radial velocity data while avoiding some caveats of the Lomb-Scargle periodogram~\citep{lomb1976,scargle1982} or its generalizations. 

Indeed it is well known that if several sources of periodicity are present in the signal, due to aliases combinations, the maximum of the periodogram might be attained at a period that does not correspond to any signal actually in the data~\citep{dawsonfabrycky2010}.
One solution to that problem is to search for several periods at once, which might be computationally costly. 

The alternative we suggest is not to search for best fitting models with one or a few periodicities, but directly for a Fourier spectrum of the true radial velocity signal.
This seemingly more complicated problem will be greatly simplified by an assumption: there are not many planets in the signal.
In other terms, the signal is sparse in the frequency domain.

\begin{figure}

\centering	
	
\includegraphics[width=12cm]{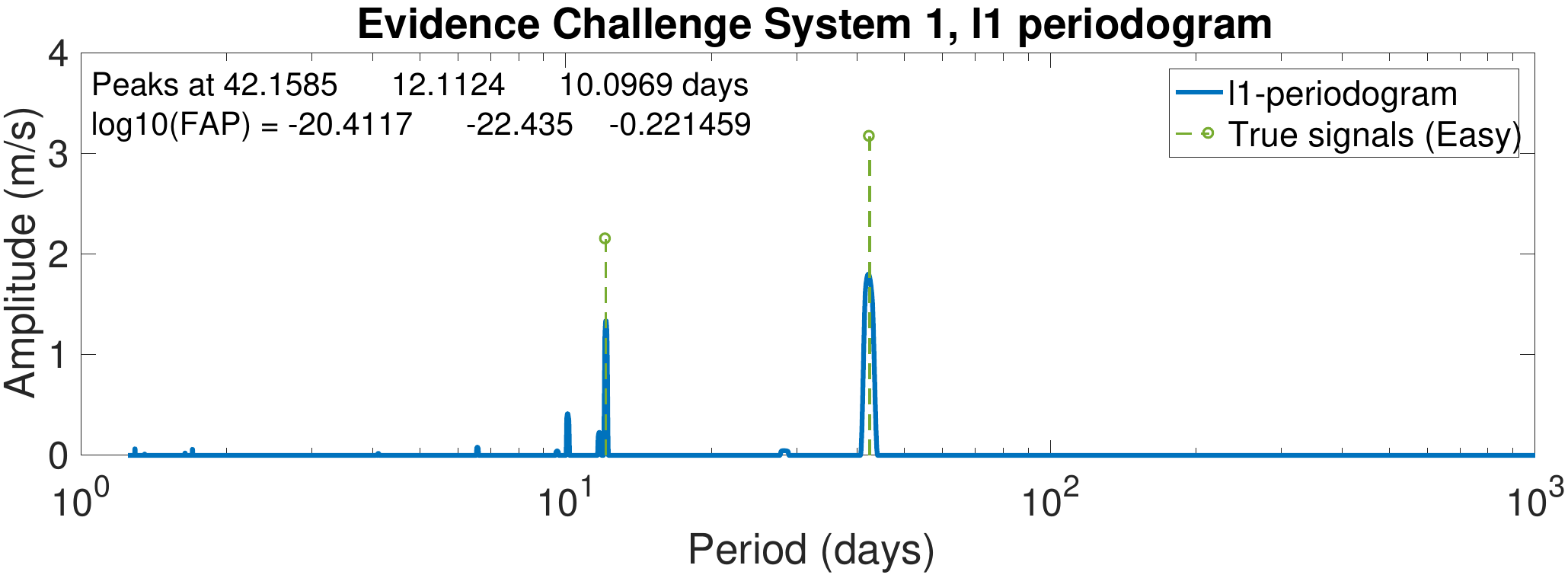} 

\includegraphics[width=12cm]{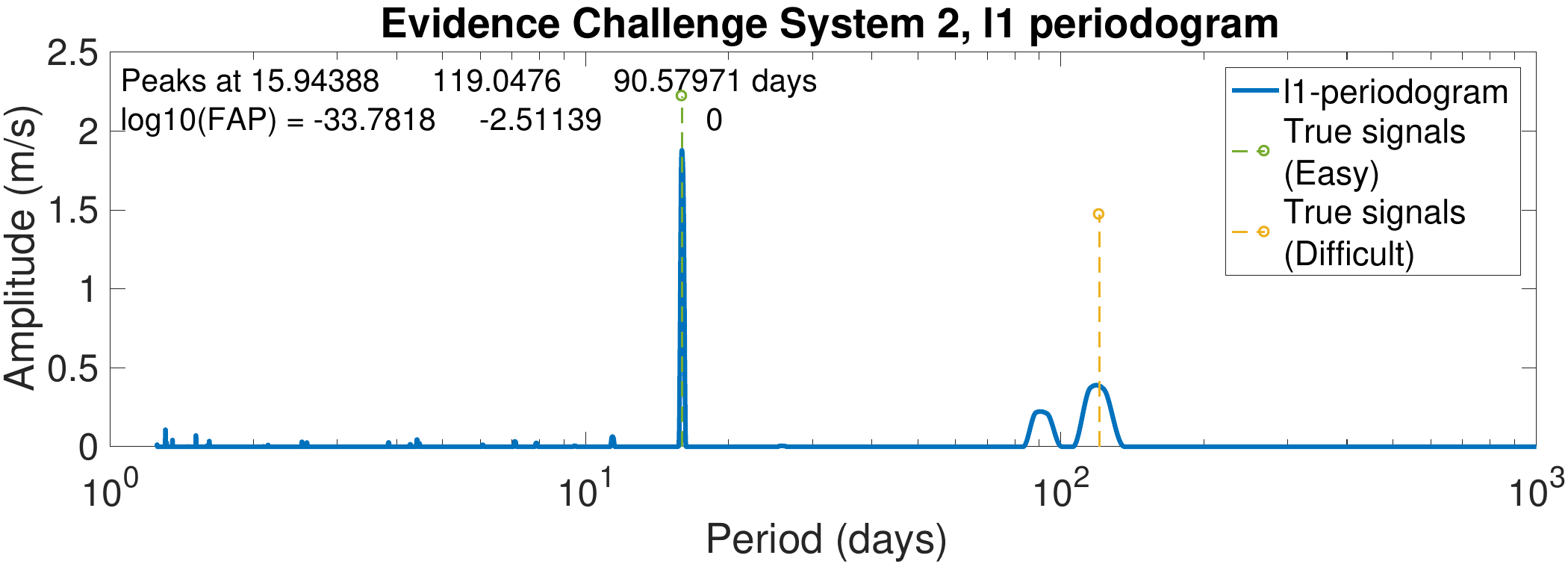} 

\includegraphics[width=7.5cm]{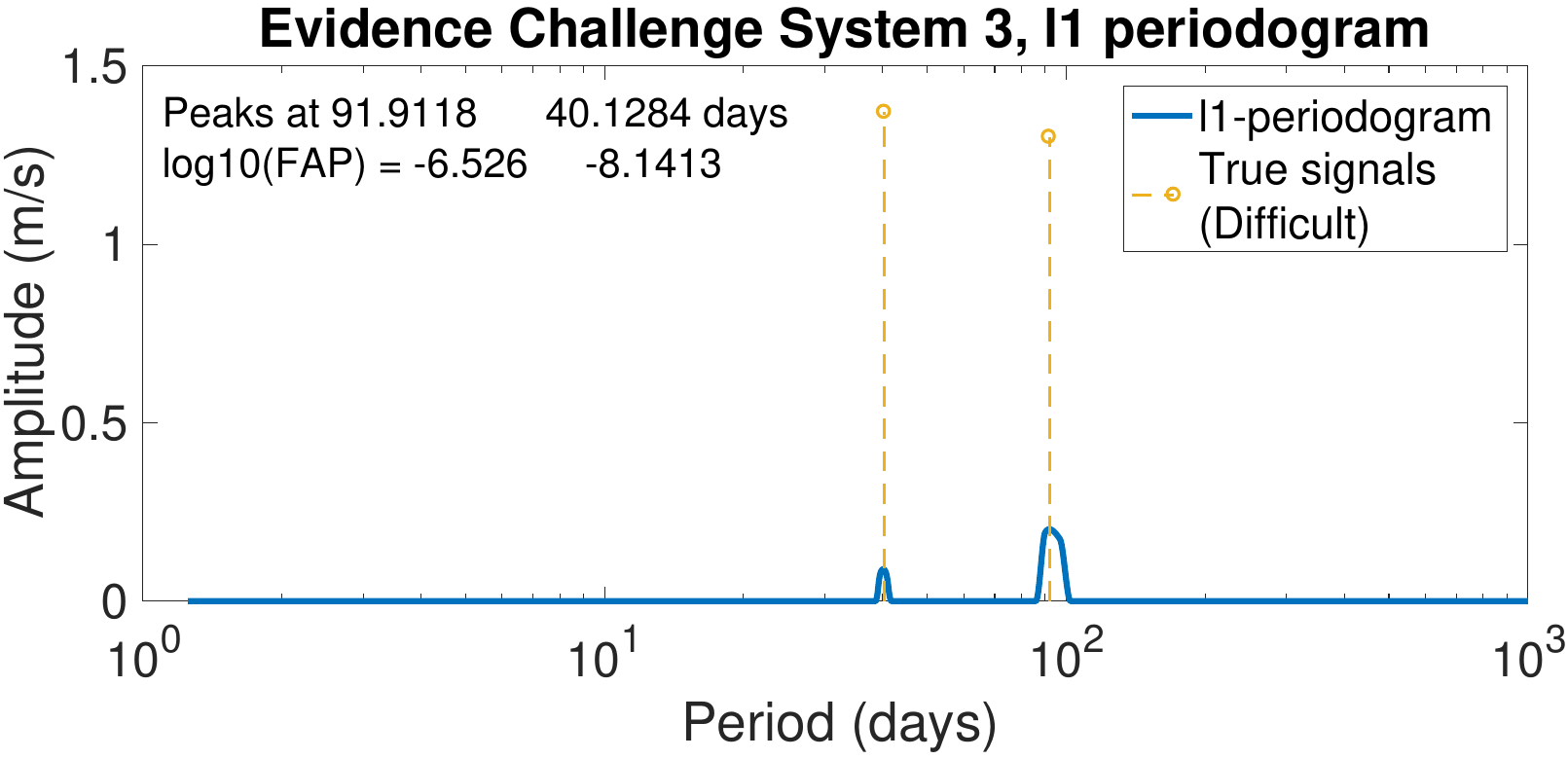} 
\includegraphics[width=7.5cm]{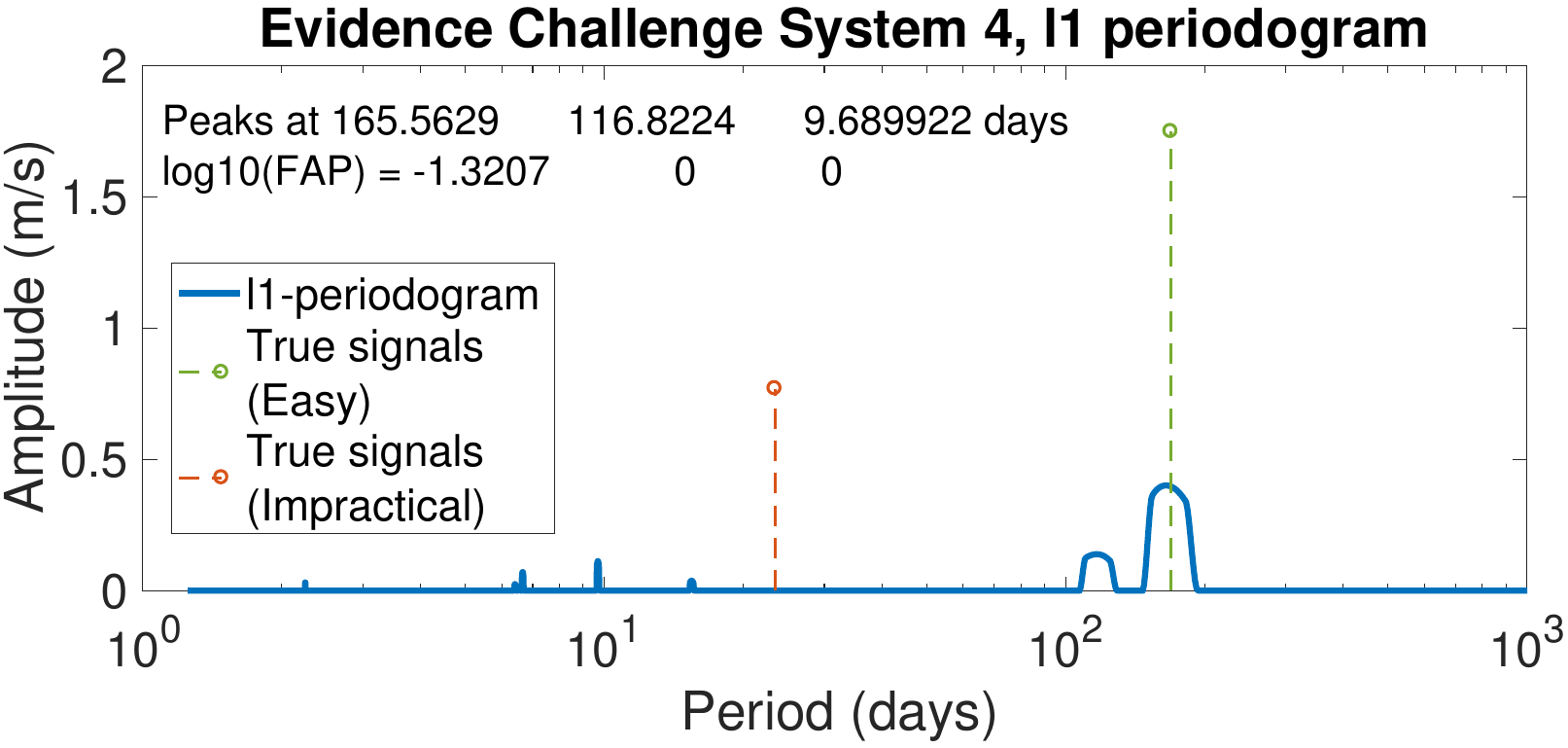} 

\includegraphics[width=7.5cm]{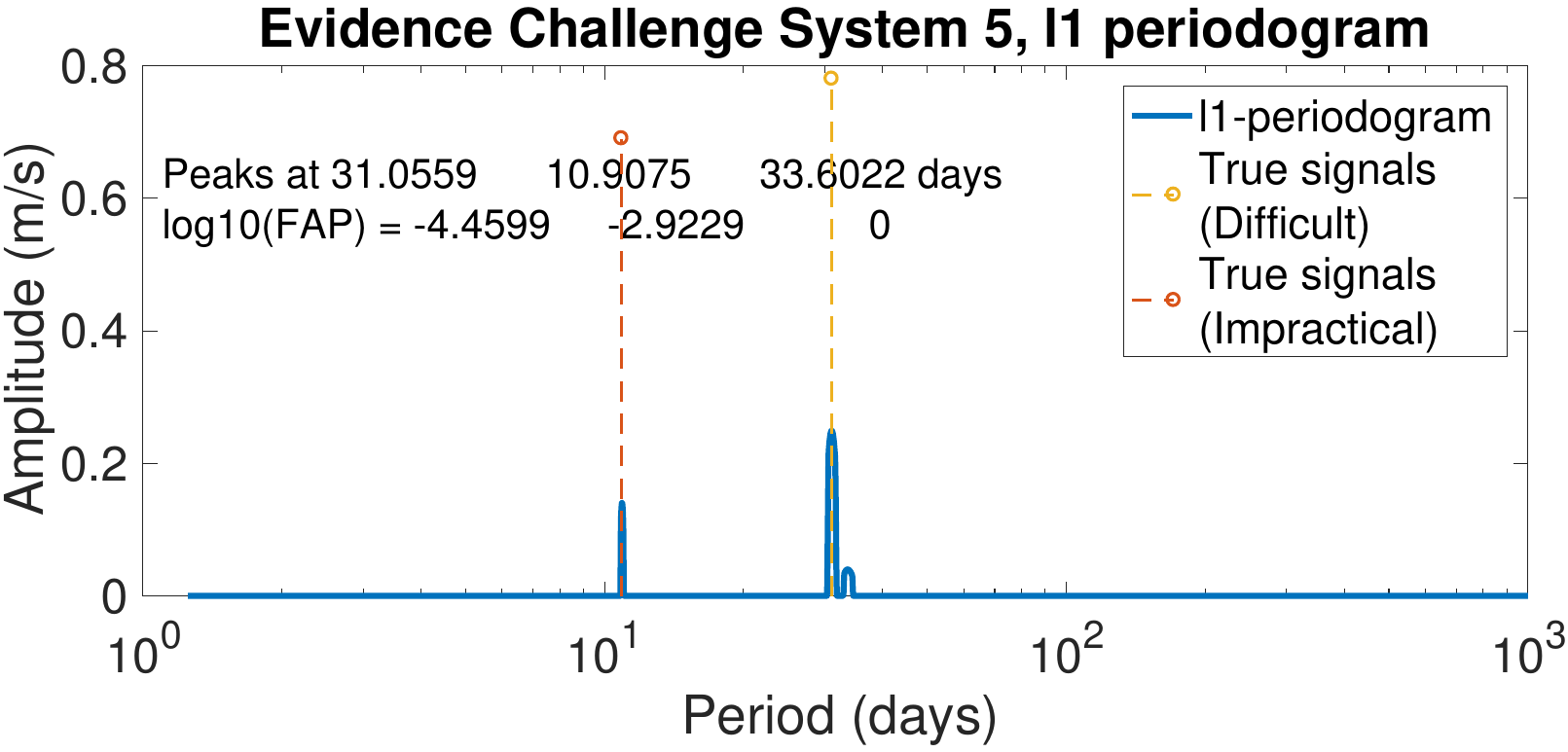} 
\includegraphics[width=7.5cm]{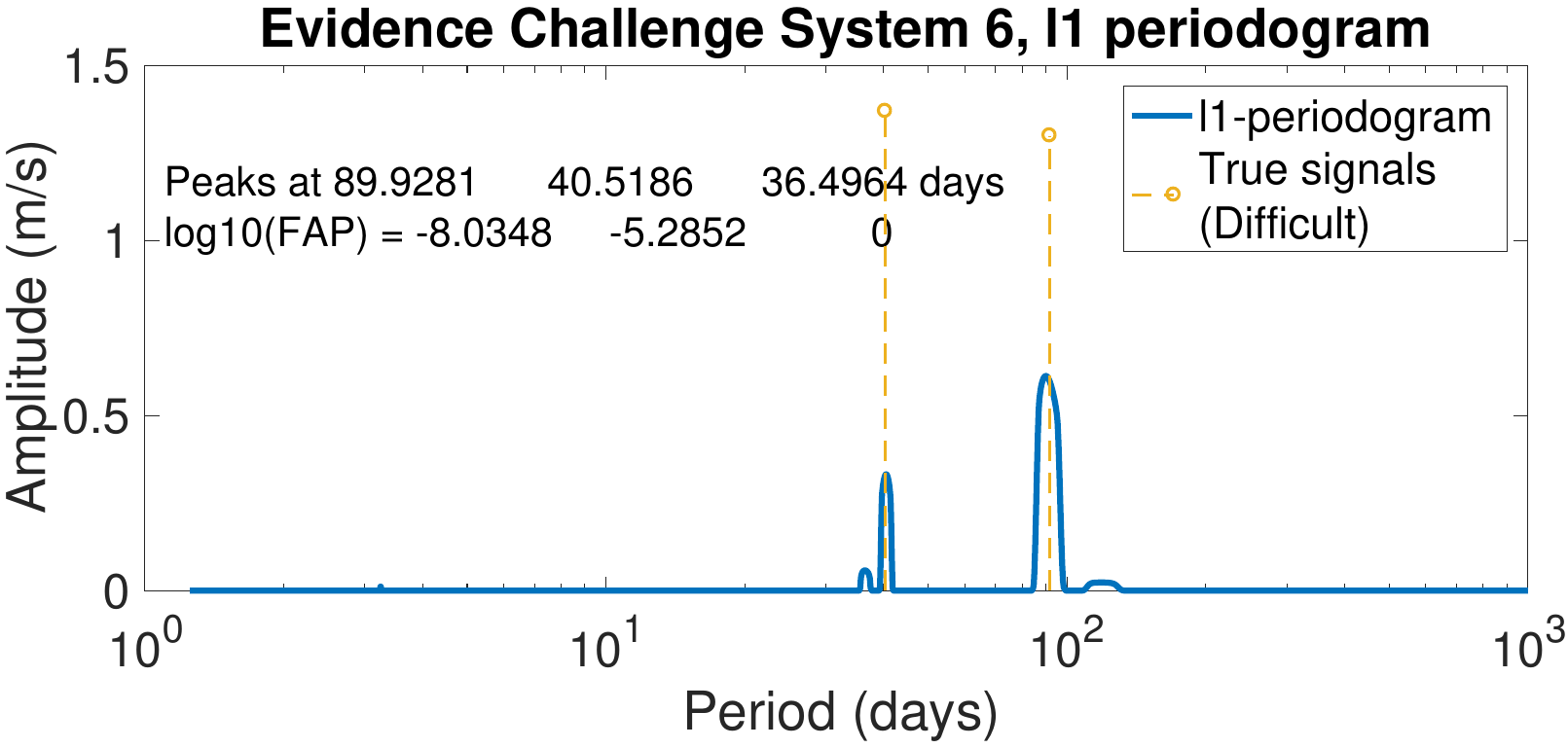} 
\caption{$\ell_1$ periodograms of the evidence challenge systems (in blue). The period and semi-amplitude of the injected signals are represented by the stems, whose color gives their detection difficulty as defined in section~\ref{sec:models}. The legend ``Peaks at...'' indicates the location of the two or three tallest peaks of the $\ell_1$ periodogram in order of decreasing amplitude. The legend ``log10(FAP)...'' gives the logarithm in base 10 of the false alarm probability of the signals at the periodicity given above. These figures were obtained before the true location of the periods was known. A version of them without the stems indicating the true signals is available on the GitHub page EPRV3EvidenceChallenge/Inputs/Hara/l1\_periodogram.}

	\label{fig:haraapp_fig}
\end{figure}

The result of our method is an estimate of the Fourier spectrum that we call $\ell_1$ periodogram.
Its plot can be read similarly to a classical periodogram, with a significance attached to each peak, but has much less peaks due to aliasing.
Figure~\ref{fig:haraapp_fig} shows the $\ell_1$ periodograms we obtain for the six systems of the evidence challenge (in blue).
The periods and semi-amplitudes of the true planets are given by the stems, with the level of difficulty of their detection in color code as defined in section~\ref{sec:models}.
For instance, on the system 1 the three main peaks are at 42.1, 12.1 and 10.01 days and have respective FAPs $10^{-20.4}$, $10^{-22.4}$ and $10^{-0.22}$, the true signals were two ``easy'' planets at 42.4 and 12.1 days.
The method is fast, that is it takes typically 5 - 10 s to run on each data set of this challenge, 20 - 30 s including the statistical significance assessment on an i7, 2.5GHz laptop processor.
After the challenge, some further work enabled us to bring these computation times on the evidence challenge datasets down to an average of 1.5 s for the $\ell_1$ periodogram calculation and 4.6 s including statistical significance assessment.
Note that more conservative values of the FAPs were obtained later, but we chose to plot figures that were publicly available before the results were unveiled.

How the plot is obtained and how the significance is computed are discussed respectively in section~\ref{sec:haraapp_basispursuit} and~\ref{sec:haraapp_statsig}. We discuss how our method fits in the present challenge in section~\ref{sec:haraapp_discuss}.

\subsubsection{Basis pursuit de-noising}
\label{sec:haraapp_basispursuit}

Let us denote by  $\vec{d_t}$ the data we would have obtained without noise, so that $\vec{d} = \vec{d_t} +\vec{n} $, $\vec{n} $ being the noise.
The variable we wish to estimate is the Fourier spectrum $x$ of $\vec{d_t}$. To obtain a finite sized variable,  we approximate $x$ it by its discretization on a fine grid of frequency equally spaced: $\vec{x} = (x(\omega_k))_{k=1..N}$ where $(\omega_k)_{k=1..N}$ span between 0 and $\Omega$ to be determined. The data then admits the following representation: $\vec{d_t} = \mathbf{A} \vec{x}$ where $\mathbf{A}$ is a $N_{\mathrm{obs}} \times 2N$  matrix whose entries are $A_{kl} = \cos \omega_k t_l$ for $l = 1..N$ and $A_{kl} = \sin \omega_k t_l$ for $l=N+1..N$, $l = 1..N_{\mathrm{obs}}$.

Obviously, $\vec{d_t}$ is unknown, we want to find an $\vec{x}$ such that $\mathbf{A} \vec{x}$ is close  to $\vec{d}$. For instance in the sense of the usual Euclidian norm, we can impose $ \|  \mathbf{A}\vec{x} - \vec{d} \|_{\ell_2}<\epsilon$ for some $\epsilon >0$,
where $\| \vec{z} \|_{\ell_2} = \sqrt{\sum_{k=1}^{2N} z_k^2 }$ for any $\vec{z} \in \mathbb{R}^{N_\mathrm{obs}}$.
 As said above, we know that the true signal contains only a few non zero frequencies (a few planets). It seems reasonable to search for an  $\vec{x}$  that satisfies the inequality and has as few non zero components as possible. Unfortunately, trying to minimize the number of non-zero components subject to a quadratic constraint is NP-hard~\citep{ge2011}. 
 
We use a proxy of the number of non-zero components of $\vec{x}$, which is the sum of the absolute values of the coefficients, $ \sum_{k=1}^{2N} |x(\omega_k)|=: \|\vec{x} \|_{\ell_1}$, also termed the $\ell_1$ norm of $\vec{x}$. So that we solve
\begin{equation}
\arg \min \|\vec{x} \|_{\ell_1}  \; \; \text{subject to}. \; \; \| \mathbf{W}( \mathbf{A}\vec{x} - \vec{d} )\|_{\ell_2} \leqslant \epsilon
\label{eq:hara_basispursuit}
\end{equation}
with $\mathbf{W} = \mathbf{\Sigma}^{-\frac{1}{2}}$, $\mathbf{\Sigma}$ being the covariance matrix of the noise.
 The quantity $\epsilon$ sets the trade-off between sparsity and agreement with observations. The minimization problem~\ref{eq:hara_basispursuit} is known as Basis Pursuit De-Noising in the signal processing literature~\citep{chen1998}. Other formulations of $\ell_1$ penalties are possible, for instance~\cite{bourguignon2007} use the Lagrange multiplier version of~\ref{eq:hara_basispursuit} for spectral estimation. Unlike the number of non zero components, the $\ell_1$ norm is a convex penalty function. Since the constraint $\| \mathbf{W}( \mathbf{A}\vec{x} - \vec{d} )\|_{\ell_2} \leqslant \epsilon$ defines a convex set, the problem~\ref{eq:hara_basispursuit} has only one local minimum and is fast to solve.

There are several algorithms written to solve~\ref{eq:hara_basispursuit}.
We selected SPGL1~\citep{vandenberg2008}.
Several parameters of the algorithms have to be tuned, such as the frequency grid width and spacing, the tolerance $\epsilon$.
In~\cite{hara2017} we provide a method to tune the algorithm parameters, we introduce the $\mathbf{W}$ matrix to take into account correlated noise and additional processing steps.
We then obtain a quantity $(x^\sharp(\omega_k))_{k=1..N}$ that can be plotted versus the frequency grid and gives an estimate of the Fourier spectrum, just like in Figure~\ref{fig:haraapp_fig} (in blue).
Note that the $\ell_1$ periodogram is used to find periodic candidates, this is not a good estimator of semi-amplitudes, which are underestimated due to the $\ell_1$ penalization in~\ref{eq:hara_basispursuit}.

Since several periodicities are searched at the same time, one can expect that the problem of aliases adding up together to give a spurious tallest peak is mitigated. Indeed, the number of misleading peaks is drastically reduced~\citep{hara2017}.
However, as in the case of the classical periodogram, peaks significances are to be determined.

\subsubsection{Significance}
\label{sec:haraapp_statsig}

We used two methods to evaluate the peaks significance, whose common feature is to test the improvement made by fitting a periodic signal  at the $n+1^{\mathrm{th}}$ tallest peak of the $\ell_1$ periodogram compared to fitting only the $n$ firsts.
For instance, on the system 1 of the Evidence Challenge (see figure~\ref{fig:haraapp_fig}, top) we compare the models with  a sinusoidal signal at 42.16 days (maximum peak) to nothing, then a model with two sines at 42.16 and 12.11 days to one signal at 42.16 day and so on.

The first way to proceed, as described in~\cite{hara2017}, is to compute the significance as if the period of the peaks had been found by a residual periodogram~\citep{baluev2008}.
These periodograms generalize the Lomb-Scargle one, and consist in comparing the likelihood of a model that constitutes the null hypothesis $H_0$ to a model with the $H_0$ model plus a sine function at a frequency $\omega$.
Here, we use the null hypothesis ``the signal contains $k$ planets at periods $P_1 ...P_k$'', and significance for an additional planet is tested. The value of the periodogram at frequency $\omega$ is  
\begin{equation}
P(\omega) =  \alpha \frac{\chi^2_{H_0, \omega} - \chi^2_{H_0}}{\chi^2_{H_0}} 
\label{eq:hara_periodogram}.
\end{equation}
where $\chi^2_{H_0}$ and $\chi^2_{H_0, \omega}$ are respectively the $\chi^2$ of the null hypothesis model, and the model with the null hypothesis plus a sinusoidal model at frequency $\omega$ and $\alpha$ is a positive constant.
To assess whether an additional periodic signal must be included in the model, one can compute the probability that the random variable  ``maximum of the periodogram'', $P_{\mathrm{max}}$, exceeds the maximum value of the periodogram of the data under the null hypothesis, that is the $p$-value
\begin{equation}
p = \mathrm{Pr}\{  P_{\mathrm{max}}\geqslant  \max\limits_\omega P(\omega) | H_0\}.
\label{eq:hara_pval}
\end{equation}
The assessment of the statistical significance of an $\ell_1$ periodogram peak can be done sequentially by using as the null hypothesis a model with sines at the $n$ tallest peaks. Denoting by $\omega_{n+1}$ the location of the  $n+1^{\mathrm{th}}$ tallest peak, we then use $P(\omega_{n+1})$ in place of $ \max_\omega P(\omega)$ in formula~\ref{eq:hara_pval}.
The values reported in figure~\ref{fig:haraapp_fig} are such $p$-values computed with formula 5 of~\cite{baluev2008}. 

The second significance testing method we used for this challenge is a Laplace approximation of the evidence at the period found, like in section~\ref{methods-ford}. We approximate the evidence of the $n$ planet model as in formula (5) of~\cite{kassraftery1995}
\begin{eqnarray}
\log \mathcal{Z}_n \approx  \log \mathcal{L}(\vec{d}|\widehat{\vec{\theta}_{n}}) + \log p_{n}(\widehat{\vec{\theta}_{n}}) + \frac{1}{2}\left( -\log |\widehat{\mathbf{I}_{n}}|  + d_{n} \log 2\pi \right)
\label{eq:hara_laplaceapprox}
\end{eqnarray}
where $d_n$ is the number of parameters of the model, $p_{n}$ is the prior on the parameters of an $n$ sines model and $\widehat{\mathbf{I}_{n}}$ is the information matrix evaluated at $\widehat{\vec{\theta}_{n}}$. The parameters $\widehat{\vec{\theta}_{n}}$ are fitted with a non-linear sinusoidal fit initialized at the periods of the $n$ tallest peaks of the periodogram. Note that the fit includes an error term in quadrature of the nominal errors in the maximum likelihood estimation. The Laplace approximation is here computed with an analytical formula we derived. 
The log odds ratio is then approximated by $\log B= \log \mathcal{Z}_{n+1} - \log \mathcal{Z}_n$. The approximated evidences and odds ratio are reported respectively in figures~\ref{fig:results-logZ} and~\ref{fig:results-logPOR}.

\subsubsection{Discussion}
\label{sec:haraapp_discuss}

Residual periodograms are robust tools with a well founded theory, but do not necessarily indicate correctly the period of the variation in the data.
The $\ell_1$ periodogram is thought as an alternative to residual periodograms, that has approximately the same computational workload but mitigates the aliasing problem (for details see~\cite{hara2017}).

Significance tests on basis pursuit solutions are a notoriously difficult problem. 
The present challenge constitutes a good test of applying FAPs or odds ratio, developed in other contexts, to test significance in our case. It seems reasonable since if there are planets, they will appear in general on the $\ell_1$ periodogram tallest peaks, and the remaining peaks will be noise. Significance tests such as FAPs or odds ratio should validate the signals until a peak due to noise is selected. The results of the challenge we obtain are consistent with this scenario.

\subsection{Nelson, Ratio Estimator (MCMC Importance Sampling)}
\label{sec:mcmcis}

Importance sampling is essentially a more general form of Monte Carlo integration to estimate $\mathcal{Z}$.
We multiply the numerator and denominator of the integrand in Equation \ref{eq-fml} by $g(\vec{\theta})$,
a distribution with a known normalization.
\begin{equation}
\mathcal{Z}=\int \frac{\mathcal{L}(\vec{\theta})p(\vec{\theta})}{g(\vec{\theta})} g(\vec{\theta}) d\vec{\theta}.
\label{eq-mcmcis-1}
\end{equation}
This does not change the value of $\mathcal{Z}$, but Equation \ref{eq-mcmcis-1} is in a convenient form such that $\mathcal{Z}$ can be estimated numerically by drawing $N$ samples from $g(\vec{\theta})$, 
\begin{equation}
\widehat{\mathcal{Z}} \approx \frac{1}{N}\sum\limits_{\vec{\theta}_i \sim g(\vec{\theta})}\frac{\mathcal{L}(\vec{\theta}_i)p(\vec{\theta}_i)}{g(\vec{\theta}_i)}.
\label{eq-mcmcis-2}
\end{equation} 

The efficiency of importance sampling depends strongly on the chosen $g(\vec{\theta})$.
Assuming our parameter space contains one dominant posterior mode, we choose a multivariate normal with mean vector $\vec{\mu}_g$ and covariance matrix $\vec{\Sigma}_g$ for $g(\vec{\theta})$.
For each model considered, we estimate $\vec{\mu}_g$ and $\vec{\Sigma}_g$ from a set of posterior samples obtained a Markov chain Monte Carlo (MCMC).

One good strategy with importance sampling is to pick a $g(\vec{\theta})$ that is heavier in the tails than $\mathcal{L}(\vec{\theta})p(\vec{\theta})$.
This makes it easier to sample from low probability parts of the posterior distribution and prevents any samples from resulting in extremely large weights.
However, the chance of sampling from the posterior mode will decrease with increasing dimensionality, which could lead to an inefficient and inaccurate estimate of $\widehat{\mathcal{Z}}$ \citep[see a discussion of the ``typical set'' in][]{betancourt2017}.
One way around this is to sample from $g(\vec{\theta})$ within some truncated subspace, $\mathcal{T}$.
This new distribution $g_\mathcal{T}(\vec{\theta})$ is proportional to $g(\vec{\theta})$ inside $\mathcal{T}$ and renormalized to be a proper probability density.
Thus, Equation \ref{eq-mcmcis-2} can be rewritten as
\begin{equation}
f \times \widehat{\mathcal{Z}} \approx \frac{1}{N}\sum\limits_{\vec{\theta}_i \sim g_\mathcal{T}(\vec{\theta})}\frac{\mathcal{L}(\vec{\theta})p(\vec{\theta}_i)}{g_\mathcal{T}(\vec{\theta}_i)}.
\label{eq-mcmcis-3}
\end{equation}
where $f$ is a factor that specifies what fraction of $\mathcal{L}(\vec{\theta}_i)p(\vec{\theta}_i)$ lies within $\mathcal{T}$.
We can estimate $f$ with the previously mentioned MCMC sample.
By counting what fraction of our posterior samples fell within $\mathcal{T}$, $f_{MCMC}$, we can rearrange Equation \ref{eq-mcmcis-3} to give us $\widehat{\mathcal{Z}}$.
\begin{equation}
\widehat{\mathcal{Z}} \approx \frac{1}{N \times  f_{MCMC}}\sum\limits_{\vec{\theta}_i \sim g_\mathcal{T}(\vec{\theta})}\frac{\mathcal{L}(\vec{\theta}_i)p(\vec{\theta}_i)}{g_\mathcal{T}(\vec{\theta}_i)}.
\label{eq:FMLfinal}
\end{equation}
There are two competing effects when choosing the size of our subspace $\mathcal{T}$.
If $\mathcal{T}$ is large (i.e. occupies nearly all of the posterior distribution), then $f_{MCMC}$ approaches 1 and we return to a basic importance sampling algorithm that may not be efficient in high dimensions.
If $\mathcal{T}$ occupies a much smaller region, then we are more likely to sample from near the posterior mode, but $f_{MCMC}$ approaches 0, making it difficult to accurately estimate $\widehat{\mathcal{Z}}$.
This necessitates carefully choosing an appropriate $\mathcal{T}$ that will provide a robust estimate for $\widehat{\mathcal{Z}}$.
\citet{guo2012} and \citet{nelson2016} provide more detailed prescriptions and investigations of this method.

Here, we compute $\widehat{\mathcal{Z}}$ for all models using small (1.5) and large (30) truncated subspace.
Our parameterization for $g(\vec{\theta})$ is $P$, $K$, $\sqrt{e}\sin\omega$, $\sqrt{e}\cos\omega$, and $\omega+M$ for each planet, and $C$ and $\sigma_J$ for the zero-point offset and jitter respectively.
We run 20 independent MCMCs per model per dataset and compute a $\widehat{\mathcal{Z}}$ value based on every MCMC.
We report the median and standard deviation for each set of 20 $\widehat{\mathcal{Z}}$ values.

\subsection{D\'iaz, Perrakis}
\label{methods-diaz}
\label{sec:perrakisis}
The Perrakis estimator is an importance sampling estimator described in detail in \citet{perrakis2014}. The importance sampling density used  is the product of the marginal posterior distributions of parameter blocks. In our case, we chose one-dimensional blocks, so that the importance sampling function is:
$$
g(\vec{\theta}) = \prod_{i=0}^D p(\theta_i|\vec{d})\;\;,
$$
so that the samples are drawn from the marginal posterior distributions:
$$
\theta^{(n)}_i \sim p(\theta_i|\vec{d})\;\;\text{for }i = {1, 2, ..., D}\;\;,
$$
for a $D$-dimensional model. This produces the estimator

\begin{equation}
\widehat{\mathcal{Z}} = N^{-1}\sum_{i=0}^{N} \frac{p(\vec{d}|\theta_1^{(n)}, \theta_2^{(n)}, ..., \theta_D^{(n)})p(\theta_1^{(n)}, \theta_2^{(n)}, ..., \theta_D^{(n)})}{\prod_{j=0}^D p(\theta^{(n)}_j|\vec{d})}\;\;.
\label{eq.perrakis}
\end{equation}

The estimate can be computed based on joint posterior samples drawn using, for example, an MCMC algorithm, but requires two additional elements: draws from the marginal posterior distributions of the parameter blocks, and an estimate of the marginal densities that appear in the denominator of Eq.~\ref{eq.perrakis}.
The former is promptly obtained by shuffling the elements of the parameter vector across MCMC samples.
This breaks the correlation between parameters and leads to samples which are drawn from the product of (independent) marginal posteriors.
More details and discussion on this is given in \citet{perrakis2014}.

As we used one-dimensional parameter blocks.
The marginal posterior densities are approximated by the corresponding normalized histogram. Of course, to obtain a precise estimate, a large posterior sample and small bin sizes are required.
However, we checked that the result does not change significantly with bin size.
This estimation could be improved by modelling the marginal distributions or using a kernel density estimation.

The resulting estimate, which we named Perrakis estimator, was previously employed in the analysis of exoplanet data in a number of articles \citep[e.g.][]{diaz2016a, diaz2016b, bonfils18}).
Here, we obtained sample of size 5000 from the importance sampling function to perform the computation of the evidence estimate, $\widehat{\mathcal{Z}}$.
The uncertainty was computed by repeating the computation 600 times.
At each time, a new sample is considered: a new subsample of the joint posterior distribution is taken and a new shuffling is performed.

The joint posterior sample was obtained using the affine-invariant ensemble sampler \citet{Goodman2010} implemented by \citet{emcee}.
For each dataset and model, we ran 300 walkers for 30000 iterations.

\subsection{Team PUC, Variational Bayes with Importance Sampling}
\label{sec:VarBayes}

Johannes Buchner used an integration algorithm based on variational Bayes (VB) and Importance Sampling.
The method is very similar to the one described in \cite{Beaujean2013} and uses their \sw{pypmc} package \citep{pypmc}.

The method proceeds as follows:

\begin{enumerate}
\item Identify likelihood maxima to guess a initial mixture. The original technique used points from several MCMC chains. Here, a single \MULTINEST\ run (see §\ref{sec:method-multinest}) is used to obtain initial posterior points. This just serves to identify an initial mixture density and does not rely on \MULTINEST\ sampling correctly. The posterior points are divided into groups based on their likelihood value and clustered further into subgroups. This is analoguous to multiple MCMC chains split into segments in \cite{Beaujean2013}.
\item Generate an initial Gaussian mixture density from the above groups. The intent is to develop a mixture that closely describes the posterior well so that importance sampling is efficient.
\item Run Variational Bayes to optimize the proposal mixture density against the posterior points.
\item Define an Importance Sampler based on the optimized mixture. Set $N$ to 1000 times the number of model parameters.
\item Loop:
\begin{enumerate}
\item Draw $N$ importance samples from the mixture and evaluate their likelihood.
\item If the importance sampling integral uncertainty is below the threshold $\sigZ<0.3$ and the effective sampling size is above 100, terminate.
\item Otherwise: Increase $N$ by a factor of 1.4. This implies that the total number of samples drawn increases exponentially.
\item Update the proposal mixture density with Variational Bayes.
\item In every third loop, the previous step is not done. Instead, the proposal mixture density is recreated from scratch (as above), but with one more point group. That group is created by starting a simple MCMC chain from the point with the highest weight, after a simple likelihood optimization.
\end{enumerate}
\end{enumerate}

Iteratively optimizing with Variational Bayes is effective in making the importance sampler efficient and improves the integration uncertainty.
However, a limitation is that the number of mixture components cannot increase.
If importance sampling discovers a new small peak, VB typically does not place a component there. To solve this, step 5e recreates the mixture from scratch (with up to 10 components).
The local MCMC run helps identifying the size of the potential new component.
In the subsequent iteration, all previous samples are used to optimized the mixture, and the number of components can shrink again (often drastically).

We also include a long run from this algorithm, where we initialise from the combination of 10 \MULTINEST\ preruns (to mitigate the problems named in the \MULTINEST\ section), higher number of importance samples, and integrate to a higher effective sampling size (20,000) before terminating.
At the cost of many likelihood evaluations, this should be safer.
For some datasets, this stringent termination criterion was never reached and the runs were terminated manually.

\subsection{Rajpaul, MCMC Nested Sampler}
\label{sec:mcmcns}
\label{sec:nestedsampling}

Nested sampling is a technique developed by \citet{Skilling2004,Skilling2006} for Bayesian model comparison via estimation of Bayesian evidence integrals.
As nested sampling produces samples from the posterior PDFs of model parameters as a trivial byproduct of the evidence integral estimation, it may be thought of as a reversal of the usual approach to Bayesian inference.
Although Skilling's original formulation was designed with Bayesian inference in mind, nested sampling is in fact a general method for numerical integration that may be applied to any continuous integrals.

Nested sampling proceeds by exploring the volume above a given likelihood threshold.
That threshold is continuously increased, such that the volume decreases by a constant factor (exponential shrinkage).
This allows nested sampling to keep track of the volume and likelihood value for making a Lebesgue integral.
At a late point, the volume is small and the likelihood flat, so that the remainder does not contribute to the integral, and the algorithm terminates.

The shrinkage of nested sampling is achieved by having e.g.\ $100$ live points sampling the prior space uniformly and then removing one.
This reduces the represented volume by $\sim 1/100$. Next, the algorithm samples a new point with a likelihood higher than the removed point.
The number of live points therefore determines the speed of the shrinkage and how coarsely the space is sampled.

The error of the integral estimate is given in \citet{Skilling2004}.
The usual implementation assumes that the bulk of the integral can be found around some shrinkage (rather than multiple); in practice this is a sufficient approximation.

Internally, nested sampling requires an algorithm for drawing a new, random point from the prior with the condition that its likelihood is higher than the current likelihood threshold.
Several general solutions for these constrained drawing algorithm exist, including those relying on local steps (e.g., MCMC, Galilean Monte Carlo, HMC, \sw{POLYCHORD} -- and those reconstructing the volume enclosed by the likelihood contour (e.g., \MULTINEST, \sw{RADFRIENDS}).
See \citet{Buchner14} for a more detailed discussion.

Here, Rajpaul implemented the MCMC sampler from \citet{Veitch2010} to generate samples within a standard nested sampling routine.
This implementation is different from \MULTINEST\ (see below) in that it replaces the clustering algorithm or ellipsoidal rejection schemes with a semi-adaptive MCMC exploration of the prior range.
In particular, the present implementation used a mixture of the following proposal schemes to draw new samples: a Student-$t$ distribution (with $\nu=2$ degrees of freedom) based on the Cholesky-decomposed covariance matrix of the live points; differential evolution using two randomly-selected points from the current live points; and affine-invariant walk and stretch moves \citep[see][]{Goodman2010}. 

The algorithm as presented by \citeauthor{Veitch2010} has two main parameters that can be adjusted: $N$, the number of live points, and $M$, the number of MCMC iterations.
By tuning $N$ and $M$, any desired level of evidence accuracy can (in principle) be achieved, albeit at the expense of increasing computational burden, with the total number of likelihood evaluations scaling linearly with both $N$ and $M$.
Based on recommendations given by \citeauthor{Veitch2010}, and to strike a balance between a reasonable computation time and (ostensible) accuracy, Rajpaul fixed $N=1000$ and $M=1000$, such that estimation of a given model's evidence would require of order $10^6$ likelihood evaluations.

Rajpaul noted a priori that his own experience was that \MULTINEST\ was typically faster and better-suited to higher-dimensional ($>10$-dimensional) problems than the above algorithm due to \citeauthor{Veitch2010}.
Nevertheless, the MCMC sampler from \citeauthor{Veitch2010} was implemented for this evidence challenge to provide a foil to the more popular \MULTINEST\ nested sampling algorithm, discussed below.

\subsection{Team PUC, MultiNest}
\label{methods-teampuc}
\label{sec:method-multinest}

Team PUC (Johannes Buchner and Surangkhana Rukdee from Pontificia Universidad Cat\'olica de Chile) employed nested sampling with the constrained drawing algorithm \MULTINEST.
{\MULTINEST}'s multi-modal ellipsoidal sampling \citep{Shaw2007,Feroz2009}, encloses the existing random points into best-fitting ellipsoids.
These are enlarged by a certain factor (inverse of the efficiency parameter).
New points are drawn from the enlarged ellipsoids, and rejected if below the likelihood threshold.
Therefore the ellipsoids reduce the space to be sampled, making \MULTINEST\ fast (in terms of number of likelihood evaluations needed).
However, if the ellipsoids accidentally cut away parameter space regions, e.g., because the enlargement is too small or the contours do not look similar to ellipsoids, the estimate can be biased.

\begin{figure}
\centering
\includegraphics[trim={2.7cm 4cm 3cm 4cm},clip,width=0.98\textwidth]{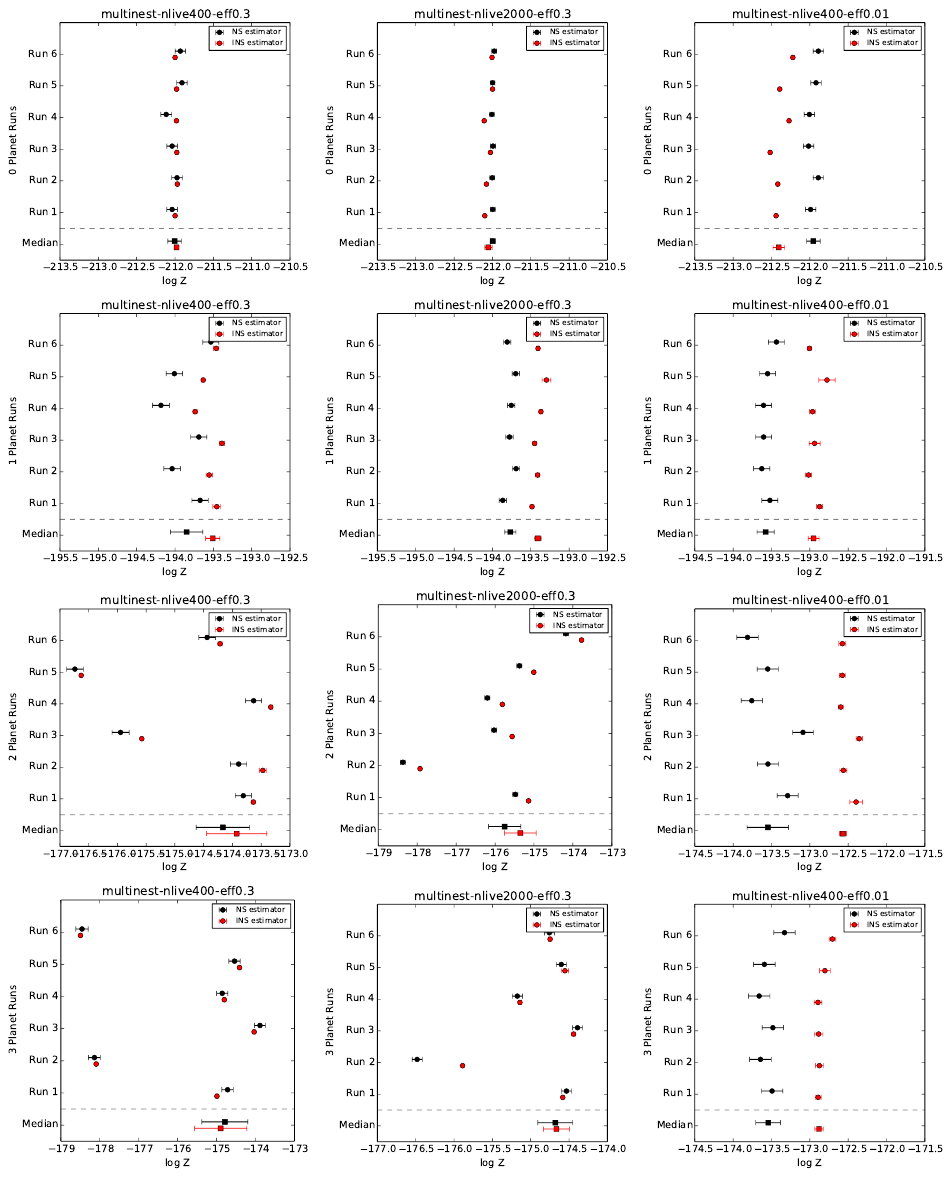}
\caption{Scattering of \MULTINEST\ \logZ\, estimates from runs against dataset 1. Panels show our three \MULTINEST\ configurations (columns) and number of planets used (rows). The Nested Sampling (NS) and Importance Nested Sampling (INS) estimates are shown in black and red, respectively. Scattering between estimates is often larger than the quoted uncertainties. Also, there are outliers. The multirun estimator (median) is shown at the bottom of each plot.}
\label{fig:evi01}
\end{figure}

\begin{figure}
\centering
\includegraphics[trim={2cm 3cm 3cm 3cm},clip,width=0.98\textwidth]{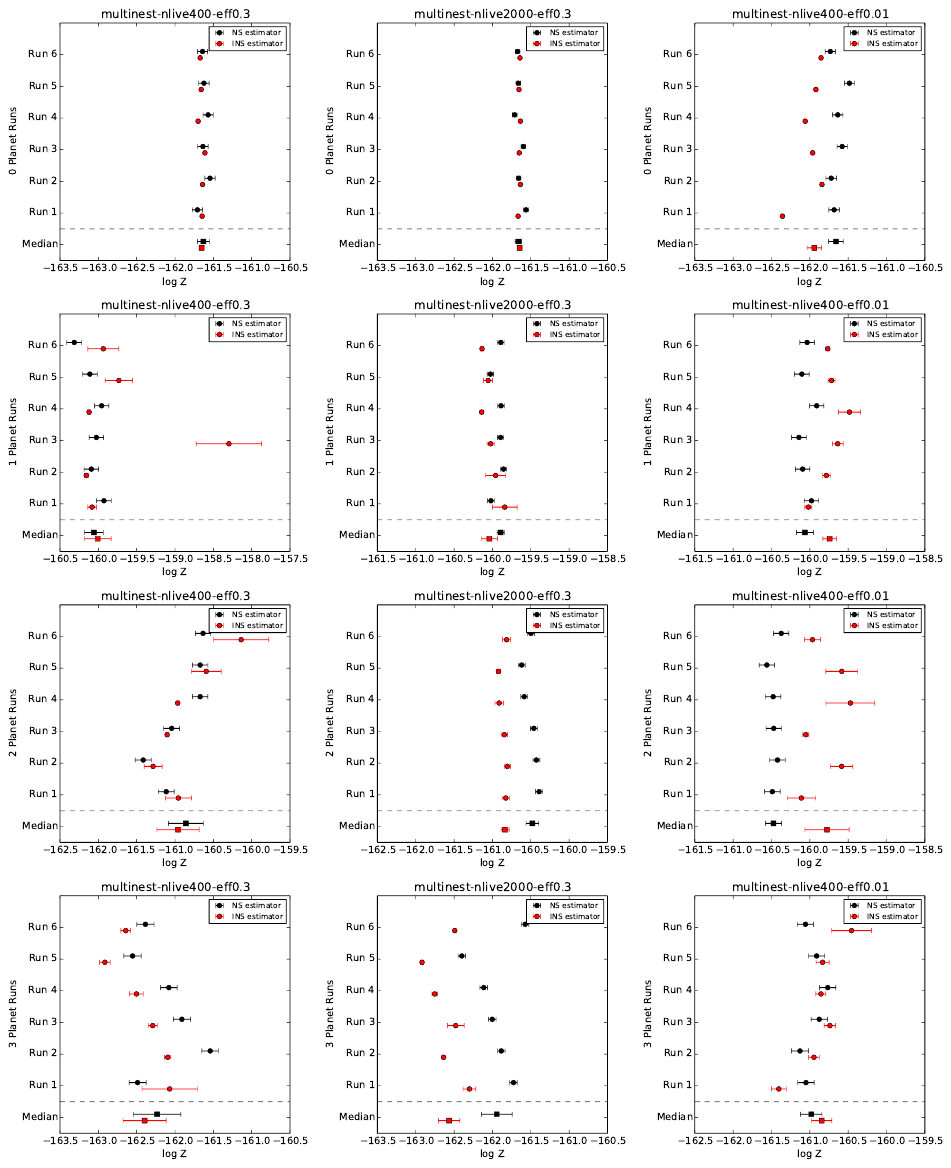}
\caption{Same as Figure~\ref{fig:evi01}, but for dataset 4.}
\label{fig:evi04}
\end{figure}

\subsubsection{Algorithm variations}

\MULTINEST\ has two parameters, the number of live points $\texttt{nlive}$ and the target efficiency $\texttt{eff}$ (inverse of the ellipsoid enlargement).
We chose a standard configuration (\texttt{multinest-nlive400-eff0.3}) and two variations, increasing either the number of live points (\texttt{multinest-nlive2000-eff0.3}) or the enlargement (\texttt{multinest-nlive400-eff0.01}). 

Importance Nested Sampling is a modification of Nested Sampling where the rejected points can improve the estimate \citep{Cameron2013,Feroz2013}.
To some degree, this also mitigates the above-mentioned issues of imperfect ellipsoid sampling.
\MULTINEST\ computes both the standard nested sampling estimator and the importance nested sampling estimator.
The results are named correspondingly (multinest-ins-nlive400-eff0.3, multinest-ins-nlive400-eff0.01, multinest-ins-nlive2000-eff0.3).

\subsubsection{Scatter between \MULTINEST\ runs}
\label{methods-teampuc-multinestscatter}

We observe that there is substantial scatter and outliers in the evidences between \MULTINEST\ runs.
Figure \ref{fig:evi01} and \ref{fig:evi04} shows the scatter and assigned errors for repeated runs of Dataset 1 and Dataset 4 respectively. Panel columns represent the three \MULTINEST\ configurations and panel rows show different number of modeled planets.
Each panel shows the comparison between Nested Sampling (NS) estimator and Importance Nested Sampling (INS) estimator for six runs. In most cases, the INS estimator gives a smaller error bar to compare to the NS estimator.
However, it sometimes shows outliers; for example, in the Dataset 4 (Figure~\ref{fig:evi04}) Run 3 with one planet Increasing $\texttt{nlive}$ from 400 (left column) to 2000 (middle column) yields smaller errors.
Decreasing the efficiency from 0.3 to 0.01 (right column) gives systematic offsets between NS and INS estimators.

Throughout, the quoted uncertainties of \MULTINEST\ are smaller than scatter between runs.
Low outliers can come from undiscovered solutions, but increasing the number of live points did not eradicate this completely.
Imperfect ellipsoids can also lead to scatter in the estimate.
Indeed, decreasing the efficiency also decreases the scatter, but at great computational cost. Using the INS estimator instead of the standard NS generally leads to overly small uncertainties.
One conclusion is that running \MULTINEST\ just once gives unreliable uncertainty estimates, which can not completely eradicated by decreasing the efficiency or increasing the number of live points.

To represent this additionally uncertainty in \MULTINEST, we define a multirun estimator.
We ran \MULTINEST\ six times and combine the evidence estimate as the median of individual estimates: $$\logZ =\mathrm{median}(\logZ_i)$$ The multirun error is defined as the median of the absolute deviations and the median individual error estimates added in quadrature:  
$$\sigZ^2=\mathrm{median}(\sigma_i)^2 + \mathrm{median}(|\logZ_i-\logZ|)^2$$
This gives appropriate errors when \MULTINEST\ is having trouble and shows substantial scatter, yet is robust against individual outliers.
The bottom of each panel of Figure \ref{fig:evi01} and \ref{fig:evi04} shows our \MULTINEST\ multirun estimators.

\subsection{Faria, Diffusive Nested Sampling}
\label{method-faria}
\label{sec:dnest}

One of the main challenges with the Nested Sampling algorithm is to generate new particles from the likelihood-constrained prior.
As described above, a number of methods have been proposed for this (and used in the current work).
However, some of those methods, and Nested Sampling in general, tend to suffer from the curse of dimensionality, with sampling efficiency decreasing rapidly with the dimension of the parameter space.
This is particularly problematic if the posterior distribution is multimodal or highly correlated.
\citet{Brewer2011} introduced a new algorithm, which they called Diffusive Nested Sampling (DNS), designed to be as flexible and general as a more standard MCMC, but also capable of efficiently exploring difficult constrained distributions.
The algorithm introduces a slight but important improvement to the classic Nested Sampling approach, in that it attempts to sample from a \emph{mixture} of successively constrained distributions, instead of using one single hard constraint at each step.

DNS starts by generating a particle from the prior (call this distribution $p_{\mathcal{L}_0}$) and evolving it with an MCMC, storing all the intermediate likelihood values.
After a given number of iterations, it finds the $1-e^{-1} \sim 63\%$ quantile of all the likelihood values, and records it as $\mathcal{L}_1$; this creates a new \emph{level} occupying about $e^{-1}$ times the mass of $p_{\mathcal{L}_0}$.
All the likelihood values lower than $\mathcal{L}_1$ are then discarded.
At this point, (classic) Nested Sampling would continue sampling from the prior constrained to $\mathcal{L}_1$ (call it $p_{\mathcal{L}_1}$).
In contrast, DNS attempts to sample from a weighted sum of the two distributions $p_{\mathcal{L}_0}$ and $p_{\mathcal{L}_1}$.
An MCMC is used to evolve the particle with this mixture of distributions as the target, and once enough samples have been obtained from $p_{\mathcal{L}_1}$, we again find the $1-e^{-1}$ quantile of all the likelihood values, and record it as $\mathcal{L}_2$.
Likelihood values smaller than $\mathcal{L}_2$ are removed. 
The particle then explores a mixture of $p_{\mathcal{L}_0}$, $p_{\mathcal{L}_1}$, and $p_{\mathcal{L}_2}$ and this process continues until a maximum number of levels is created.

Once all the levels have been obtained, the particle simply continues to explore the mixture of all the levels until the algorithm is terminated.
In order to create the mixture of distributions, we need to provide a weighting scheme for each component.
Simple uniform weights for all distributions would work, albeit inefficiently.
\citet{Brewer2011} proposed exponentially-decaying weights with a scale length $\Lambda$, which describes how far (down in likelihood) the particle is able to go in order to explore more freely.
When the desired number of levels has been created, the weights can be changed to uniform, and further samples are drawn from all the component distributions.
The algorithm can then continue to sample for as long as required, with the evidence and posterior samples converging to their true values.
Each time a new level is created, its constrained distribution covers about $e^{-1}$ times as much prior mass as the last distribution. Therefore, the $X$-value of the $k$th level can be estimated as $\exp(-k)$.
However, as the levels are being created, their actual $X$-values can be modified from this theoretical expectation.
This means that the weight of each distribution is actually different and the exploration is thus not completely correct.
The $X$-values can nevertheless be corrected. 
At a given level $k$, the values of the likelihood will be higher than the upper level's likelihood cut-off a fraction $X_{k+1} / X_k$ of the time.
Thus, we can use the actual fraction of samples in which this happens as an estimate of the true ratio of the $X$-values for consecutive levels.

In summary, the DNS algorithm is essentially an application of the Metropolis-Hastings algorithm to a distribution other than the posterior.
Changing the target distribution improves upon other MCMC algorithms by providing the value of the evidence in one single run and being less sensitive to the presence of complicated features in the posterior.
Classic Nested Sampling also shares these advantages, but DNS improves upon the classic algorithm by alleviating the problem of sampling from the likelihood-constrained prior.
Because the target distribution used by DNS always includes the prior distribution as one of the components of the mixture, sampling from posteriors with substantial multimodality is still possible and even efficient.

\subsubsection{Details} 

In this work, Faria used the DNS algorithm implemented in the \sw{DNest4} package \citep{Brewer2016}.
The specific application of \sw{DNest4} to the exoplanet problem is implemented in a new open-source package called \texttt{kima} (Faria et al. in prep.).
The code allows to calculate the posterior distribution for the orbital parameters, and the value of the evidence for a model $\mathcal{M}_n$ with $n$ planets.

The DNS algorithm has a few options, which need to be set for each run.
We set the scale length $\Lambda$ to $25$ and require $500$ samples from the consecutively constrained distributions before creating a new level.
The maximum number of levels is determined automatically by \sw{DNest4} \citep[see][]{Brewer2011}.
For all the simulated datasets, we obtained $100\,000$ samples from the DNS target distribution.
This corresponds to different numbers of posterior samples for each dataset, and for each model.

In the DNS algorithm, there is no explicit global search step as the algorithm is always free to explore the full prior volume.
This means that once the settings mentioned above are fixed, the results were computed automatically for all datasets, without any dataset-dependent input.

For the analysis with constrained priors for the orbital period, the prior pdf was set to 0 outside of the provided period bounds.
Inside the bounds, the prior is still a Jeffreys between $1.25$ and $10^4$ days.

The error we report for the evidence value is calculated from one single run, by probabilistic re-assignment of X-values to the samples, as in standard nested sampling \citep[see][]{Brewer2011}.
These errors are likely to be over-optimistic.





\bibliographystyle{aasjournal}
\bibliography{refs}



\end{document}